# Origin, Experimental Realization, Illustrations, and Applications of Bessel beams: A Tutorial Review

A. Srinivasa Rao[1,2,3*]

[1]Graduate School of Engineering, Chiba University, Chiba, 263-8522, Japan
[2]Molecular Chirality Research Centre, Chiba University, Chiba, 263-8522, Japan
[3]Institute for Advanced Academic Research, Chiba University, Chiba, 263-8522, Japan
E-mail: *asvrao@chiba-u.jp, sri.jsp7@gmail.com

**Abstract**
Over the past 36 years much research has been carried out on Bessel beams (BBs) owing to their peculiar properties, *viz* non-diffraction behavior, self-healing nature, possession of well-defined orbital angular momentum with helical wave-front, and realization of smallest central lobe. Here, we provide a detailed tutorial review on BBs from their inception to recent developments. We outline the fundamental concepts involved in the origin of the BB. The theoretical foundation of these beams was described and then their experimental realization through different techniques was explored. We provide an elaborate discussion on the different kinds of structured modes produced by the BB. The advantages and challenges that come with the generation and applications of the BB are discussed with examples. This tutorial review provides reference material for readers who wish to work with non-diffracting modes and promotes the application of such modes in interdisciplinary research areas.

## 1. Introduction

Since the inception of laser sources, various kinds of structured modes have been brought into existence in laser beams by modulating their intensity in the transverse cross-section in a controlled manner under transverse phase engineering. In addition to wavelength, the ability to control and modulate the amplitude, phase, and polarization distributions of structured modes enhanced their applications in modern science and technology. The well-developed and advanced experimental configurations have generated several types of structured modes. However, only a few structured modes got attention owing to their tremendous applications in science and they have become well-known structured modes in optical society [1-10]. These structured modes based on their nature can broadly classified into two types. The first type of modes is fundamental structured /generalized Gaussian modes: Laguerre Gaussian (LG) modes, Hermite-Gaussian (HG) modes, and Ince-Gaussian (IG) modes, etc. [11-15]. These modes are eigenmodes and have stable propagation with self-similarity. In addition, the superposition of these eigenmodes with the same mode number produces eigenmodes and with different mode numbers creates non-eigenmodes. The characteristics of eigenmodes and non-eigenmodes can be understood through the investigation of their dynamical properties in the presence of focusing. Under focusing conditions, the eigenmodes preserve their shape in their throughout propagation. However, non-eigenmodes have propagation-dependent shape. For example, non-eigenmodes have the same shape at their beam waist and far-field while it has a different shape in the near-field regime. On the other side, the second category of beams experimentally produced by means of the self-interference of laser beams along their propagation. For instance, Bessel Beams (BBs) and Airy beams, Bottle beams, etc. [16-18]. These beams have their characteristic shape within the overlapping region of self-interference. Also, these modes have completely different dynamical properties as compared with Eigen and non-Eigen modes. It is worth noticing that self-interference modes have different shapes at their focusing position with reference to their far-field position.

The variety of structures created in the cross-section of laser beams has enlarged the applications of light in fundamental and applied science fields. In particular, LG beams and BBs are superior over other structured modes and gained much interest in applications due to their helical wave-front which is quantified with an Orbital Angular Momentum (OAM). These beams are used in a wide variety of scientific disciplines ranging from fundamental to applied sciences. There are multiple review articles that have been written on LG beams that provide complete information on them. In the case of BBs, there are very few review articles that were written and focused on only certain concepts of BBs. Also, the past few decades have witnessed substantial progress towards the generation and applications of BB. Timely, it is necessary to have a full-fledged tutorial review on BBs. The motivation and aim of the present tutorial review is to provide a complete conceptual analysis of BBs.

All the discussions carried out in this tutorial on BB are on an ad hoc basis and the details are organized as follows. The fundamental aspects of BB are discussed by ray optics and wave optics in the first five sections and thereafter we utilized these discussions to understand the illustrations of BB in the next four sections. First, in section 1, we provided the objective of our tutorial and gave a brief introduction to the BB to show how it is different and important from other structured modes created in laser beams. Section 2 is dedicated to the theoretical analysis of BB where we provide a detailed analysis of the theoretical background and its structure. In section 3, we summarize all the experimental techniques used to generate BBs with their favorability and drawbacks. We provide the suitability of each experimental configuration for a desired application in the method of comparative study. In section 4, we outlined the possible tunabilities in the properties of BB by certain amendments made in its generation techniques. Section 5 provides an intriguing explanation for the origin of BB properties and outlines the basic concepts involved in their experimental realization. In section 6, we provide a brief analysis of different kinds of BBs and the generation of other structured modes from them. Non-diffracting and self-healing vector modes created in BBs are discussed in section 7. In section 8, we discussed the frequency tuning of BB which plays a pivotal role in the wavelength-dependent applications of BB. In section 9, finally, we end up the tutorial by reviewing various applications of BB in different branches of science. In section 10, we provided conclusions and notable points drawn from our tutorial review.

## 2. Theoretical background

Diffraction is an intrinsic property of light owing to its wave nature and this phenomenon occurs when a wave encounters an obstacle. The coherent unidirectional light provided by laser sources has beam-like intensity and propagates long distances without much spread in their beam radius. In experimental laboratories, it is always necessary to control these laser beams' directions with optical elements and to further confine them into a small volume for multiple scientific applications. However, the dynamical properties of the optical beams are governed by the law of diffraction and the minimum spot size maintenance of the beam will be restricted. The optical field initially confined to a finite area of radius $w$ in a transverse plane will be subject to diffractive spreading as it propagates outward from that plane in free space. The characteristic distance beyond which diffractive spreading becomes increasingly noticeable is $\pi w_0^2/\lambda$, the Rayleigh range ($w_0$ is the beam waist of the optical field which is the smallest size in its propagation). Hence for this reason it is commonly thought that any beamlike field must eventually undergo diffractive spreading as it propagates, and it is not possible to confine the optical field in small cross-sectional area for a long distance of propagation. However, Durnin, in his seminal paper, theoretically discovered and experimentally demonstrated that an optical beam with a transverse intensity profile in the form of a Bessel function of $0^{th}$ order is immune to the effects of diffraction [19,20]. Unlike most of the laser beams that spread upon propagation, the transverse distribution of these BBs remains constant. The diffraction-free mode solutions of the Helmholtz equation can contain the Bessel functions. The generalized monochromatic plane wave solutions of the Helmholtz wave equation

$$\left(\nabla^2 - \frac{1}{c^2}\frac{\partial^2}{\partial t^2}\right)E(r,t) = 0, \tag{2.1}$$

($c$ is the velocity of light and $r^2 = x^2 + y^2$) is in the form of

$$E(r,t) = U(r)V(z)\exp(-i\omega t). \tag{2.2}$$

By the method of separation of variables, the differential Eq. 2.1, with its generalized solution can be disintegrated into

$$\left(\nabla_\perp^2 + k_r^2\right)U(r) = 0, \tag{2.3}$$

$$\frac{\partial^2 V(z)}{\partial t^2} + k_z^2 V(z) = 0, \tag{2.4}$$

here, $\nabla_\perp^2 = \frac{\partial^2}{\partial x^2} + \frac{\partial^2}{\partial y^2}$, and $k_r^2 + k_z^2 = k^2 = (\omega/c)^2$.

From Eqs. 2.2, 2.3, and 2.4; the exact solution of Eq. 2.1 for $z \geq 0$ is

$$E(r,z,t) = \exp(ik_z z - i\omega t)\int_0^{2\pi} A(\phi')\exp[ik_r r\cos(\phi'-\phi)]d\phi'. \tag{2.5}$$

Further simplification of Eq. 2.5 leads to the generalized plane wave solution which contains the $l^{th}$ order Bessel function of the first kind $J_l$ is

$$E_l(x, y, z) = \exp(ik_z z - i\omega t) J_l(k_r r) \exp(il\phi). \tag{2.6}$$

Here, $l$ is the order of the BB. The last term in Eq. 2.6 represents the helical wave-front of the higher order BB and it is one for $0^{th}$ order. Eq. 2.6 clearly reflects that it will be a simple plane wave solution for $k_r = 0$ and for $0 < k_r \leq \omega/c$, it is non-diffracting BB. The key parameter of BB is its radial propagation vector $k_r$. The central lobe size of BB depends on the $k_r$. The $k_r$ plays a pivotal role in the experimental BB generation which is well discussed in the next coming section. The experimental results of BB generation will be compared with theoretical predictions through $k_r$. The order of the Bessel function is also called OAM of BB. Milonni and Eberly showed that the $E_l(x,y,z)$ field results from the symmetric convergence of a set of plane waves whose wave vectors indicated with $k$ lie on a cone making an angle $\theta$ with respect to the $z$-axis [21].

Next, we can acquire the position-independent intensity of BB from Eq. 2.6 is

$$I_l^{BB}(x, y, z) = J_l^2(k_r r). \tag{2.7}$$

The intensity distribution of ideal BB in its $xz/yz$-plane (propagation plane) from Eq. 2.7 is presented in Fig. 2.1 (phase profiles of BB provided in the insets of corresponding images). The intensity of the ideal BB is independent of its longitudinal position coordinate, $z$. Hence, its intensity is constant throughout propagation. The $0^{th}$ order BB has maximum intensity on the beam axis with low-intensity side rings. The successive bright and dark fringe pattern in the transverse direction formed by dint of periodic phase change between 0 and $\pi$ radians in the radial direction.

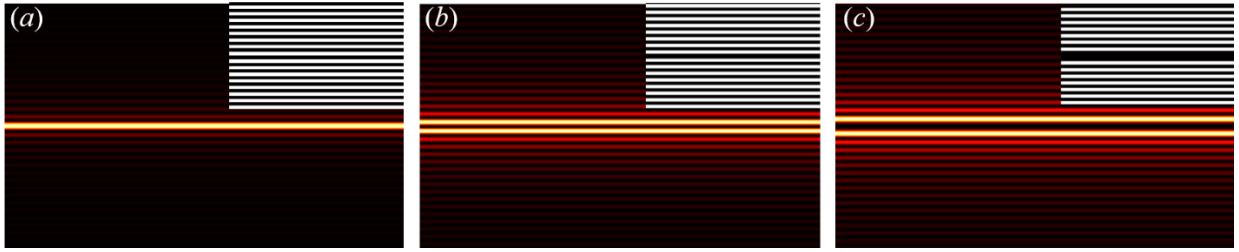

Fig. 2.1. The intensity distribution of the ideal Bessel beam in its propagation plane (longitudinal cross-section in $xz$-plane) for (*a*) $0^{th}$ order, (*b*) $1^{st}$ order, and (*c*) $2^{nd}$ order (corresponding phase profiles are provided in their insets). Here, radial propagation vector $k_r = 1.0307 \times 10^3$ cm$^{-1}$.

The transverse intensity distribution of BB in its $xy$-plane (cross-section plane) is given in Fig. 2.2 for its first three orders. While the $0^{th}$ order BB has a central bright spot, the higher order modes have a central dark core due to phase singularity formed owing to their helical wave-front. This dark core size increases with its order. Higher order BBs are also often termed non-diffracting vortex modes or OAM modes. The change in the transverse phase as a function of the order of BB can be seen in the inset of Fig. 2.2. In $0^{th}$ order BB, the half-width of the central peak is approximately $k_r^{-1}$, and its transverse skirt of the distribution decays as $r^{-1}$ [17]. As shown in Fig. 2.2(d), even though the intensity profile is sharply peaked at the center lobe, the amount of energy in each ring is approximately equal to that contained in the central maximum. It would therefore require an infinite amount of energy to create a BB over an entire plane. Thus, the ideal BB has infinite energy with an infinite range of spatial extension and non-diffraction propagation as shown in Fig. 2.1. However, BB with such kind of properties cannot be realized at experimental laboratories.

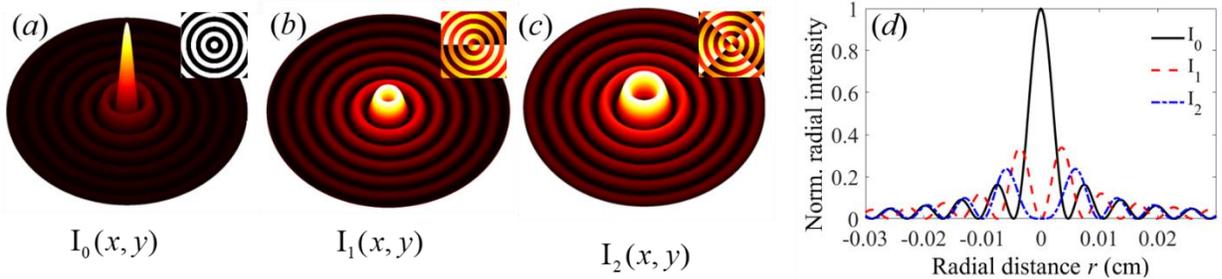

Fig. 2.2. The intensity distribution in the transverse cross-section ($xy$-plane) of the ideal Bessel beam: (*a*) $0^{th}$ order, (*b*) $1^{st}$ order, and (*c*) $2^{nd}$ order Bessel beams. The transverse phase distributions are shown in the insets. (*d*) Line profile of Bessel beams (intensities are normalized with the peak intensity of $0^{th}$ order Bessel beam). The radial propagation vector used here is $k_r = 1.0307 \times 10^3$ cm$^{-1}$.

After the inception of BB by Durnin, multiple theories and experiments were developed on it to further understand and enhance its applications. For instance, the development of the theory for pulsed BB [22], demonstration of spatiotemporal BBs [23], general description of circular symmetric BBs of arbitrary order and their transverse profile investigation [24], analysis and design of BB launchers using a finite inward cylindrical traveling wave aperture field distribution [25], deep investigation of Quantum-mechanical properties of BBs [26], on the validity of localized approximations for BBs: all N-BBs are identically equal to zero [27], Goos–Hänchen and Imbert–Fedorov shifts of a BB while it is propagating through dielectric interface [28], and study on energy characteristics of the superposition of TE- and TM-polarized electromagnetic BBs [29].

In the initial period of BBs demonstration by Durnin et al., soon thereafter another popular kind of BB was introduced by F. Gori et. al. in 1987 with the consideration of Gaussian envelop and it is called as Bessel-Gaussian Beam (BGB) [30]. Later multiple theoretical calculations and experiments were performed to experimentally realize and characterize the BGB [31-33]. The optical amplitude of BGB at $z = 0$ has the form of

$$E_l(x,y,z) = J_l(k_r r)\exp\left(-\frac{r^2}{w_0^2}\right)\exp(il\phi). \tag{2.8}$$

Here complex amplitude is considered to be one and $w_0$ is Gaussian spot size at $z=0$. At any arbitrary position $z$, the state of the BGB can be obtained in a paraxial approximation by Fresnel integration [30]. After Fresnel integration of Eq. 2.8 with the straightforward analytical calculations, the expression for the position-dependent BGB is given by

$$E_l(x,y,z) = \frac{w_0}{w(z)}\exp\left\{i\left[\left(k - \frac{k_r^2}{2k}\right)z - \arctan\left(\frac{z}{z_r}\right) - \omega t\right]\right\} J_l\left(\frac{k_r r}{1 + \frac{iz}{z_r}}\right)$$

$$\exp\left\{\left[-\frac{1}{w^2(z)} + \frac{ik}{2R(z)}\right]\left(r^2 + \frac{k_r^2 z^2}{k^2}\right)\right\}\exp(il\phi). \tag{2.9}$$

Here, $z_R$ is Rayleigh range, $w(z) = w_0(1+z^2/z_R^2)$, and $R(z) = z(1+z_R^2/z^2)$. Further, the intensity distribution of BGB is given by

$$I_l(x,y,z) = \frac{w_0^2}{w^2(z)} J_l^2\left(\frac{k_r r}{1 + \frac{iz}{z_r}}\right)\exp\left\{-\frac{2}{w^2(z)}\left(r^2 + \frac{k_r^2 z^2}{k^2}\right)\right\}. \tag{2.10}$$

The major difference between BB and BGB is the Gaussian envelop which is shown as the last term in Eq. 2.10. The intensity distribution in the cross-section for BB depends on $k_r$, and for BGB it depends on $k_r$ and $w_0$. The intensity distribution of BGB for different Gaussian beam waist is plotted for fixed $k_r = 1.0307\times10^3$ cm$^{-1}$ in Fig. 2.3.

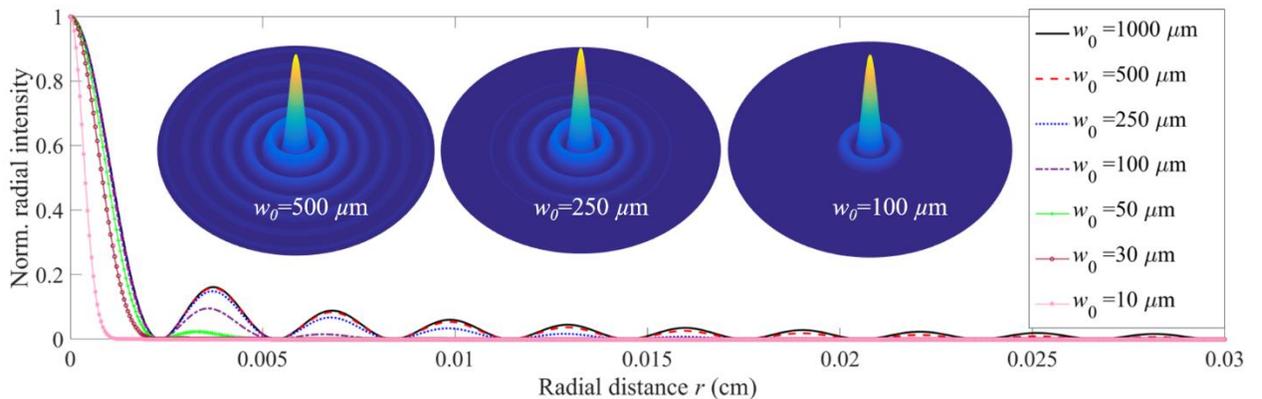

Fig. 2.3. Normalized line profile of Bessel Gaussian beam for different sizes of Gaussian beam waists at $z=0$ (the intensity distribution in the cross-section of Bessel Gaussian beam for various Gaussian spot sizes are provided in the insets).

The optical energy in the BGB is unequally distributed in its rings. The optical energy contained in each ring of BGB decreases with increasing their radius. As a result, the accumulation of optical energy in the BGB is close to the optical axis as compared with BB. The beam cross-sectional intensity distribution of BGB transforms into BB while increasing the Gaussian beam waist and on the other side, it transforms to a Gaussian beam while decreasing the Gaussian beam waist. For $k_r = 1.0307 \times 10^3$ cm$^{-1}$, the BGB transforms to BB around $w_0=1000$ µm and it is converted into Gaussian mode for $w_0$ is around 30 µm. From this analysis, we can infer that the cross-sectional intensity distribution of BGB can be controlled through $k_r$ and $w_0$. It is worth noticing that this scenario is true for higher order BGBs i.e., higher order BGB transform to higher order BB for larger $w_0$, and for smaller $w_0$, they modify into LG modes with zero radial index. There are ample comparative studies on BB, BGB, and LGB with reference to the conventional Gaussian beam in the perspective of advantage that are well documented in the literature [34-37]. Even though individually these two beams have their own advantages and disadvantages, overall, the combination of these modes has seen multiple advantages in the scientific and industrial applications.

## 3. Generation of Bessel beams by various techniques

The major issue with the BB is its experimental realization. As we discussed in the preceding section, to achieve the ideal BB, we need infinite optical energy which is practically not possible. However, we can generate quasi-BBs having a finite range of propagation and spatial extent in the laboratories through various experimental configurations.

Before going to the experiments, we can have some discussion on the possible ways to produce the BBs. The wave nature of light has two most popular phenomena: diffraction and interference [38-40]. Diffraction of light beam always induces divergence while it is propagating in the free space. The diffraction effects increase with decreasing the confined area of the light beam. Hence, the focused laser beam of any shape has a plane wave-front at the beam waist and divergence on either side of it. The low diffraction propagation range called Rayleigh length decreases with decreasing the beam waist size. So, it is not possible to generate a non-diffracting laser beam with its spot size in the range of a few wavelengths by means of traditional focusing diffractive optical elements. The self-diffraction phenomenon can be seen in the focused Gaussian beam as depicted in Fig. 3.1(*a*). Under the focusing conditions, all the individual waves in the beam have the same direction of propagation at the beam waist ($z=0$), i.e., $k(r) = k_z$. As the beam crosses its waist position, the waves get diffracted. The angular divergence of individual waves increases while moving away from the beam axis. For example, marginal rays have a larger divergence than the paraxial rays. The divergence can also be understood with propagation vector of individual waves as follows. For $z \neq 0$: on-axis waves' *k*-vector has the condition $k(r=0) = k_z$. The off-axis waves' *k*-vector follows the condition $k(r>0) = k_r + k_z$, $k_r$ is the radial *k*-vector. As we move from the paraxial ray region to the marginal ray region, $k_r$ increases.

On the other side, the interference of light can produce maxima and minima/zero intensities depending on the respective constructive and destructive interference owing to internal phase differences. The intensity of the light can be confined in the constructive interference region with zero intensity at destructive interference. Here, destructive interference acts as a constraint and guides the light in the free space in a constructive way. This criterion can be used to generate non-diffraction laser beams in the laboratory. It was first used by Durnin in his first seminal paper for an experimental demonstration of BB [17].

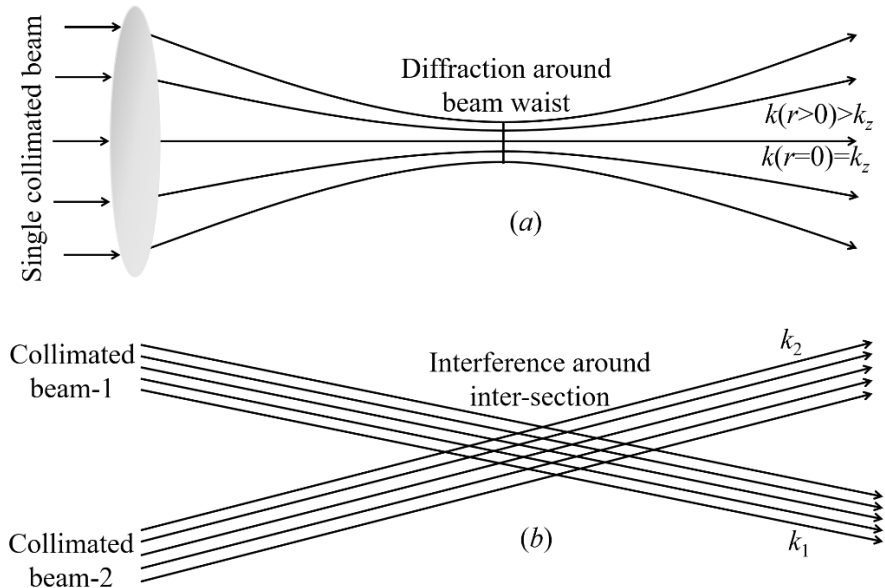

Fig. 3.1. (*a*) Self-diffraction of focused Gaussian beam and (*b*) interference of two beams at their intersection to produce non-diffraction light propagation.

To understand the BB generation, first of all, we can draw two collimated laser beams from a single light source and then interfere with them by crossing each other as shown in Fig. 3.1(*b*). The interference pattern formed in the overlapping region of two beams. The interference fringe width and shape depend on the phase difference provided between the two beams. In a similar way, the BBs can be easily realized in experimental laboratories by creating two-dimensional (2D) interference with cylindrical symmetry. The interference must take place along the propagation to maintain the same transverse profile.

The Generic method of BB generation in the experimental laboratory is shown in Fig. 3.2. The principle behind the generation of BB is wave-front division interference [41]. The optical phase retarder divides the incident laser beam wave-front radially in cylindrical symmetry to produce a circular symmetric interference pattern in the beam cross-section. The interference is carried out along the propagation direction to produce a circular fringe pattern which is directed along the beam axis. The range of BB is governed by the self-interference region of the laser beam. Thus, the generation of BB from the conventional Gaussian beams is basically enforcing the individual optical waves radially inward without diffraction effects in a long range of propagation. As a result, we can obtain the long-range micro-size central small spot surrounded by multiple diffraction-free rings. Moreover, optical energy transformation takes place between the central spot and diffraction-free rings in the radial direction throughout propagation.

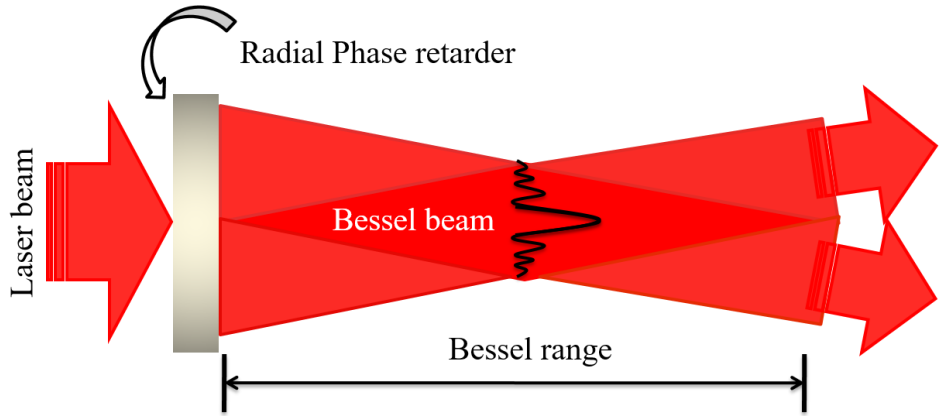

Fig. 3.2. Schematic diagram of 0$^{th}$ order Bessel beam generation in the experimental laboratory (*xz*/*yz*-plane).

The propagation properties of electromagnetic waves of a collimated Gaussian beam while it is converted into BB under radial retardation are illustrated in Fig. 3.3. The collimated Gaussian beam has the propagation/momentum vector *k* along the beam propagation direction ($k=k_z$). The radial retardance $e^{-ikr}$ provided by the diffractive optical element produces BB. Then the *k*-vector has all its Cartesian components of $k_x$, $k_y$, and $k_z$. Moreover, the conversion of collimated Gaussian beam into BB can be understood as optical waves propagating in a cylindrical shape are converted into a conical shape [Fig. 3.3(*b*)]. This optical cone makes an angle, $\theta = \tan^{-1}(k_r/k_z)$ with respect to the optical axis and it is termed the opening angle of the cone. The opening angle, also sometimes called the Bessel cone angle is a constraint of the BB. In the experiments, the Bessel cone angle can be always acquired from the parameters of diffractive optical elements.

From Fig. 3.2 and 3.3, we can also understand that the BB generation is nothing but the longitudinal projection of pump modes' transverse intensity along the propagation (optical axis of the beam). Thus, the longitudinal dimension and intensity distribution of BB depends on the transverse intensity distribution of the pump beam and the aperture size of diffractive optical elements used in the experiment. Noted to the statements given in the introduction, the BB has evenly distributed energy/power between its rings. The increase in the number of rings of BB increases its non-diffraction range at the cost of central lobe power which is prominent for many applications.

The possible BBs which are synthesized in the experimental laboratories through various techniques can be theoretically investigated and understood by the Fresnel diffraction integral of optical fields in the presence of diffractive optical elements [42-44]. Therefore, the transformation of a collimated Gaussian beam into BB in the presence of diffractive optical elements under the stationary phase method can be expressed as

$$E(r,\phi,z) = \frac{1}{i\lambda z}\exp\left[ik\left(z+\frac{r^2}{2z}\right)\right]\int_0^R\int_0^{2\pi} A(r',\phi')\exp\left(\frac{-r'^2}{w_i^2}\right)\exp\left(ikr'^2/2z\right)$$
$$\exp\left(-ikr'r\cos(\phi-\phi')/z\right)r'dr'd\phi'. \qquad (3.1)$$

Here, $w_i$ is the spot size of the input Gaussian beam, $R$ is the aperture radius of the diffractive optical element and function $A(r', \phi')$ represents the diffractive optical element.

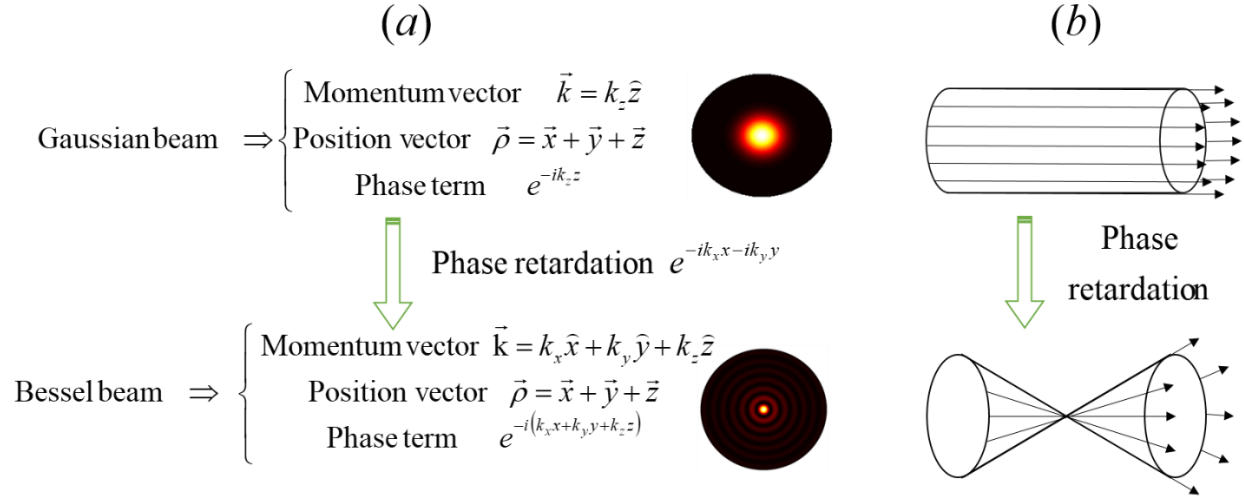

Fig. 3.3. Generation of Bessel beam from a collimated Gaussian beam: (*a*) the evolution of dynamical properties of the optical field while it is converting from Gaussian mode to Bessel mode and (*b*) the *k*-vectors in the cylindrical shape propagation are turned into a conical shape.

*3.1. Annular aperture*

The first experimental realization of a non-diffraction beam is an annular aperture-based experimental configuration. The way to generate non-diffraction beams through an annular aperture technique can be visualized as depicted in Fig. 3.4. Before going to discuss the experimental generation of conventional 2D non-diffraction beam through an annular aperture, we would like to introduce 1D non-diffracting beam for better understanding the generation of BB. As shown in Fig. 3.4(*a*), a collimated Gaussian beam can pass through two pin holes to create two coherent collimated laser sources. We can use a convex lens as a Fourier lens to Fourier transform these coherent laser sources. To carry out the Fourier transformation, the pin holes must be at the front Fourier plane of the lens. The Fourier transformed laser sources create a 1D non-diffraction beam in the back focal plane. The 1D interference fringes formed in the plane of coherent sources and the fringe direction is perpendicular to the bisecting line of interfering beams. The interference fringes are non-diffracting in the plane of incidence owing to their dark fringes resulting from destructive interference. The optical fields get diffracted in the conventional way in the direction perpendicular to the plane of incidence. Hence, it is a 1D non-diffraction beam. The intensity distribution in the non-diffracted direction follows the cosine function. Hence, these beams are often called Cosine beams [45-48]. In addition to 1D Cosine beams, we can also generate 2D Cosine beams whose propagation is two-dimensionally non-diffracted. However, these beams have completely different features with respect to BBs [49-51]. While 2D BB have circular rings with central bright spot, 2D Cosine beam have an array of square-type optical needles. For the understanding of Cosine beams, the intensity and phase of 1D and 2D Cosine beams are presented in Fig. 3.5. Depending upon the host laser beam shape, we can generate different kind of unique Cosine beams. We can further generate a variety of structured modes by the superposition of 1D cosine beams and the results can be found elsewhere [51].

In the case of 2D BB generation, the collimated Gaussian beam of spot size at an annular aperture $w_i$ is passed through an annular aperture to create an annular beam [Fig. 3.4(*b*)]. The annular beam is Fourier transformed with a convex lens. After the Fourier lens, the annular beam propagates in a conical shape to produce circular interference fringes along the cone axis. The 2D interference produces a non-diffracting intensity pattern along the cone axis with the transverse intensity distribution in the form of the Bessel function. Hence, it is termed as 2D BB or simply BB. The BB generation can be understood step by step as shown in Fig 3.4(*c*) where the transverse cross-sectional view is presented at each step to clearly visualize the generation of BB.

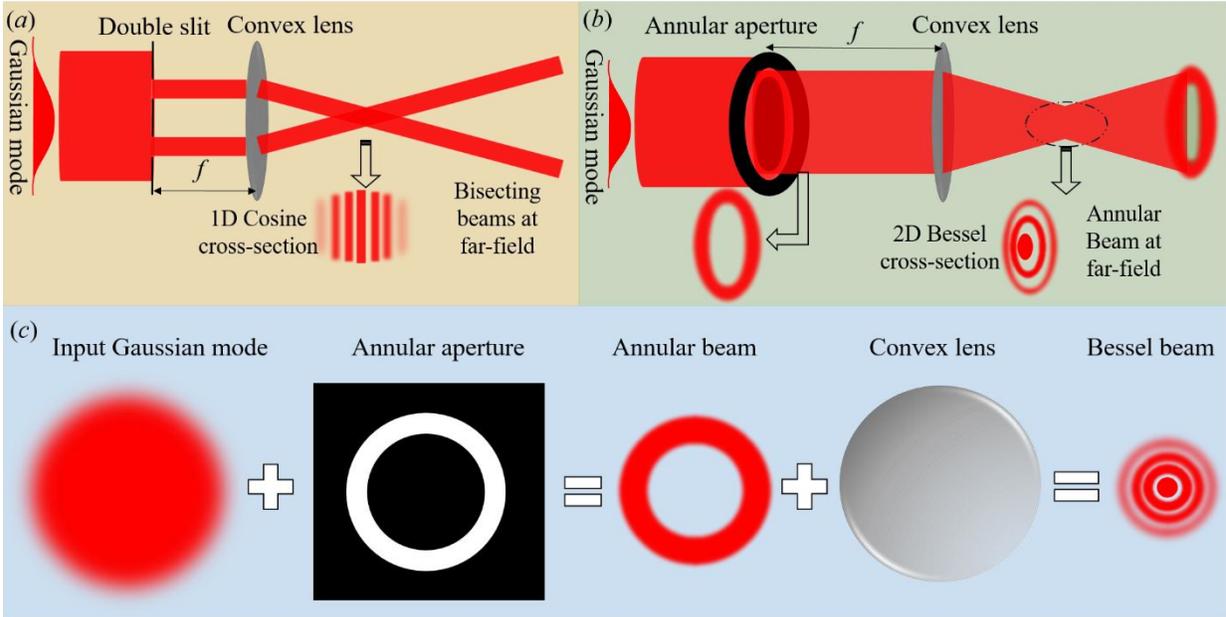

Fig. 3.4. Schematic diagram of generation of non-diffraction beams through the Fourier transformation of aperture beam: (*a*) generation of 1D non-diffraction beam (Cosine beam) through the Fourier transformation of double slit, (*b*) generation of 2D non-diffraction beam (Bessel beam) by Fourier transformation of annular slit, and (*c*) transverse cross-sectional view of Bessel beam generation from collimated Gaussian beam.

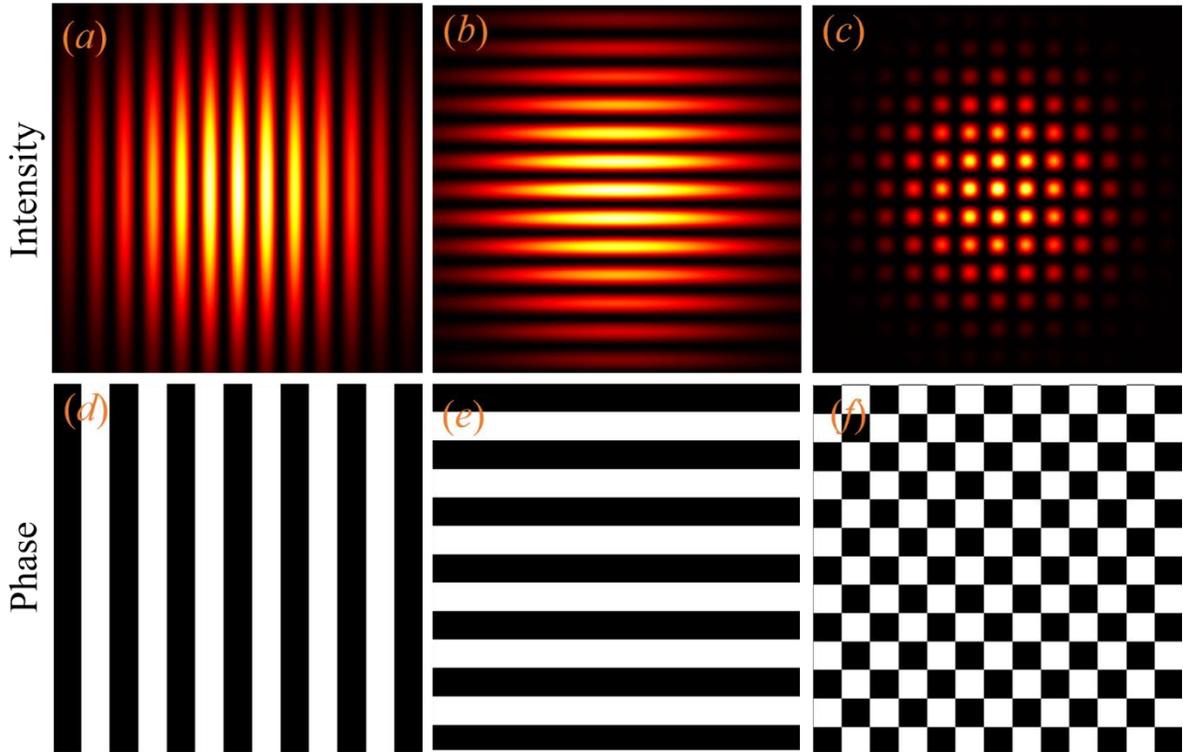

Fig. 3.5. The intensity and phase distributions in the transverse cross-section of Cosine beams are given in respective first and second rows: 1D *x*-Cosine beam (first column), 1D *y*-Cosine beam (second column), and 2D Cosine beam (third column).

The parameters of the annular aperture experiment and the region where the resultant output formed as BB are depicted in the ray diagram of Fig. 3.6. Here, the annular slit width is considered to be Δ, and its inner, outer, and mean radii are designated with $r_1$, $r_2$, and $r_{12}$ respectively. The intensity distribution modulated by the annular aperture is in the shape of an annular ring and it can be expressed in the mathematical form as [52]

$$I(r') = \frac{P}{\pi w_i^2}\left(1 - \frac{r_1}{r'}\right)\exp\left[-2\left(\frac{r'-r_1}{w_i}\right)^2\right], r' \geq r_1, \qquad (3.2)$$

where $P$ is optical power within the Gaussian mode. The angular beam size gets spread over its propagation with a diverging angle equal to $\lambda/\Delta$ [43]. While the annular beam passes through a lens, it sees the phase retardation of

$$\phi' = \exp\left(\frac{ikr'^2}{2f}\right). \qquad (3.3)$$

Next, substituting the amplitude of the optical field resulting from the annular aperture (from Eq. 3.2) and phase retardance acquired from the lens given by Eq. 3.3 in Eq. 3.1, we can obtain the mathematical expression for the amplitude of BB in the overlapping region as [52]

$$E(r,\phi,z) = \frac{kE_0}{z}\int_0^R \left(1 - \frac{r_1}{r'}\right)^{1/2}\exp\left[-\left(\frac{r'-r_1}{w_i}\right)^2\right]\exp\left[ik\left(\frac{r'^2}{2f} - z - \frac{r'^2+r^2}{2z}\right)\right]J_0\left(\frac{krr'}{z}\right)r'dr'. \qquad (3.4)$$

For a quantitative study, the longitudinal dimension of BB can be represented with Bessel range, and is given by $z_{range} = z_{max} - z_{min}$. Here, $z_{min}$ and $z_{max}$ are the onset and offset positions of BB formed in the experiment. The BB has its characteristic properties only within the Bessel range. In the case of the annular aperture method, the Bessel range formed in the two beams overlaying region is limited to $z_{range} = \lambda f^2/\Delta r_{12}$ [43] provided that the lens has a larger aperture than the transverse extends of the optical field on the lens as shown in Fig. 3.6. The center of the Bessel range is equal to the focal length of the Fourier lens. From this, we can understand that one can tune the position and range of BB along the optical axis by changing the focal length of the Fourier lens. After the BB range, the optical energy propagates in the cone shape with an annular intensity distribution. Let the annular beam cone angle be assumed to be $2\theta$. Here, the cone axis is the same as the BB axis. Then the radial and longitudinal propagation vectors are given by $k_r = k\sin\theta$ and $k_z = k\cos\theta$ respectively. Here, the propagation vector in free space $k = 2\pi/\lambda$, the opening angle $\theta = d/(2f)$, $d=2r_{12}$ is the annular slit mean diameter, $f$ is the focal length of Fourier lens, and $\lambda$ is the wavelength of light. It is also noted that for larger annular aperture width, the Fourier lens produces a focused hollow beam. However, by decreasing the annular aperture width focused hollow beam can be transformed into BB. The range of the BB increases with decreasing the annular aperture width at the cost of huge optical power loss. Nevertheless, this power loss can be reduced by replacing the Gaussian beam with a Hollow Gaussian (HoG) beam in front of the annular aperture. The mobility of the annular aperture in the generation of the hollow beam and BB can be simultaneously used for high transverse resolution imaging and volumetric imaging [53,54].

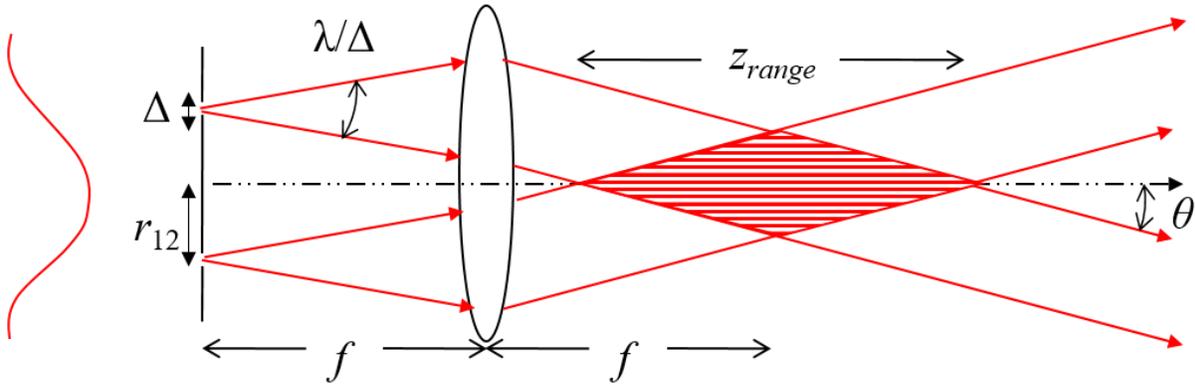

Fig. 3.6. The ray diagram of generation of Bessel beam by use of an annular aperture technique: Annular slit width is $\Delta$ and its inner, outer, and mean radii are $r_1$, $r_2$, and $r_{12}$ respectively.

In the annular aperture-based BB generation, the amplitude along the propagation is modulated by the diffraction envelope of the annular slit. These modulations are negligible within the finite output aperture $R$ (mode radius at input-facet of the lens), provided that the width of the annular slit $\Delta \ll \lambda f/R$. From the geometrical optics, $z_{max} = R/\tan\theta = R[(k/k_r)^2-1]^{1/2}$. This expression for $z_{max}$ predicts accurately for $k_r$ in the range $k \geq k_r \geq 2\pi/R$.

Advantage:

1. The position and range of BB along the propagation can be tuned by controlling the annular aperture mean radius and focal length of the Fourier lens.
2. This technique is cost-effective and can be easily demonstrated at undergraduate laboratories.
3. By controlling the width of the annular aperture, we can generate a hollow beam as well as BB from a single experiment for bio-imaging.

Disadvantage:
1. Very low mode conversion efficiency attributed to the blocking of the central part of the Gaussian beam.
2. This technique produces diffraction oscillations in the central lobe intensity along the propagation.
3. This technique cannot be used to generate higher order BBs.
4. This technique is not suitable for ultrashort laser sources.

### 3.2. Spatial light modulator

The spatial light modulator (SLM) is a liquid crystal-based diffractive optical element, and it is interfaced with a computer. By projecting computer-generated holograms on the SLM, we can create various diffractive gratings. The intensity of the laser beam passes through this liquid crystal display modulated by the computer-generated hologram. Thus, by projecting holograms of LG, HG, and BBs on SLM, we can transform the conventional Gaussian beam from the laser cavity to LG, HG, and BBs respectively [55-61].

To investigate the role of SLM in the generation of BB, first let us assume that a unit-amplitude plane wave illuminates a circular hologram, projected on SLM, of radius $R$ that is characterized by the complex-amplitude transmission function [62]

$$t(r',\phi') = \begin{cases} B(\phi')C(r') & r' \leq R \\ 0 & r' > R \end{cases}. \quad (3.5)$$

Where $B(\phi')$ has azimuthal phase factor $\exp(-il\phi')$ which determines the order of the BB. The second term $C(r') = \exp(-i2\pi r'/r_0)$ is a radial retardance and it converts the plane wave-front into a conical shape to produce BB. The constant $r_0$ determines the central lobe size and thickness of Bessel rings like the Gaussian beam spot size $w$ in the Gaussian function $\exp(-r^2/w^2)$. The field distribution behind the hologram of SLM can be obtained by substituting Eq. 3.5 into Eq. 3.1 as

$$E(r,\phi,z) = \frac{1}{i\lambda z}\exp\left[ik\left(z+r^2/2z\right)\right]\int_0^R r'dr'\exp\left(-i2\pi r'/\beta\right)\exp\left(ikr'^2/2z\right)$$
$$\int_0^{2\pi} d\phi' B(\phi')\exp\left(-ikr'r\cos(\phi-\phi')/z\right). \quad (3.6)$$

Here, the constant $\beta = r_0 k$. The first integral can be evaluated with the method of stationary phase [63]. Further, the integration over azimuthal angle $\phi'$ can be expressed in terms of the Fourier series of $B(\phi')$,

$$B(\phi') = \sum_{l=-\infty}^{\infty} b_l \exp(il\phi'). \quad (3.7)$$

Here, $b_l$ is the Fourier coefficient. Then evidently, after substituting Eq. 3.7 in 3.6

$$E(r,\phi,z) = \frac{1}{i\lambda z}\exp\left[ik\left(z+r^2/2z\right)\right]\sum_{l=-\infty}^{\infty} b_l \exp(il\phi)\int_0^R f_l(r')\exp(ik\mu(r'))dr', \quad (3.8)$$

here [64],

$$\int_0^{2\pi} d\phi'\exp(il\phi')\exp\left(-ikr'r\cos(\phi-\phi')/z\right) = 2\pi i \exp(il\phi)J_l(krr'/z), \quad (3.9a)$$

$$\mu(r') = r'^2/2z - \frac{2\pi r'}{\beta}; f_l(r') = 2\pi r'(-i)^l J_l(krr'/z), \quad (3.9b)$$

$$\int_0^R f_l(r')\exp(ik\mu(r'))dr' \propto \frac{f_l(r_c')\exp(ik\mu(r_c'))}{\sqrt{k\mu^{(2)}(r_c')}}, r_c' = \frac{k_r z}{k}. \quad (3.9c)$$

The intensity distribution of BB after the hologram is in the form of

$$I(r,\phi,z) \propto z\left|\sum_{l=-\infty}^{\infty} 2\pi(-i)^l b_l \exp(il\phi)J_l(2\pi kr/\beta)\right|^2, \tag{3.10}$$

the Bessel offset position $z_{max}$ for SLM-based hologram is given by $z_{max} = r_0 R/\lambda$.

Further, the complex transmission function $t(r', \phi')$ of the hologram used in the theoretical calculation to achieve the BB can be experimentally realized either by on-axis or by off-axis holograms. The on-axis hologram for arbitrary BB is constructed by the product of the azimuthal phase factor and radial phase factor as depicted in Fig. 3.7 [65, 66]. The resultant transmittance function for the on-axis hologram is given by

$$T_O(r',\phi') = \exp[i(l\phi' - 2\pi r'/r_0)]. \tag{3.11}$$

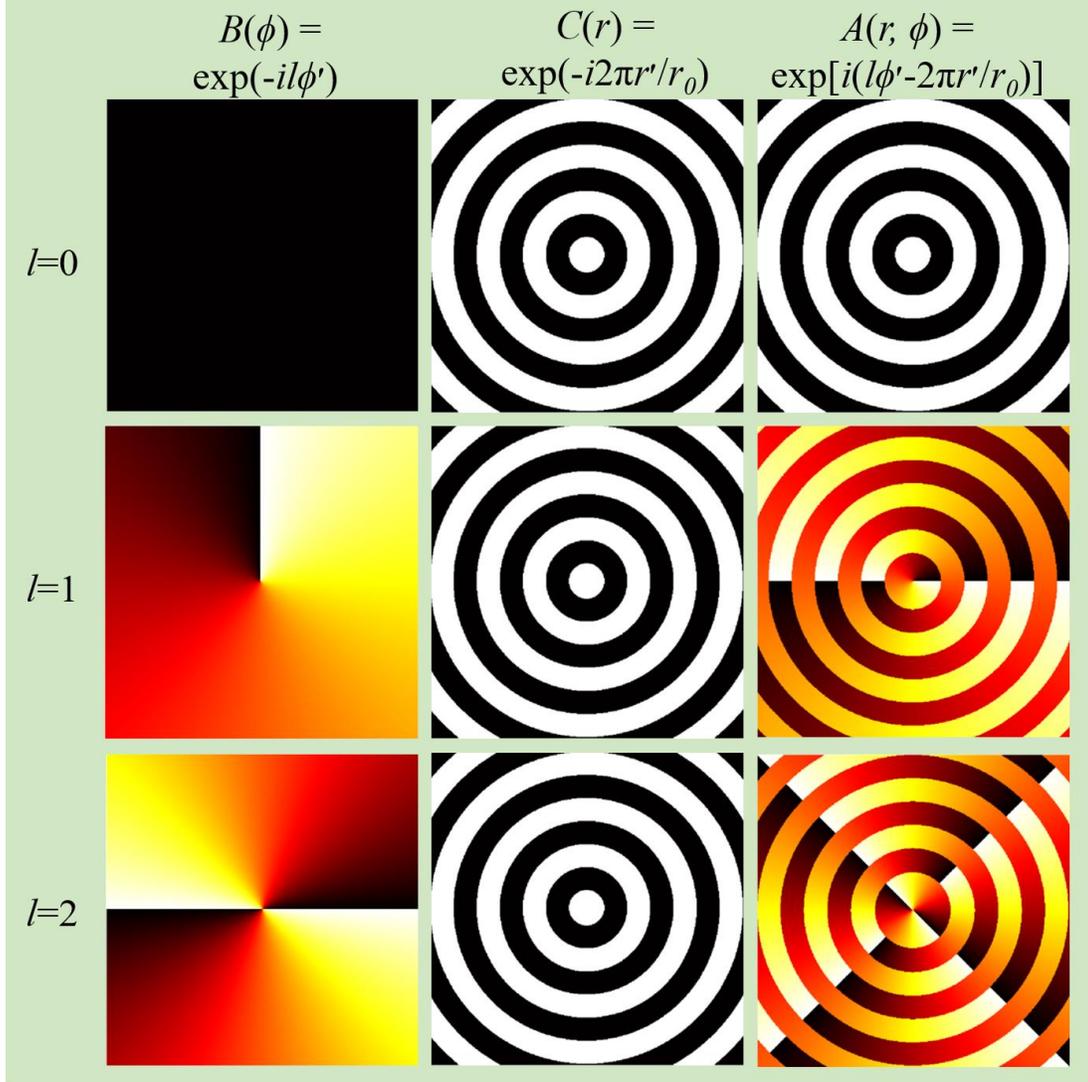

Fig. 3.7. On-axis hologram of Bessel beam of 0$^{th}$, 1$^{st}$ and 2$^{nd}$ orders in the respective 1$^{st}$, 2$^{nd}$ and 3$^{rd}$ row.

Besides, the combination of an on-axis hologram with horizontal linear grating can be used to produce an off-axis hologram. The off-axis hologram of $T(r', \phi')$ can be realized with the carrier-frequency method introduced by Burch. If the amplitude and phase of $B(\phi')$ are designated with $a(\phi')$ and $\psi(\phi')$, respectively, then the transmission function $T(r', \phi')$ is given by

$$T(r',\phi') = 1/2\{1 + a(\phi')\cos[\Psi(r',\phi')]\} \tag{3.12}$$

where

$$\Psi(r',\phi') = 2\pi\nu r'\sin\phi' + \Psi(\phi') - 2\pi r'/r_0.$$

Here, $v$ denotes the carrier frequency (in the $x$ direction), which separates the diffraction orders produced by the hologram. In this form, the amplitude $a(\phi')$ and the phase $\psi(\phi')$ are thus essentially coded into the visibility variations and the positional distortions, respectively, of a cosinusoidal grating [67,68]. The Burch holograms for different orders of BBs produced with Eq. 3.12 are given in Fig. 3.8.

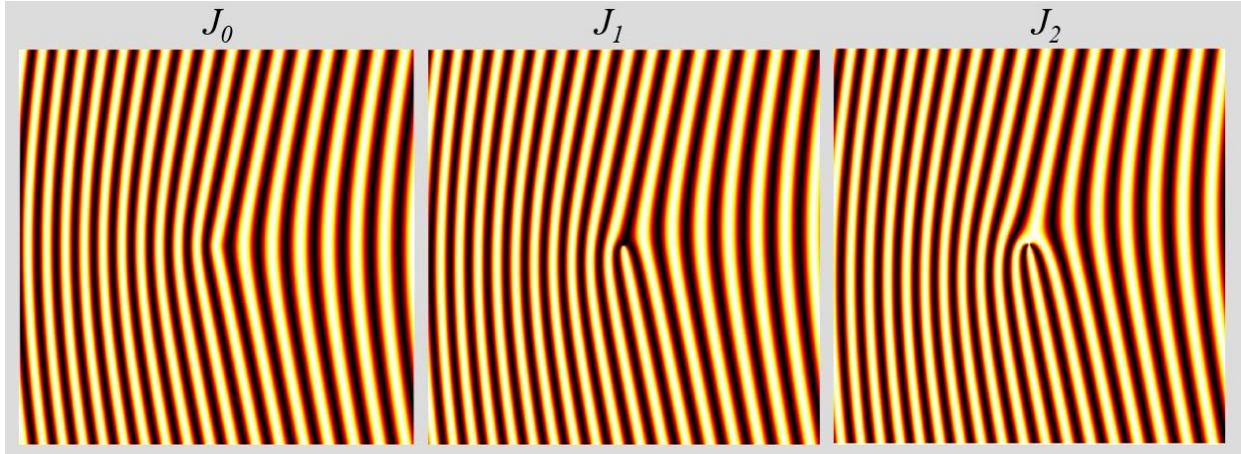

Fig. 3.8. Burch hologram of Bessel modes: $J_0$, $J_1$, and $J_2$.

In addition to Burch hologram, we can use binary–amplitude-coded holograms to generate BBs. The hologram transmittance function is given by the formula

$$T(r',\phi') = \begin{cases} 0 & 0 \leq (1/2)\{1+\cos[\Psi(r',\phi')]\} \leq q \\ 1 & q \leq (1/2)\{1+\cos[\Psi(r',\phi')]\} \leq 1 \end{cases}. \quad (3.13)$$

Where the inequality parameter $q = (1/\pi)\sin^{-1}[a(\phi')]$. In binary-amplitude coding, the phase of the function $B(\phi')$ is still stored in the locations of the fringes, but the amplitude information is now recorded in the variations of fringe widths instead of visibility. The main effect of binarization is the redistribution of energy among the various diffraction orders of the hologram [69-73]. The fringe pattern in the binary hologram for various Bessel orders is presented in Fig. 3.9. As shown in Fig. 3.10, there is a third kind of hologram called Complex-amplitude modulation hologram which is formed by the product of complex-amplitude of the BB with off-axis hologram [74-79]. It is very important to note that the SLM can be directly used with an off-axis hologram to generate any structured mode without any calibration. However, in the case of an on-axis hologram, one must calibrate the SLM to achieve high-quality structured modes [80].

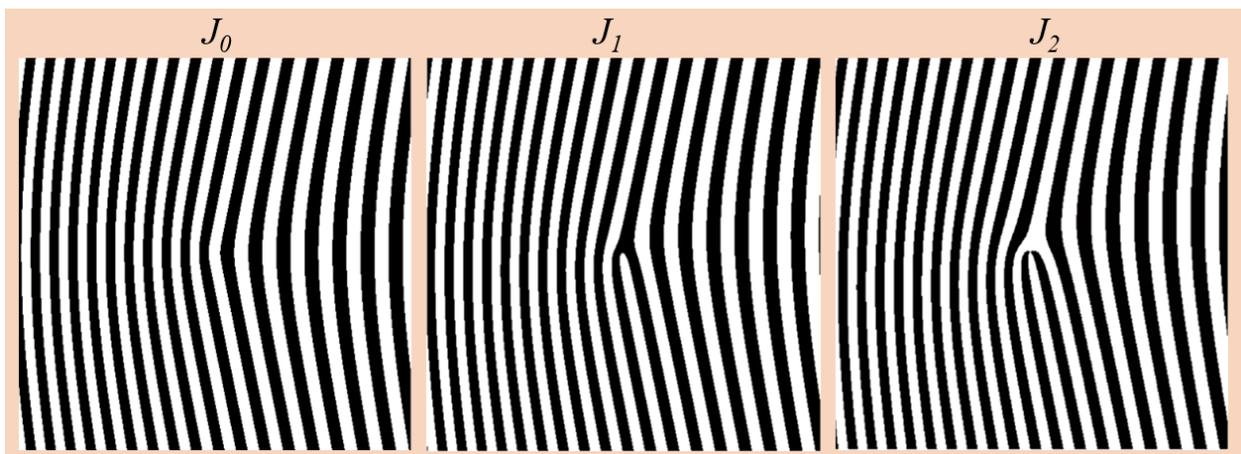

Fig. 3.9. Binary hologram of Bessel modes: $J_0$, $J_1$, and $J_2$.

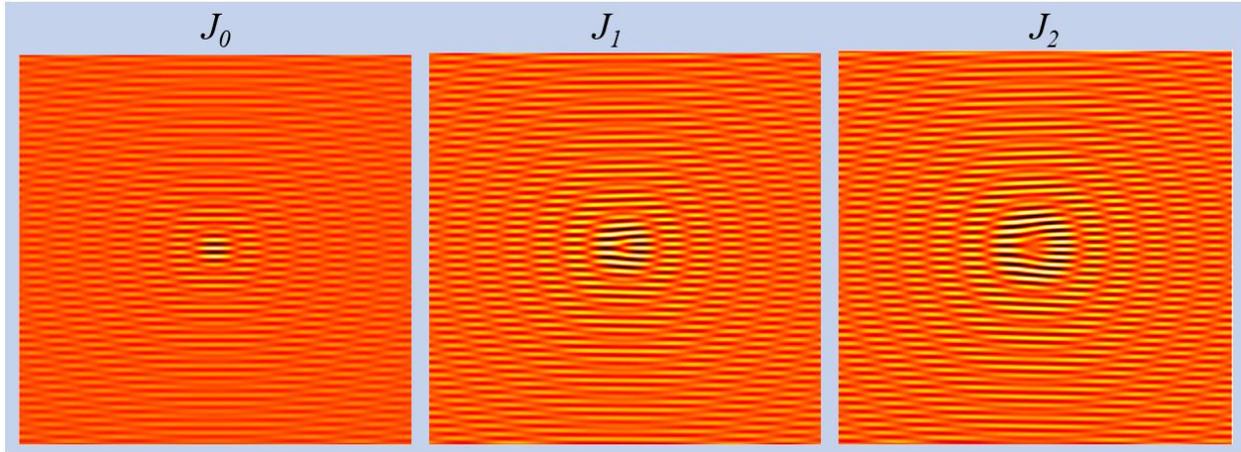

Fig. 3.10. Complex-amplitude modulation hologram of Bessel modes: $J_0$, $J_1$, and $J_2$.

Similar to the annular aperture, the aperture of holograms in the SLM produces intensity modulation in the BB along the propagation as shown in Fig. 3.11. The modulation frequency and depth increase while moving away from the aperture. The on-axis intensity modulations along the axis of BB are presented for various ratios of aperture radius $R$ and Gaussian beam spot size $w$. The on-axis modulations decrease with increasing the hologram aperture with reference to Gaussian spot size. Hence, the aperture diffraction effect on BB can be reduced by maintaining the incident Gaussian beam spot size two times less than the aperture radius with compromising the resolution of BB. It is also noted that in the annular aperture, the on-axis intensity modulations cannot be removed. However, here we can successfully minimize these effects by reducing the pump mode size.

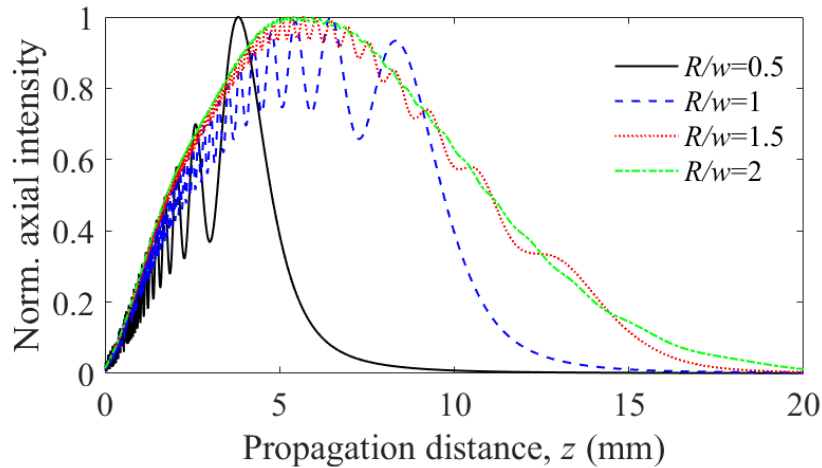

Fig. 3.11. Line profile of 0[th] order Bessel beam, which is generated with SLM hologram, along the longitudinal axis (z-axis) for different aperture parameter ratio, $R/w$.

Advantage:
1. SLM is a universal diffractive optical element for the generation of various structured laser beams. Hence, it is a readily available device to generate BB.
2. SLM alone can produce all orders of BB and BGB.

Disadvantage:
1. SLM cannot be used to generate high-power BBs due to its low damage threshold.
2. SLM can be used for a certain range of wavelengths of the electromagnetic spectrum.
3. SLM based BB generation is expensive compared with most of the other techniques.

### 3.3. Axicon

Axicon is a conical shape lens that has seen tremendous applications in fundamental and engineering-based optical science experiments [81-83] and one of its major applications in recent decades is the Bessel beam generation. As like in our previous discussions on aperture-based Cosine beams and BB generations (section 3.1), we can also generate these beams with the aid of axicon. The BB can be generated in a very simple way by introducing conical shaped axicon in the path of a perfectly collimated Gaussian beam [84] and similarly, we can also produce a Cosine beam with a triangular-shaped axicon termed Fresnel biprism which is readily available in undergraduate labs [48]. The schematic diagram of Cosine beam generation using a Fresnel biprism is shown in Fig. 3.12(*a*). Fresnel biprism is a triangular-shaped prism and it divides the wave-front of incident collimated circular Gaussian beam into two semicircular Gaussian parts. The interference formed in the overlapping region of these two semicircular Gaussian parts produces a non-diffraction 1D Cosine beam. The generated 1D Cosine beam has non-diffraction property in the plane of incidence attributed to interference and diverging property in the plane of perpendicular due to no interference. At the far-field, the output results in two beams. As shown in Fig. 3.12(*b*), the collimated Gaussian beam transformed into a conical shape while it is propagating through an axicon. The linearly varying thickness of the axicon in its radial direction transforms the plane wave-front into a conical shape. Thus, the wave-front division takes place in the radial direction with azimuthal symmetry. As a result, it produces circular interference fringes with their center along the axis of the axicon. The optical axis of the BB coincides with the axis of the axicon, and the far-field distribution is an annular shape.

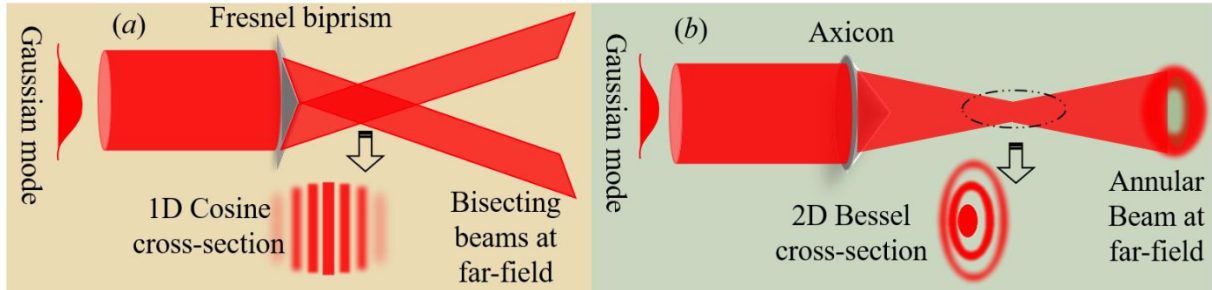

Fig. 3.12. Schematic diagrams of (*a*) 1D Cosine beam generation using Fresnel biprism and (*b*) generation of Bessel beam based on axicon.

Any arbitrary order BB can be generated by pumping a suitable order collimated Gaussian vortex given by Eq. 3.14 [85] to the axicon. While the azimuthal phase variation of BB is provided by Gaussian vortex mode, its radial phase variation is attributed to the axicon.

$$\mathrm{E}(r',\phi') = \sqrt{\frac{2^{l+1}P}{\pi w^2}} \left(\frac{r'}{w}\right)^l \exp\left(\frac{-r'^2}{w^2}\right)\exp(il\phi'), \qquad (3.14)$$

here, $P$ is the optical power within the laser beam. Also, $w$ and $l$ are respective Gaussian spot size and OAM of the vortex beam. The phase retardance provided by the axicon is exp(-*ikr*). After substituting the amplitude of the Gaussian vortex beam and phase retardance in Eq. 3.1, the theoretical expression for the optical field amplitude of BB generated by axicon is given by [86,87]

$$\mathrm{E}(r,\phi,z) = \frac{1}{i\lambda z}\sqrt{\frac{2^{l+1}P}{\pi w^2}} \exp\left[ik\left(z+\frac{r^2}{2z}\right)\right] \int_0^R r'dr' \left(\frac{r'}{w}\right)^l \exp\left(\frac{-r'^2}{w^2}\right)\exp(-ik_r r')\exp\left(\frac{ikr'^2}{2z}\right)$$

$$\int_0^{2\pi} d\phi' \exp(-il\phi')\exp\left(\frac{-ikr'r\cos(\phi-\phi')}{z}\right). \qquad (3.15)$$

After straightforward integration with the integral formulae given in Eq. 3.9 (a-c), Eq. 3.15 is simplified into

$$\mathrm{E}(r,0,\phi) \propto \frac{(-i)^{l+1}2\pi r_c}{\lambda\sqrt{kz}} \exp\left[ik\left(z+\frac{r^2}{2z}\right)\right]\left(\frac{r_c}{w}\right)^l \exp\left(\frac{-r_c^2}{w^2}\right)J_l(krr_c/z)\exp\left(i\frac{-ik_r^2 z}{2k}\right). \qquad (3.16)$$

Here, $r_c$ is the critical point. There is only one critical point in this integral and it is given by $r_c = k_r z/k$ [87]. The intensity distribution of BB is given by

$$\mathrm{I} \propto \left(\frac{z}{z_{max}^G}\right)^{2l+1} \exp\left(\frac{-2z^2}{z_{max}^{G\;2}}\right)J_l^2(k_r r). \qquad (3.17)$$

The relation between the radial *k*-vector of BB and the properties of axicon is given by $k_r = k\sin[(n-1)\alpha]$. Here, $\alpha$ and $n$ are the respective opening/base angle and refractive index of the axicon as shown in Fig. 3.13. For a smaller axicon angle, the radial $k$-vector can be considered as $k_r = k(n-1)\alpha$. The apex angle of the axicon ($\gamma$) is in terms of the opening/base angle of the axicon given by $\gamma = 180-2\alpha$. The $z^G_{max}$ of BB in Eq. 3.17 is given by

$$z^G_{max} = \frac{kw}{k_r} = \frac{w}{(n-1)\alpha}. \qquad (3.18)$$

The $z^G_{max}$ is the range of BB formed in the presence of a Gaussian beam as a pump source. The Bessel range can be increased either by increasing the Gaussian beam spot size at the axicon or by decreasing the axicon angle. The opening angle of the optical cone formed by the axicon is $\theta = (n-1)\alpha$. The apex angle of the optical cone formed after the axicon is $2\theta$. Arbitrary BB generation can be clearly visualized with a ray diagram as illustrated in Fig. 3.13. The Gaussian beam-pumped axicon can generate $0^{th}$ order BB [88, 89]. Furthermore, with the addition of another diffractive optical element, which can produce a Gaussian vortex with arbitrary order, to the axicon we can produce higher order BBs. For example, a Gaussian beam can be passed through a parallel aligned spiral phase plate (SPP) and axicon to produce higher order BBs [90]. Past one-decade multiple vortex laser sources were successfully demonstrated from visible to mid-infrared wavelength region [91-93]. By inserting an axicon at the outlet of these laser sources we can produce arbitrary scalar and vector BBs with high conversion efficiency [94].

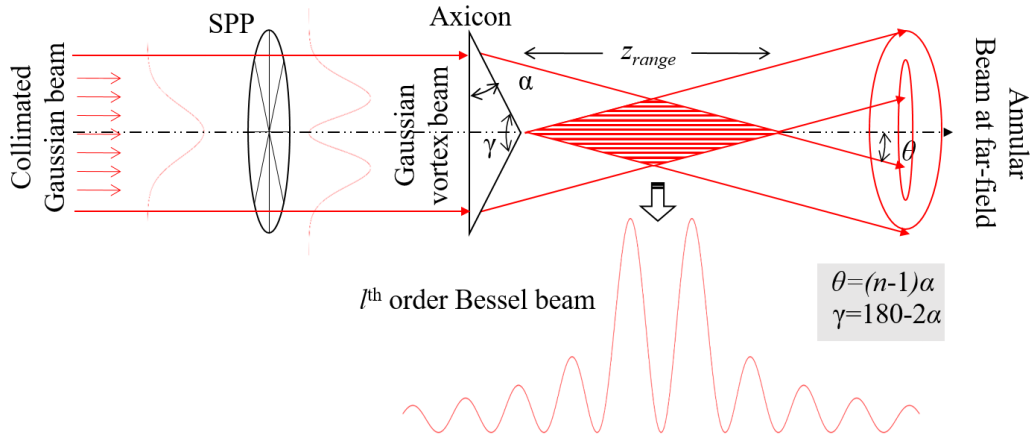

Fig. 3.13. Ray diagram of arbitrary Bessel beam Generation in the presence of axicon and spiral phase plate.

As shown in Fig. 3.14, the ideal axicon-based BB, in the presence of a Gaussian pump, has no rapid on-axis intensity oscillation generally associated with the annular aperture method. Here, the simulated BB shown is obtained by considering the following parameters: axicon angle 1°, the refractive index of the axicon is 1.5, the wavelength of light used is 1064 nm and its spot size at axicon is 1.5 mm. Under the Gaussian pump, the axicon produces BB from its tip itself, so the $z^G_{min}=0$ and the $z^G_{max}$ is given by Eq. 3.18. Thus, the Bessel range acquired in the presence of a Gaussian pump is $z_{range}= z^G_{max}$. In the present calculations, the maximum range and peak positions of the BB are $z^G_{max}$=17.2 cm and z($I_{max}$)=8.6 cm respectively. In axicon-based BB generation, the alignment of the illuminating beam with the axicon axis is quite similar to the alignment of the lens. However, BB generation is very sensitive to axicon alignment. Any oblique illumination of the pump beam onto the axicon results in an element of astigmatism being introduced and this leads, not to an on-axis spot with a set of concentric circles, but rather to a cheque board type pattern [95-97]. As like a conventional lens, the axicon also has a focal length. The focal length of an axicon is where on-axis intensity reaches its maximum [98], For the Gaussian pump, the focal length of the axicon is given by $f_a= z^G_{max}/2$.

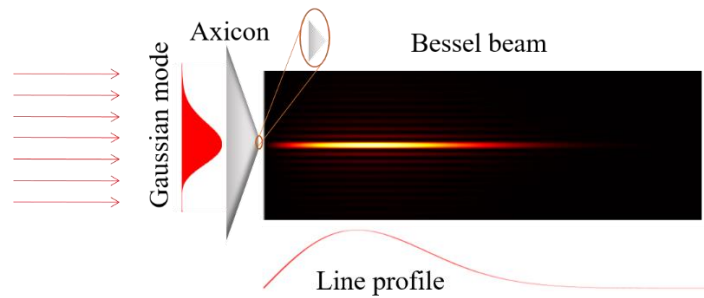

Figure. 3.14. Bessel beam generation through the pumping of Gaussian beam to the ideal axicon.

Higher order BBs generated through an axicon in the presence of a Gaussian vortex pump are shown in Fig. 3.15. In higher order BB generation, the BB formation takes place after a particular distance from the axicon, and it depends on the inner radius of the pump vortex mode. The inner and outer radii of the Gaussian vortex increase with increasing its order [99] and correspondingly the BB formation position along the axicon axis shifts away from its tip. So, the onset and offset positions of BB have to be expressed in terms of its pump Gaussian vortex radii as $z^l_{min}$ =$r_1/$ $(n-1)α$, and $z^l_{max}$ =$r_2/$ $(n-1)α$. The range of BB is $z^l_{range}$= $(r_2- r_1)/(n-1)$ $α$. The peak position of BB in the propagation direction in terms of Gaussian vortex order $l$ is given by $z^l(I_{max}) = z^G_{max} (2l+1)^{1/2}/2$. Further understanding, the properties of the Gaussian vortex beam and BB were calculated for their different orders and the results are shown in table 3.1. According to the definition given to the focal length of the axicon by reference [98], the focal length of the axicon for an arbitrary order of BB is $f_a= z^G_{max} (2l+1)^{1/2}/2$.

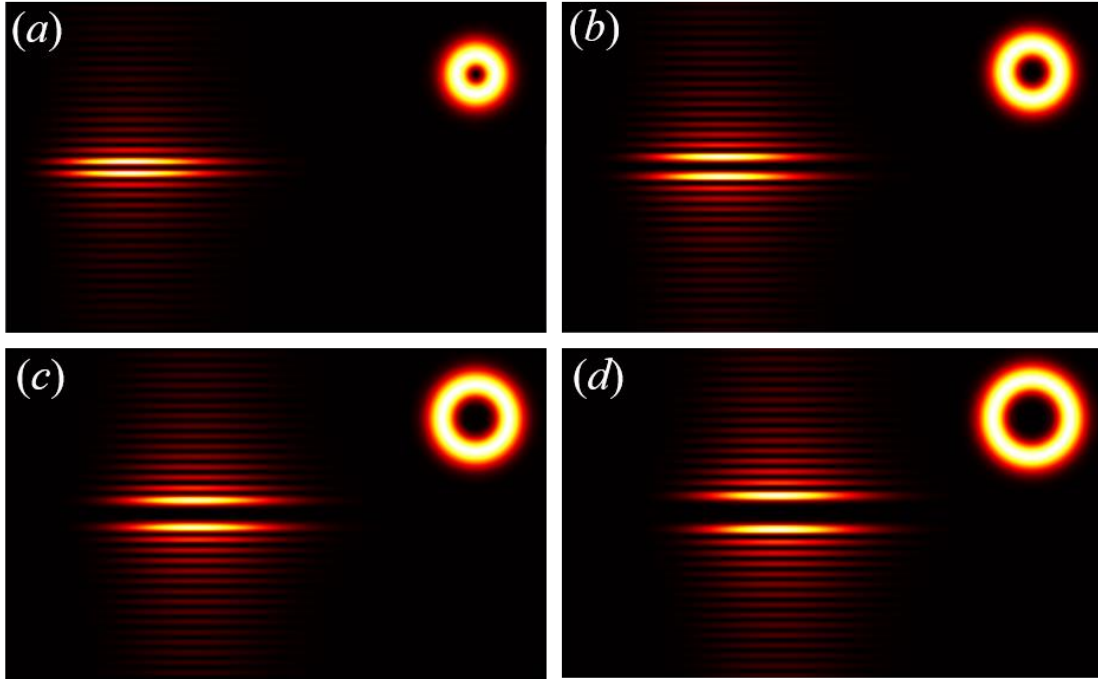

Fig: 3.15. Theoretically generated Bessel beams of order (*a*) *l*=1, (*b*) *l*=2, (*c*) *l*=3, and (*d*) *l*=4, from an axicon of opening angle α=1⁰ and Gaussian beam spot size *w*=1.5 mm. The transverse intensity profile of pump Gaussian vortex modes used in the numerical calculations are shown in the insets of their corresponding Bessel beams.

Table 3.1: The properties of the Gaussian vortex beam and Bessel beam: $r_1$ and $r_2$ are the inner and outer radii of the Gaussian vortex beam. $z_{min}$, $z_{max}$, $z_{range}$, and $z(I_{max})$ are the respective onset, off-set, range, and peak intensity positions of Bessel beam.

| OAM | Gaussian vortex properties | | BB properties | | | |
|---|---|---|---|---|---|---|
| $l$ | $r_1$ (mm) | $r_2$ (mm) | $z^l_{min}$ (cm) | $z^l_{max}$ (cm) | $z^l(I_{max})$ (cm) | $z^l_{range}$ (cm) |
| 0 | 0 | 1.5 | 0 | 17.2 | 8.6 | 17.2 |
| 1 | 0.6 | 2.66 | 2.9 | 25.9 | 19.3 | 23 |
| 2 | 1.16 | 3.25 | 6.9 | 30.5 | 25.9 | 23.6 |
| 3 | 1.62 | 3.72 | 10.3 | 34.1 | 31.4 | 23.8 |
| 4 | 2.01 | 4.12 | 13.3 | 37.2 | 35.6 | 23.9 |

As shown in Fig. 3.14, the ideal axicon has a conical shape with a sharp tip at its apex. The radial phase retardation provided by the axicon produces BB with no on-axis intensity modulation along the beam propagation. However, it is not possible to manufacture a perfect axicon with a sharp edge at its apex [100-104]. For a comparative understanding of the defect axicon with reference to the ideal axicon, both types of axicon shapes in the longitudinal cross-section are given in Fig. 3.16. The quality of the axicon tip depends on the manufacturing techniques. The imperfections at the axicon tip were well studied with several types: round-tip, blunt-tip, and oblate-tip [105-108]. Such kinds of tips generate a refracted beam that interferes with the quasi-BB created behind the axicon as depicted in Fig. 3.17. The focused beam from the tip of the axicon interferes with the optical cone which produces the BB.

Hence, we could see the interference of two kinds of beams along the optical axis. In turn, an undesired intensity modulation occurs that significantly degrades the unique properties of the experimentally generated BB. Especially, we could see on-axis intensity modulation along the optical axis. Generally, we use a conventional Gaussian beam as a pump source for the axicon to generate the BB. Gaussian beam has its peak intensity at the center. Hence, even though the defect at the axicon tip is small as compared with the axicon, we could see a noticeable effect of the axicon tip defect on the intensity distribution of BB.

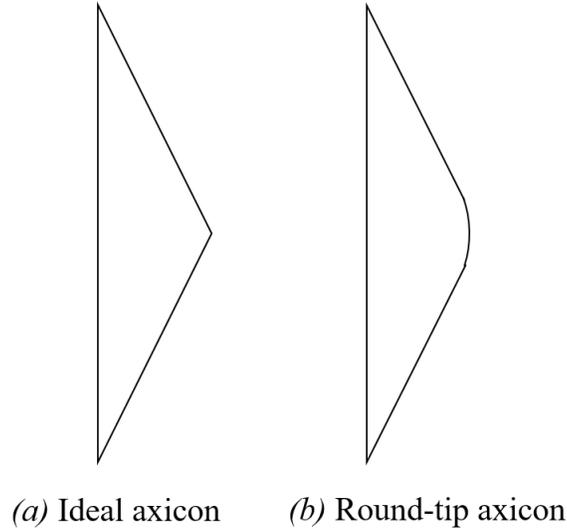

*(a)* Ideal axicon    *(b)* Round-tip axicon

Fig. 3.16. Schematic diagrams of (*a*) ideal axicon and (*b*) axicon with round-tip/blunt shape at apex.

The round-tip effect is predominantly seen in the $0^{th}$ order BB due to the Gaussian beam pumping. However, higher order BBs generated with an axicon are free from the round-tip of the axicon. Because the higher order BBs were generated via pumping of Gaussian vortex beam to the axicon, and Gaussian vortex beams have a central dark core. Hence, round-tips have no role in the generation of higher order BBs.

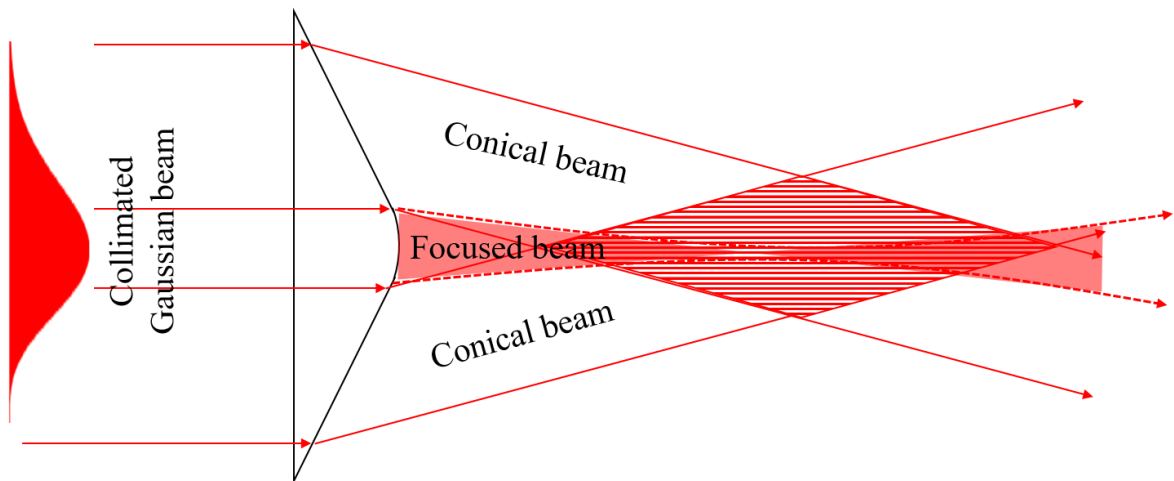

Fig. 3.17. Ray diagram of Bessel beam generation by using a round-tip axicon under Gaussian beam illumination.

Selcuk Akturk and his group assumed that the axicon with defect at its tip with hyperbola because it provides an excellent approximation of a real manufacturing axicon tip. It gives a flat face at $r = 0$, and asymptotically approaches the real sloped profile for larger $r$ [98]. They have defined a radius $r_{hyp}$, under which the axicon thickness is hyperbolic. The radial thickness variation of axicon mathematically given by

$$d(r') = \begin{cases} c_1 - R_{hyp}\tan^2(\alpha)\sqrt{1 + \dfrac{r'^2}{R_{hyp}^2 \tan^2(\alpha)}}, & r' \leq r_{hyp} \\ c_2 - r'\tan^2(\alpha), & r' > r_{hyp} \end{cases} \quad (3.19)$$

Here, $R_{hyp}$ is the curvature of the hyperbola, $c_2$ is a constant determined by the full radius of the axicon, and $c_1$ is a constant, used to match the two profiles at $r = r_{hyp}$. The ideal axicon phase retardance $\exp(-ikr')$ must be replaced with $\exp[-ikd(r')]$ in Eq. 3.19 to realize the BB, generated by apex defected axicon. In another work, the path difference provided by the defect axicon is derived by O. Brzobohaty et.al., as shown in Eq. 3.20 [106] by assuming the shape of the axicon surface is a hyperboloid of revolution of two sheets.

$$d(r') = -\sqrt{a^2 + \dfrac{r'^2}{\tan^2(\gamma/2)}}. \quad (3.20)$$

Here the variable $a$ specifies the defect at the tip of the axicon. The smaller the parameter $a$, the more closely the axicon approaches the ideal sharp shape of the axicon. The oblate tip axicon thickness variation is approximated by the simple polynomial of radial coordinate given by [107, 109]

$$d(r') = \begin{cases} -0.083 r'^2 + 0.023 r'^3, & 0 \leq r' \leq 1\,\text{mm} \\ 0.038 + 0.098 r', & r' > 1\,\text{mm} \end{cases} \quad (3.21)$$

The thickness of the round-tip axicon in the radial direction is compared with the ideal axicon given in Fig. 3.18. The thickness variation of a real axicon is the same as the ideal axicon for radial positions far from its apex position, however, around the apex position, its thickness variation deviates from the ideal axicon.

The BB, generated by the apex defect axicon, has on-axis intensity modulation along the propagation. The origin of these on-axis intensity modulations can be understood as follows. The optical waves close to the optical axis see the apex defect of the axicon and have very small convergence and their $k$-vector along the optical axis ($k_z$) can be successfully assumed to be $k$. As we know the optical cone formed by the axicon has the $k$-vector along the optical axis $k_z = k\cos\theta$. Therefore, the two co-propagating waves with different wave-vector lengths interfere on the optical axis and create a periodic modulation of the axial beam intensity with a period $\lambda/(1-\cos\theta)$. The depth of this modulation decreases at axial positions placed farther from the axicon by means of the decreasing intensity of the diverging spherical waves provided by the round-tip of the axicon.

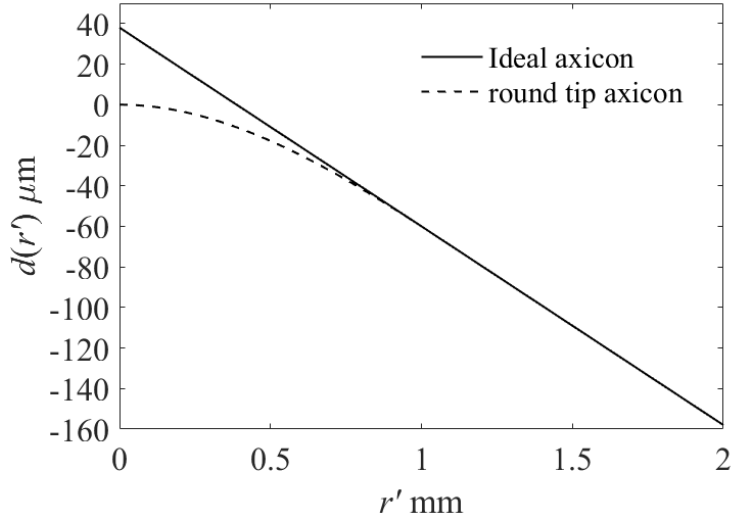

Fig. 3.18. Comparison of thickness variation of round-tip axicon with respect to ideal axicon.

Manufacturing high-quality axicons without defects at their apex position is very difficult. To overcome this manufacturing defect, several methods were proposed to avoid the effect of the apex defect of axicon on BB. In that one of the techniques is the combination of an annular aperture and an axicon which can be used to avoid the round-tip effect on BB [Fig. 3.19]. By inserting an annular aperture before the axicon, the light illuminated on the axicon tip can be successfully removed [110]. One can also insert the aperture next to the axicon to stop the focused light from the axicon tip. In either case, the main drawback of this technique is the maximum pump power loss in the mode conversion. Further, the diffraction effects of the annular aperture degrade the BB quality. Hence, this technique was generally not very useful for the generation of high-quality BB.

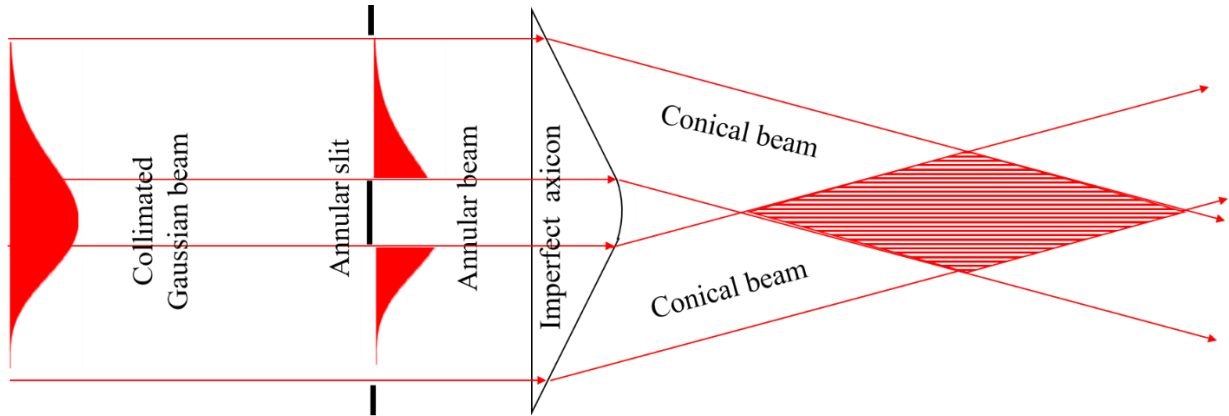

Fig. 3.19. Ray diagram of Bessel beam generation with round-tip axicon in the presence of annular slit.

Another technique to generate high-quality BB with axicon is by the way of spatial filtration of the beam in the Fourier plane [105,106,111]. It improves the spatial form of BB and removes the undesired modulation. The experimental setup generally used for the spatial filtration of BB is schematically depicted in Fig. 3.20. The incident Gaussian beam on the axicon transformed into BB with on-axis intensity modulation. The first lens of focal length $f_1$ will do the Fourier transformation of the BB. The back focal plane of the first lens represents the Fourier plane where the spatial-frequency spectrum of the BB is formed. The low-frequency components correspond to the round-tip of the axicon present in the paraxial region. The spatial filter (opaque circular obstacle placed in the back focal plane of the first lens) inserted at the Fourier plane blocks the low spatial frequencies. The transmitted frequencies are further transformed into a BB by the inverse Fourier transformation carried out by the second lens of focal length $f_2$. The BB formed after spatial filtering will not have any on-axis intensity modulations. As compared with the previous technique it is effective with a low cost of optical power. In this technique selection of the size of the optical blocker at the exact Fourier plane is tricky. We could also still see some low-intensity modulation in the BB due to the edge diffraction effects of optical blocker used for spatial filtering.

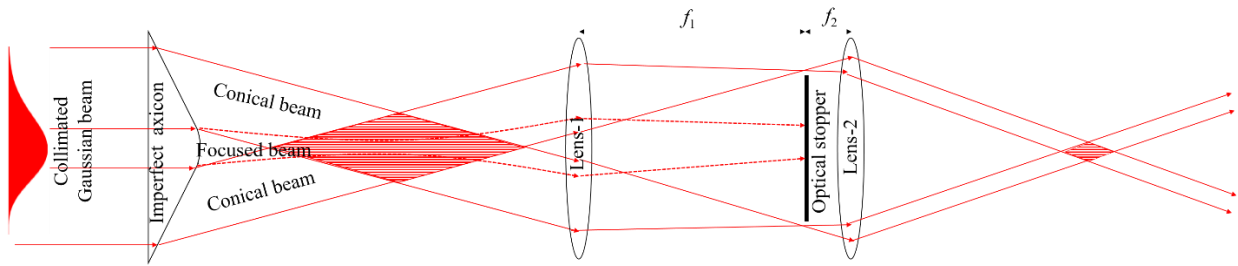

Fig. 3.20. Ray diagram of BB generation from round-tip axicon under spatial filtration.

The third technique of removing the on-axis intensity modulation due to the round-tip of the axicon is the HoG/annular beam-pumped axicon. The schematic diagram of BB generation with axicon under the HoG beam pump is shown in Fig. 3.21. The HoG beam is a doughnut-shaped beam with no helical wave-front [112] and it can be easily synthesized with SLM or SPP [113-117]. As shown in the below Fig. 3.21, first we can pass the Gaussian beam through the SPP of topological charge $l$ to produce a Gaussian vortex beam of order $l$ and further this beam can pass through the second SPP of topological charge, $-l$. The second SPP creates azimuthal phase retardation in the opposite direction to the first SPP. Hence, the resultant azimuthal phase of the optical beam is zero with doughnut shape intensity distribution called HoG beam. The HoG beam while it passes through the axicon, will not experience manufacturing defects present at the tip of the axicon owing to its dark core [118]. Hence, by using the HoG pump, we can completely prevent the undesired on-axis intensity modulations while other techniques couldn't prevent but only alleviate the intensity modulations. This technique is better than the previous two techniques owing to the fact of no power loss and diffraction effects due to obstacles.

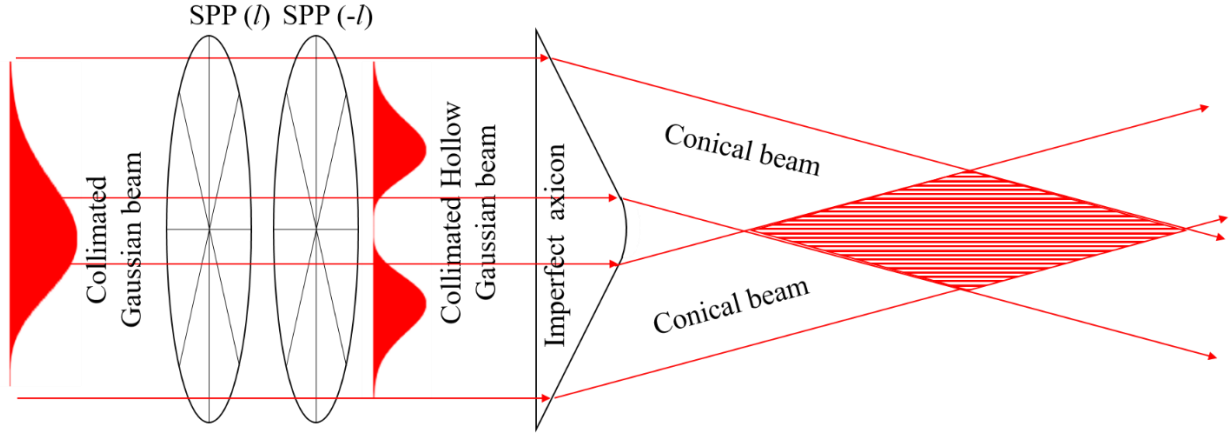

Fig. 3.21. Ray diagram of Bessel beam generation with a round-tip axicon which is pumped with hollow Gaussian beam.

There is an alternative way to remove the round-tip effect of the axicon on the BB without modulating the pump beam by immersing the axicon in a suitable liquid [119]. Through the immersion of conventional glass axicon in index-matching liquid, we can minimize its aberration effect on the quality of BBs as shown in Fig. 3.22. The focused rays from the round-tip of the axicon have the same angle at the liquid, glass, and air interfaces. It leads to the reduction of the focusing efficiency of the axicon tip. As a result, we could reduce the on-axis intensity of focused rays from the round-tip and finally, we can decrease the on-axis oscillation depth in the BB. The phase retardation provided by the axicon of refractive index $n_a$ immersed in a liquid of refractive index $n_l$ is given by

$$\phi(r') \propto \tfrac{2\pi}{\lambda} r' \tan\left[(\alpha(n_l - n_a))\right]. \qquad (3.22)$$

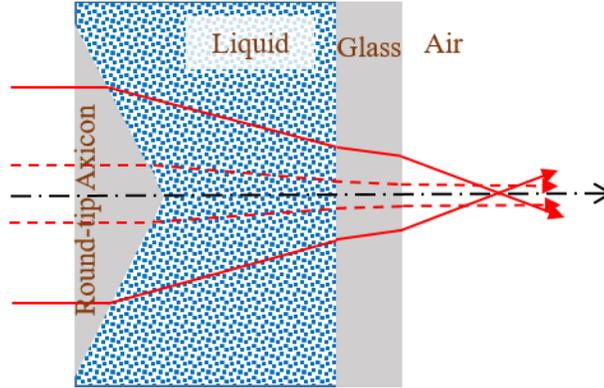

Fig. 3.22. Ray diagram of Bessel beam generation via liquid immersed axicon. Here rays corresponding to the full line are diffracted with the axicon phase without any defect and dashed line rays are diffracted from round-tip of the axicon.

Aberrations that occurred in the BB due to refraction of the axicon can also be avoided with the use of a reflective axicon [120, 121]. The differences between refractive and reflective axicons in terms of their functioning can be easily understood. For example, the refractive axicon is similar to the plano-convex lens and the reflective axicon works like concave mirror. While the refractive axicon is a positive axicon, the reflective axicon is a negative axicon. As we have seen till now in this section, the refractive axicon axis is always parallel and coincides with the incident beam axis, however, the reflective axicon is obliquely illuminated with the pump beam. The refractive axicons got more attention than the reflective axicons in the BB generation due to their simplicity in alignment. The BB generation with reflective axicon is illustrated with a ray diagram in Fig. 3.23. The reflective axicon provides improvements regarding damage threshold, chromatic aberrations, and group velocity dispersion as compared with the refractive axicons. The round-tip effect on the BB can be removed with these axicons [122,123]. However, their alignment and access of BB from them are quite difficult as compared with the conventional axicons. In the reflective case, it must be necessary to engineer the surface to generate off-axis beams to avoid potential distortion due to the oblique illumination of the pump beam. The phase retardation provided by the reflective axicon is given by [122]

$$\phi(x', y') = \tfrac{2\pi}{\lambda} \tan(\theta/2)\sqrt{\cos^2(\psi)x'^2 + y'^2}. \qquad (3.23)$$

Here ψ is the incident angle of the beam on the axicon and θ is the conical angle of BB. The on-axis intensity distribution of the BB under reflective axicon was obtained in ref. 122 as

$$I(z) = \frac{8\pi P_0 z \sin^2(\theta)}{\lambda w^2} \exp\left[-2\left(\frac{z\sin(\theta)}{w}\right)^2\right] \quad (3.24)$$

Where $P_0$ is the peak power of the incident Gaussian beam on the axicon. The central lobe size of $0^{th}$ order BB $r_b$ and Bessel range $z_{range}$ for Gaussian beam pumped reflective axicon are given by respective Eq. 3.25(*a*) and 3.25(*b*) [123].

$$r_b = \frac{1.75\lambda}{4\pi\alpha}, \quad (3.25a)$$

$$z_{range} = \frac{w}{2\alpha}. \quad (3.25b)$$

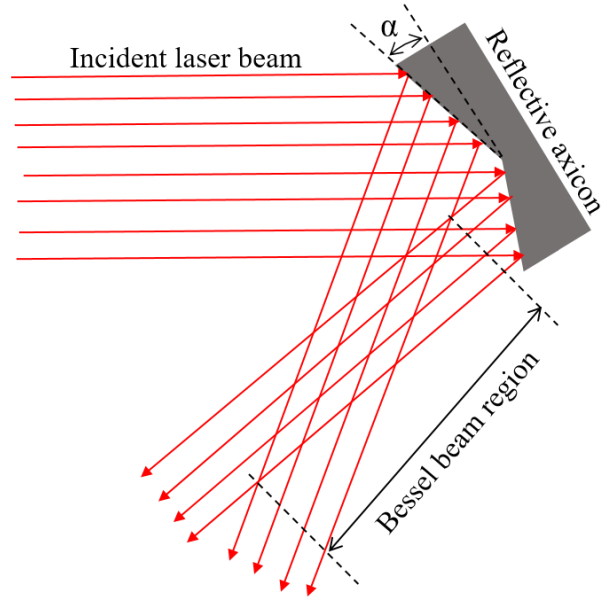

Fig. 3.23. Ray diagram of Bessel beam generation with reflective axicon.

It is worth noticing that at any wavelength in the electromagnetic spectrum, we can generate high-power BB with axicon by manufacturing it from a suitable transparent material [124-126].

Advantage:
1. Axicon has a high damage threshold so it can be used for synthesizing high-power BB.
2. Experimental configuration is very simple compared with other techniques. It can be easily integrated into any experimental setup for the applications of BB.
3. The conversion efficiency of BB from the pump beam is very high as compared with other techniques.
4. Wide-range electromagnetic transparent axicon can be easily manufactured to generate wavelength versatile BB.
5. To avoid the round-tip defect, chromatic aberrations, and group velocity dispersion in the BB, reflective axicons are better than refractive axicons.

Disadvantage:
1. Axicon alone can produce only $0^{th}$ order BB. For higher order BB generation, we need to pump the axicon with Gaussian vortex which is generally acquired by SPP/SLM, or directly from the vortex laser cavity.
2. In the case of Gaussian beam pumped axicon, the BB has on-axis intensity modulation along the propagation due to unavoidable manufacturing defect at the axicon tip.

### *3.4. Fabry-Perot resonator*

The Fabry-Perot resonator is formed by two highly reflective mirrors for the wavelength of interest, placed parallel to each other in the opposite direction. The constructive and destructive interference that occurred within the standing waves formed in the resonator produces circular symmetric spatial-frequency distribution as an output. Selectively choosing a single ring from the Fabry-Perot resonator and its Fourier transformation produces BB

[43,127,128]. The schematic ray diagram of BB generation from the Fabry-Perot resonator can be visualized in Fig. 3.24.

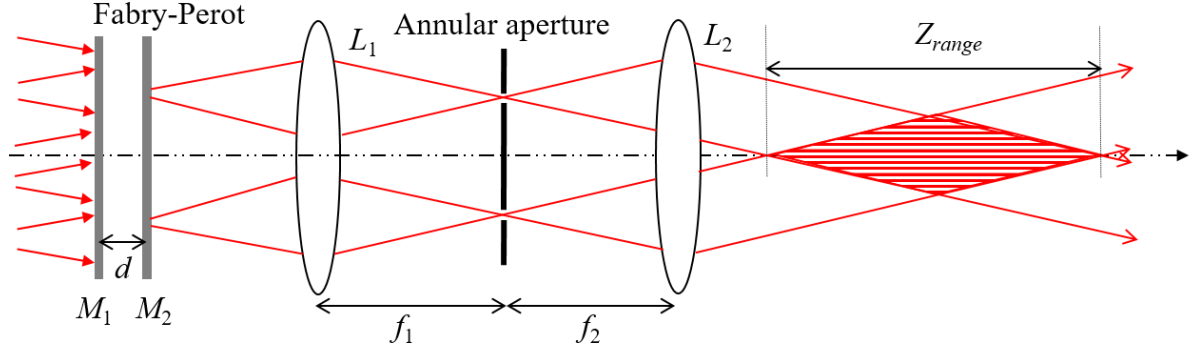

Fig. 3.24. Ray diagram of Bessel beam formed in Fabry-Perot resonator. $M_1$ and $M_2$ are the mirrors of the Fabry-Perot resonator. The combination of lenses $f_1$ and $f_2$ with an annular aperture is used for spatial filtering and for the generation of the Bessel beam. The separation between the Fabry-Perot mirrors is $d$.

The circular fringes delivered by the Fabry-Perot resonator formed by $M_1$ and $M_2$ mirrors, collected by the lens $L_1$. An annular aperture with selective radius and width is placed at the focal length $f_1$ of the first lens to open the window for the required circular ring and opaque for the rest of them. The second lens $L_2$ is placed next to the annular aperture at a distance equal to its focal length $f_2$ to produce BB. The generated BB properties can be expressed in terms of the properties of lenses and the Fabry-Perot resonator [127]. The central bright spot of the $0^{th}$ order BB is provided by

$$r_0 = 0.383 \left(\frac{f_2}{f_1}\right)^2 (\lambda d)^{1/2}. \quad (3.26)$$

The Bessel range over which we can have the BB properties after the second lens is given by

$$z_{range} = \left(\frac{Fd}{\pi\sqrt{R}}\right)\left(\frac{f_2}{f_1}\right)^2. \quad (3.27)$$

Where $F=\pi R^{1/2}(1-R)$ is the finesse of Fabry-Perot and $R$ is the reflectance of mirrors: $M_1$ and $M_2$. Even for fixed Fabry-Perot cavity configuration, we can control the BB properties for any application by changing the focal length of two lenses $L_1$ and $L_2$.

Advantage:
1. Similar to the annular aperture method, this technique is cost-effective, and it can be easily used in undergraduate laboratories for demonstrating the BB to the students.
2. This technique produces very less on-axis intensity modulations in the BB.

Disadvantage:
1. This technique cannot be used for most of the commercial applications due to its low conversion efficiency of optical power from pump beam to BB.

### *3.5. Digital micro-mirror device*

The digital micro-mirror device (DMD) is a product of micro-mechanics controlled by a computer similar to SLM [129,130]. It is a reflective type of SLM. The DMD contains numerous micro-mirrors as actuating components to switch small portions of light on and off. Recently, the DMD has been applied in laser beam shaping, to generate various structured modes [131-138]. Here, millions of DMD pixels provide numerous control freedoms for tailoring the wave-front of light to produce high-quality BBs. A detailed analysis of the DMD is provided elsewhere [132]. The required hologram can be projected on the DMD screen with the interfaced computer in a similar way to SLM to generate the BB. It is also noted that continuous hologram works for DMD by reason of it is capable of modulating the light intensity through sequential modulation of the width of the electric addressing signal. In this technique, the output signal is not linearly proportional to the grayscale in SLM. However, this relationship is in terms of the gamma curve. In order to minimize the distortion in the output beam of DMD, linear correction is required to correct the projected holograms [139].

Advantage:

1. The DMD is potentially better than the liquid crystal SLM in speed, spectrum sensitivity, and polarization modulation.

Disadvantage:
2. This technique is expensive.
3. The experimental setup is bulky and computer interfacing is required as in the case of SLM.

*3.6. Bessel beam generation with fiber*

Fiber-based BB generation can fulfill the rapidly growing demands of compact and flexible BB delivery in microscopic or submicroscopic applications, which have not been attained in prior bulk diffractive optical elements. As discussed in the previous subsections, the BBs generation with diffractive optical elements involves space-consuming optics. To overcome this limitation, efforts have been made to create BBs from optical fibers through the special design of their structure. As shown in Fig. 3.25, the diffractive structure created within or on the outlet tip of the fiber produces suitable BBs for micro-optics. Below we have given some of the techniques used to generate BBs from the fiber.

Bessel-like beams were generated by engraving a plasmonic lens consisting of a subwavelength slit-metallic groove structure on the cleaved end facet of a composite optical fiber using the focused ion beam milling process [140]. Higher order BBs were demonstrated by directly fabricating multi-element structure onto a single-mode optical fiber facet with the aid of Three Dimensional (3D)-direct laser writing [141]. Higher order cladding mode excitation with a long period fiber grating was used to experimentally generate the $0^{th}$ order BB. The number of rings, width, and propagation distance of the central lobe of BB were accurately controlled by means of experimental parameters [142]. Demonstration of $0^{th}$ order BB was carried out through the fabrication of a deep-seated negative axicon with micrometer dimensions inside a selective optical fiber tip by chemical etching of the fiber tip in hydrofluoric acid under the influence of capillary action [143]. A compact all-fiber BB generator was designed and demonstrated using a hollow optical fiber and coreless silica fiber based on a self-assembled polymer lens [144]. Fully controllable $0^{th}$ and higher order BBs were demonstrated through a photonic structure that uses stacked miniaturized optical elements 3D printed in a single step on the fiber facet [145].

Advantage:
1. It is very compact compared with other diffractive optical elements.
2. It can be easily integrated into any commercial experimental setup for BB applications.

Disadvantage:
1. It is very difficult to manufacture fiber with a diffractive structure to produce BB.
2. The quality of the Bessel mode generated in this technique is not better than the same mode generated by conventional techniques.
3. In most cases, it is not possible to generate an order tunable BB from a single fiber.

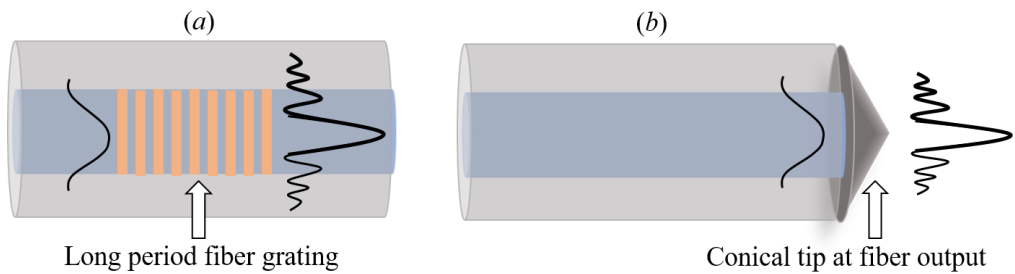

Fig. 3.25. (*a*) Long period fiber grating in the single mode fiber transforming the Gaussian mode into Bessel mode, and (*b*) Conical shaped tip at the output facet of fiber producing Bessel mode.

*3.7. Direct generation of Bessel beams from the laser*

Till now, in this section, we discussed BBs generation in a passive way (generation of BBs outside of the laser cavity). In this subsection, we are going to discuss the active way of BBs generation which is the process of directly generating the BBs from the laser cavity. Conventional laser cavity under without any intra-cavity diffractive optical elements support the generalized Gaussian modes: LG, HG, and IG modes. By providing circular, rectangular, and elliptical symmetry under proper gain control, we can produce respective LG, HG, and IG modes directly from the laser cavity [146-148]. In a similar way, we can also produce BBs directly from the laser cavity by introducing suitable phase retardance of BB in the laser cavity. Several optical resonators for BB generation have been proposed by Durnin and Eberly [149]. Some of the configurations used to generate BBs and BGBs directly from the laser cavity are presented in Fig. 3.26. As shown in Fig. 3.26 (*a*) and (*b*), we can produce BBs by using reflective or refractive axicon as one of the end mirrors or as an intra-cavity diffractive optical element in the laser cavity. A. N. Khilo et. al. proposed a simple linear laser cavity in the presence of an intra-cavity reflective axicon at close to the output coupler to produce $0^{th}$ order BB and BGB [150]. While the plane-plane laser cavity produces BB, the plane-concave laser cavity supports the BGB. The relation between the laser cavity parameters and BB parameters given by the pair of Eqs. 3.28($a$) and 3.28($b$) as

$$\alpha = \frac{\zeta}{1+4L^2/(kW_0^2)^2}, \qquad (3.28a)$$

$$R = L + \frac{(kW_0^2)^2}{4L}. \qquad (3.28b)$$

Here, $R$ and $\alpha$ are mirror radius and axicon parameter (base angle of axicon) respectively and $L$ is the cavity length. The cone angle and width in the BB are $\zeta$ and $W_0$ respectively. J. Rogel-Salaza et. al. proposed a laser cavity, formed by refractive/reflective axicon and plane mirror, to produce BBs [151]. They have used the idea of BB is viewed as a transverse standing wave formed in the interference region of incoming and outgoing conical waves, to develop their theory and condition for producing maximum gain in the cavity in terms of cavity length $L$ with $L=2a/\alpha$. The parameter $a$ is the radius of the axicon. For similar experimental laser cavities, another two groups used the condition for BB generation is [152,153]

$$L = \frac{a}{2(n-1)\alpha}. \qquad (3.29)$$

Here, $n$ is the refractive index of the axicon. Further, thin-disc [154,155] and unstable [156,157] laser cavity configurations are also theoretically demonstrated for the generation of BB. The cavity configuration shown in Fig. 3.26 ($c$) was used by Uehara and Kikuchi to produce nearly diffraction-free beams. This scheme was based on the transformation of a ring field into a Bessel light beam by means of an intra-cavity Fourier-transform lens. One of the resonator's mirrors was a plane, and the other was a plane mirror with an annular aperture. This scheme is nothing but intra-cavity annular aperture BB generation [158]. A. Onae et. al. produced nearly diffraction-free beams from a $CO_2$ laser with cavity configuration shown in Fig. 3.26 ($d$). Here, they used a concave mirror and an output mirror with high reflectivity (99.4%) in the marginal part and low reflectivity (94.5%) in the central part to produce BBs [159]. J. K. Jabczynski used a confocal resonator with an annular active medium to generate BB [160]. Also, there are some more techniques were used to generate BB and BGB directly from the laser cavity [161-165].

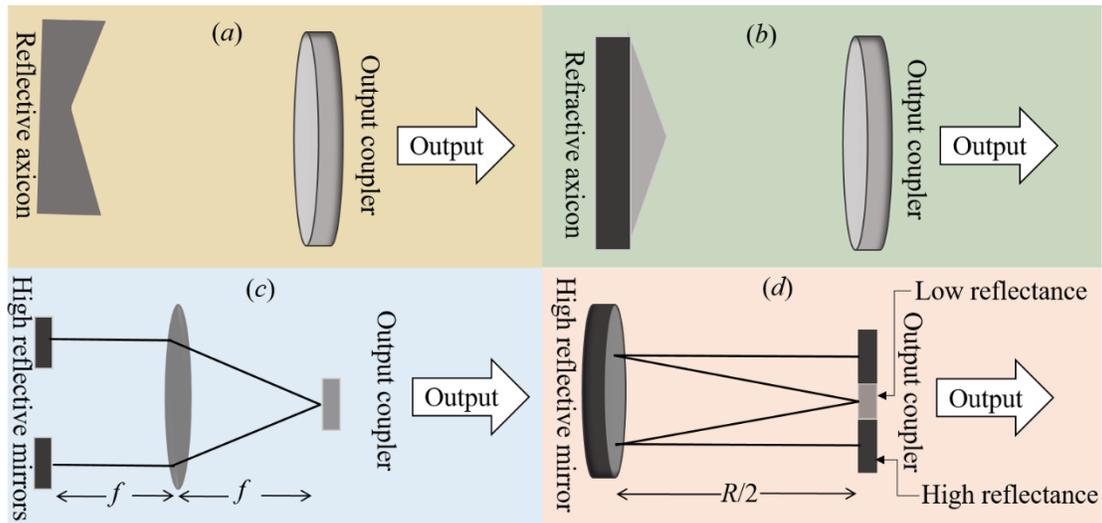

Fig. 3.26. Laser cavity configurations demonstrated for direct generation of Bessel beam from the laser cavity: ($a$) reflective axicon-based laser cavity, ($b$) refractive axicon-based laser cavity, ($c$) plane-plane laser cavity with intra-cavity lens aperture, and ($d$) plano-concave laser cavity with non-uniform transmittance output coupler.

Advantage:
1. It is very compact and cost-effective as compared with the passive method.
2. It produces high-quality BBs.
3. This scheme provides the possibility of realizing intra-cavity frequency conversion of BBs.

Disadvantage:
1. It cannot be easily integrated into any commercial experimental setup for BB applications.
2. This technique may not be useful for tunable order BB generation.
3. By this technique we cannot produce high-power BBs.
4. Some of the laser configurations used for BB generation cannot be used to generate other structured modes.

*3.8. Other techniques used for the generation of Bessel beams*
In addition to the most popular techniques used to generate BBs which we discussed in previous subsections, there are multiple methods that were used to generate BBs. One such technique is the Fresnel zone plate in which its phase retardance varies in the radial direction [166-167]. The Fresnel zone plate can easily modulate an incident plane wave into a conical shape to produce BB. Metasurfaces consisting of subwavelength-spaced phase shifters have been demonstrated to fully control the optical wave-front to produce wavelength and order tunable BBs with high mode purity [168-173]. Also, the BBs were generated using an inward Hankel aperture distribution, leaky-wave modes, and a radial line slot array [174-176].

**4. Tunable Bessel beams for applications**
In the previous section, we explored various methods to generate BBs in great detail. These methods under Gaussian mode pumping have suffered from limitations in generating user-friendly BBs for applications. More refined methods are needed for the generation of intended BB with high quality. Moreover, constructive amendments to these methods are desirable to make them more powerful. Especially from the application point of view of BBs, $0^{th}$ order BB have seen much interest over higher order BBs owing to its high on-axis intensity. In the properties of $0^{th}$ order BB, on-axis intensity modulation is very imperative for modern science applications. In experimentally realized BBs, the intensity along the propagation depends primarily on the transverse intensity distribution of the input beam and the transmittance function of the Bessel mode converter. For example, the collimated Gaussian beam illumination on an axicon produces the Bessel beam with its longitudinal axial intensity governed by linear and Gaussian function of $z$ i.e., BB intensity is proportional to $z \exp(-z^2/z_{max}^2)$ [87]. Over the past three decades, by due consideration of pump beam transverse intensity distribution and structure of Bessel mode converter as the tuning parameters, several techniques have been proposed and successfully utilized to generate various kinds of on-axis intensity modulations in BBs with tunable range, variable peak intensity position. The generation of BBs with a tunable axial intensity within the accessible range of spatial frequencies can have multiple advantages over their straightforward generation, especially in light-matter interaction. Some of the techniques presented in the past 3 decades for Bessel beams' intensity modulation are discussed below.

*4.1. Diffractive optical elements*
R. Piestun and J. Shamir [177] proposed a method based on the projections-onto-constraints-sets algorithm to design diffractive elements, which include phase and amplitude, for the control of beam propagation in a 3D region. Based on the iterative method for designing the diffractive element, they could control the intensity of BB along its propagation in the range of 4 m. However, this technique is very difficult to use to generate smooth on-axis intensity distribution along BB propagation in a required shape.

Also, there is another method based on fluidic axicon was used to generate BBs with dynamic reconfiguration. The optical cone angle created by fluidic axicon is given by [178]

$$\theta_f = \sin^{-1}\left\{\frac{n_{PDMS}}{n_{air}}\sin\left[\sin^{-1}\left(\frac{n_{PDMS}}{n_{air}}\sin\alpha\right)-\alpha\right]\right\}. \qquad (4.1)$$

Here $n_{air}$ and $n_{PDMS}$ are refractive indices of air and polydimethylsiloxane respectively and α is the base angle of the axicon. The spatial properties of delivered BB can be easily controlled by simply changing the refractive index of fluid filled in the axicon. The tunability of on-axis intensity depends on the available optically transmissive fluids' refractive index. The fluidic axicon can be used for high-power laser beams. Although this technique can reduce the on-axis oscillations due to the axicon tip but not completely.

Another technique of BB generation is from doublet-lens axicon which is made from spherical surfaces [179]. The lens axicons can be designed in such a way that they can minimize oblique aberrations and produce the required on-axis intensity. This prevents the off-axis broadening and distortion of the focal line that happens with an ordinary axicon. The lens axicon can also be compensated for chromatic aberrations, allowing for broadband illumination. In addition to the above advantage, it has a disadvantage in the generation of tunable peak intensity position and on-axis intensity of BB.

The tunable peak intensity position, on-axis intensity, and range of the BB can be successfully achieved by SLM-based holographic grating [180,181]. Here, the properties of BB will be controlled in a required direction by adjusting the initial parameters used in the hologram construction. This technique can efficiently convert the maximum optical power of the pump beam to BB. However, this technique cannot be used for high-power laser beams whose power is above the damage threshold power of SLM.

*4.2. Apodization*

The on-axis intensity of BB can be modulated in the required shape by using an apodizing filter of transmittance function $T(r)$ before the diffractive optical element. After insertion of apodizing filter, the expression for BB in $z>0$ is given by [182]

$$E(r,z) = A\exp(-ik/z)\exp\left[ik\left(z+r^2/2z\right)\right]\int_0^R T(r')J_0(k_r r')\exp\left(ikr'^2/2z\right)J_0(krr'/z)r'dr'. \quad (4.2)$$

The apodizing filter transmittance function $T(r')$ is given by

$$T(r') = \begin{cases} 1 & 0 \leq r' \leq \epsilon R \\ t(r') & \epsilon R < r' \leq R \end{cases}. \quad (4.3)$$

Here, $0 \leq \epsilon \leq 1$ and $t(r')$ is the aperture edge function which is smoothly decreasing in the edge region. Based on this technique, A. J. Cox and Joseph Danna [182] could produce constant on-axis intensity in the BB of range $z_{range}= 30$ mm by utilizing $t(r') = \frac{1}{2}\{1+\cos[\pi(r'-\epsilon R)/(R-\epsilon R)]\}$ and $t(r') = \exp[-(r'-\epsilon R)^2/w^2]$.

The BB with constant intensity along the optical axis can also be produced by the combination of apodizing transmittance $t(r') = 1/(r')^{1/2}$ and two phase-only holographic elements [183]. Here, the desired axial intensity distribution of BB can be produced by selecting the proper apodization of the intensity from the first holographic element. The apodizing filter can control the longitudinal intensity distribution of BB in a required shape at the cost of input laser power.

*4.3. Spatial spectrum*

T. Cizmar and K. Dholakia [184] have experimentally demonstrated BB with constant and ramp shape on-axis intensity along the propagation by analyzing the Fourier transformation between on-axis spatial frequency function $U(r=0, z)$ and spectrum of spatial frequencies in $k$-space $S(k_r, z=0)$, provide by Eq. 4.4.

$$U(r=0,z) = \int_0^k k_z S(k_r = \sqrt{k^2 - k_z^2}, z=0)\exp(ik_z z)dk_z. \quad (4.4)$$

The key information provided by this Fourier transformation is that for any azimuthally independent field, its on-axis evolution is linked with its spatial spectrum by a one-dimensional Fourier transform. Hence, one can produce BB with an arbitrary on-axis field dependence within the range of possible spatial frequencies in $k$-space. To further avoid the on-axis oscillations in BB due to the hard-aperturing of the spatial spectrum they have introduced a Gaussian amplitude envelope around the spatial spectrum centered at the value of $k_{r0}$.

Also, an apodized annular-aperture logarithmic axicon was used to produce constant on-axis intensity [185,186]. The phase retardation acquired in this apodization is

$$\varphi(r') = -\frac{R^2 - \rho^2}{2(z_{max} - z_{min})}\ln\left(z_{min} - \frac{z_{max} - z_{min}}{R^2 - \rho^2}\rho^2 + \frac{z_{max} - z_{min}}{R^2 - \rho^2}r'^2\right) + \text{const.} \quad (4.5)$$

Here $\rho \leq r \leq R$ and $\rho$ is the radius of the masking opaque disk centered within the axicon of radius $R$. The intensity distribution in BB after substituting the phase retardation in Fresnel diffraction integral is given by

$$I(r,z) = \left(\frac{2\pi}{\lambda z}\right)^2 \left|\int_0^R \exp\left\{ik\left[\varphi(r') + r'^2/2z\right]\right\}J_0(krr'/z)r'dr'\right|^2. \quad (4.6)$$

It is also noted that this method can control the longitudinal intensity distribution of BB in a required shape at the cost of input laser power.

*4.4. Annular aperture*

A simple and cost-effective technique for the generation of tunable range and position of the BB is by the use of an annular aperture with either the simple lens or axicon [187,188]. As shown in Fig. 4.1(*a* & *b*), the Fourier transformation of an annular aperture beam with a normal convex lens can produce the tunable range and peak position of BB. The onset and offset positions of BB in terms of annular aperture radii are acquired as $z_{min}=r_1/\tan\theta$ and $z_{max}=r_2/\tan\theta$ respectively. Here, $r_1$ and $r_2$ are the respective inner and outer radii of the annular aperture and $\theta=d/(2f)$ with $d$ as an annular slit diameter. The range of the BB is controlled with $r_2 - r_1$ while the peak position of BB depends on the mean radius of the annular aperture. We can also generate similar tunable BB by the combination of an annular aperture with an axicon [Fig. 4.1(*c*)]. Even though we could generate BB with tunable range and position by the annular aperture technique, it was not successfully used due to high optical power loss in the mode conversion and on-axis oscillations due to aperture diffraction. Moreover, we cannot generate high-peak power tunable BBs with this technique.

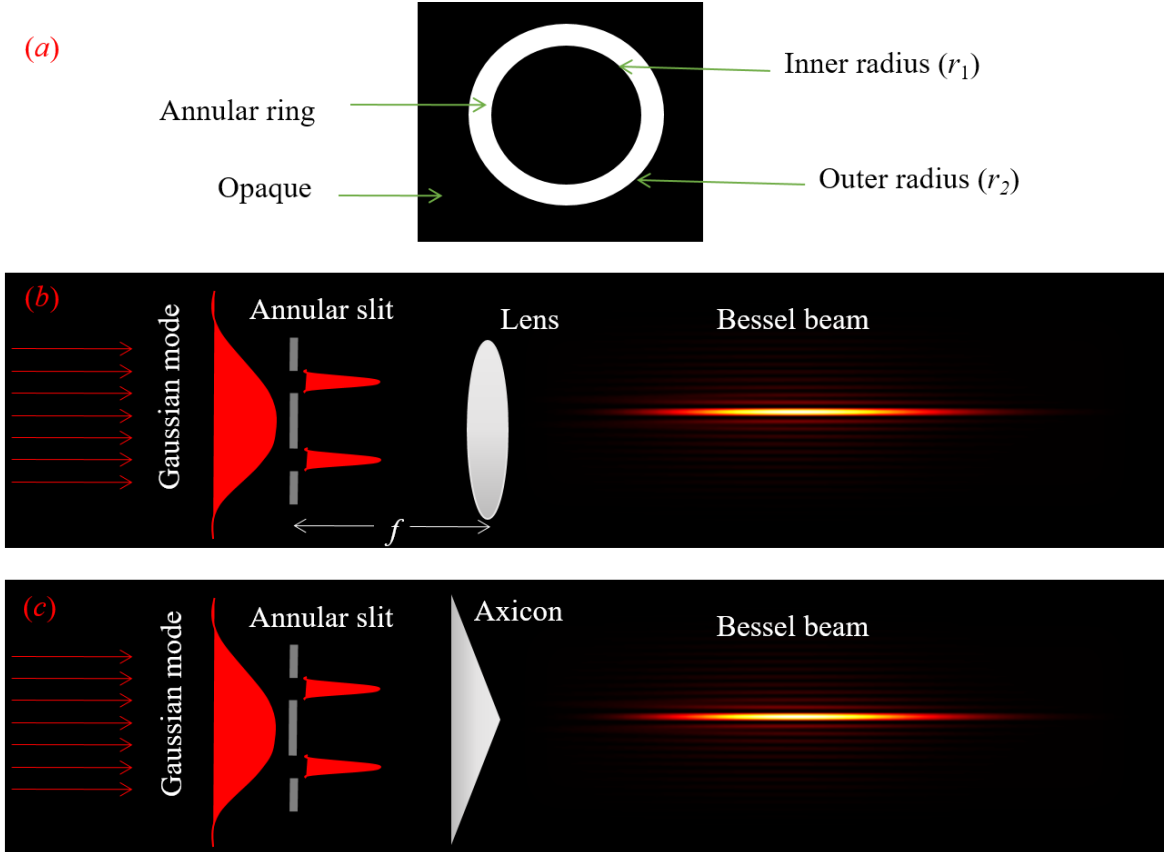

Fig. 4.1. The tuning of position and range of Bessel beam by an annular aperture: (*a*) annular aperture structure, (*b*) tunable Bessel beam generation in Durnin experimental configuration, and (*c*) combination of annular aperture and axicon for generation of tunable Bessel beam.

### 4.5. Pump mode intensity modulation

As discussed in the experimental section, the Gaussian mode pumped axicon produces high-power BB with high mode conversion efficiency with costing inevitable on-axis intensity modulations. Additionally, we cannot tune the shape and position of BB. On the other side, an annular aperture provides a tunable size and range of BB with very low mode conversion. By using a HoG beam as an input source to the axicon, we can get the advantages of both annular aperture and axicon without any power loss in the mode conversion. To achieve this, first of all, the Gaussian beam has to be converted into a HoG beam either by using two identical SPPs or by the computer-generated holograms [115-118]. Both the diffractive optical elements provide the maximum conversion efficiency. The experimentally generated optical field amplitude of the HoG beam of order *m* is in the form of [113,118]

$$E_m(r',0,\phi') = \left(\sqrt{\frac{P2^{2m+1}}{\pi w^2 (2m)!}}\right)\left(\frac{r'}{w}\right)^{2m} \exp\left(\frac{-r'^2}{w^2}\right). \qquad (4.7)$$

Here, *P* is the optical power of the HoG and *w* is the Gaussian spot size of the collimated HoG beam at axicon. After inserting the beam amplitude of HoG and axicon phase in Eq. 3.1 and by using a straightforward integration, the optical field amplitude of $0^{th}$ order BB in terms of HoG beam order *m* is obtained as

$$E(r,0,\phi) \approx \left(\sqrt{\frac{2^{2m}}{w^2(2m)!}}\right)\sqrt{\frac{z}{z_{max}^G}}\left(\frac{z}{z_{max}^G}\right)^{2m} \exp\left(\frac{-z^2}{z_{max}^G{}^2}\right) J_0(k_r r)\exp\left[ik\left(z+r^2/2z\right)\right]\exp\left(i\frac{-ik_r^2 z}{2k}\right) \quad (4.8)$$

and its intensity is given by

$$I \propto \left(\frac{2^{2m}}{w^2(2m)!}\right)\left(\frac{z}{z_{max}^G}\right)^{4m+1} \exp\left(\frac{-2z^2}{z_{max}^G{}^2}\right) J_0^2(k_r r). \quad (4.9)$$

The onset position ($z^m_{min}$) and offset position ($z^m_{max}$) of BB are given by $z^m_{min} = r_1/(n-1)\alpha$ and $z^m_{max} = r_2/(n-1)\alpha$ respectively. Here, $r_1$ and $r_2$ are the inner and outer radii of the HoG beam. The BBs' onset and offset positions increase with increasing the order of the HoG beam with reference to the axicon position. As shown in Fig. 4.2, we can experimentally generate high peak power BB with tunable position, range, and on-axis intensity distribution by HoG-pumped axicon.

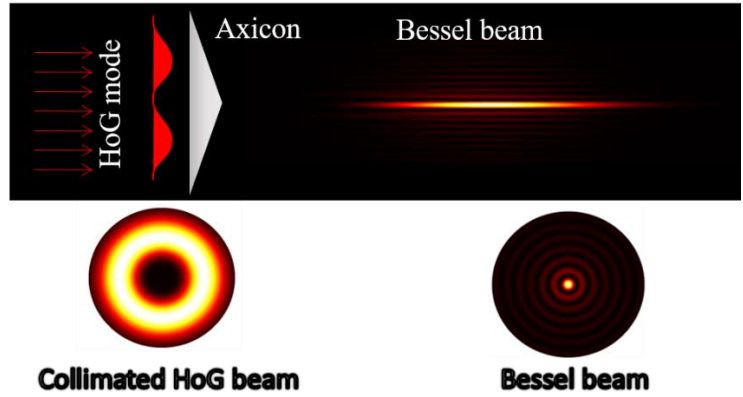

Fig. 4.2. The mode conversion of a Hollow Gaussian beam of 2nd order to a 0th order Bessel beam in the presence of an axicon.

As mentioned above, a single HoG mode pumped axicon can provide user-friendly BBs with a tunable on-axis position. However, it has a limitation in the modulation of on-axis intensity distribution. This constraint can be easily overcome by simply pumping the axicon with two superposed HoG modes with orthogonal polarizations. The amplitude of two superposed HoG beams is given by

$$E_{m_1,m_2}(r',0,\phi') = \left(\sqrt{\frac{P2^{2m_1+1}}{\pi w_1^2(2m_1)!}}\right)\left(\frac{r'}{w_1}\right)^{2m_1}\exp\left(\frac{-r'^2}{w_1^2}\right)\hat{x} + \left(\sqrt{\frac{P2^{2m_1+1}}{\pi w_1^2(2m_1)!}}\right)\left(\frac{r'}{w_1}\right)^{2m_1}\exp\left(\frac{-r'^2}{w_1^2}\right)\hat{y}. \quad (4.10)$$

The intensity of the resultant BB is given by

$$I \propto \left(\frac{z}{z_{max\,1}^G}\right)^{4m_1+1} \exp\left(\frac{-2z^2}{z_{max\,1}^G{}^2}\right) J_0^2(k_r r) + \left(\frac{z}{z_{max\,2}^G}\right)^{4m_2+1} \exp\left(\frac{-2z^2}{z_{max\,2}^G{}^2}\right) J_0^2(k_r r). \quad (4.11)$$

From Eq. 4.11, we can understand that the range of BB can be easily increased by pumping the axicon with superposed HoG modes. In superposed HoG pumping, the onset position of the BB is determined by the inner radius of the lower order HoG beam while the offset position is provided by the outer radius of higher order HoG beam. As shown in Fig. 4.3, we can extend the Bessel range with continuous tunability via pumping two superposed HoG beams to a single axicon. We can smoothly control the on-axis intensity of BB in a required shape by providing suitable weight factors to the two HoG beams. For example, the relative weight factor of both the HoG beams is considered as one for Figs. of 4.3(*a*), 4.3(*b*), and 4.3(*d*). In Fig. 4.3(*c*) we have used it as 27/1 for HoG of orders $m_1/m_2$ to generate constant on-axis intensity distribution in the BB. Based on the requirement, we can properly select the order and weight factors of HoGs to generate suitable BB. Here, the weight factors of modes are straightforwardly related to the optical power present in the corresponding modes.

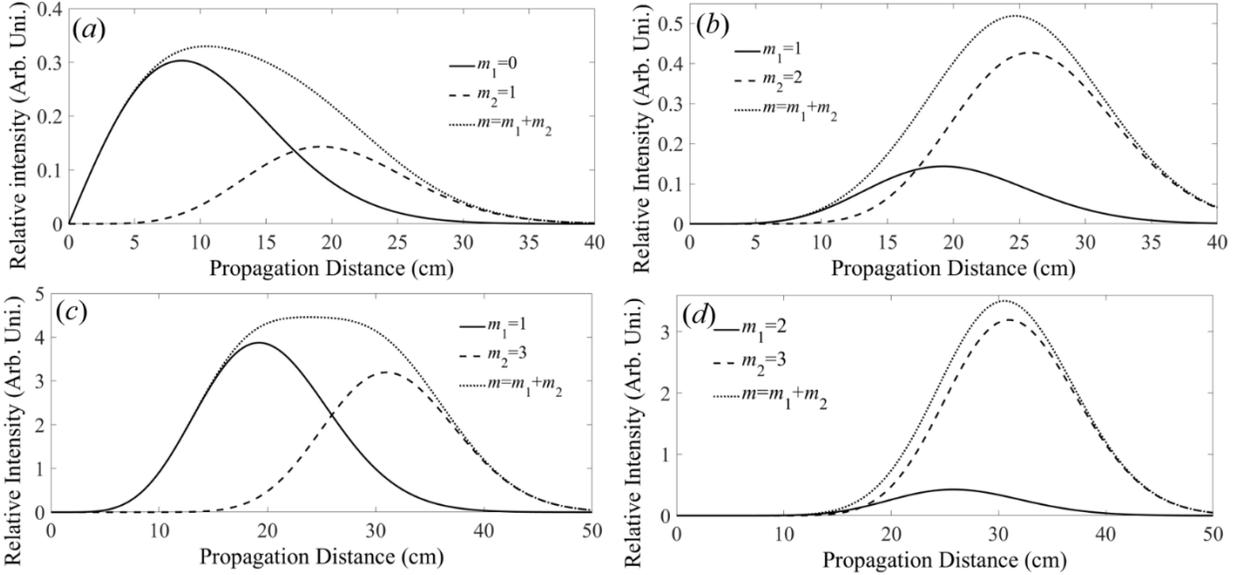

Fig. 4.3. Modulating and extending of on-axis intensity distribution in the Bessel beam along its propagation by pumping the superposed hollow Gaussian modes of different orders to a single axicon.

As shown in Fig. 4.4, we can create an optical potential dipole by using the superposition of HoG modes as a pump source to the axicon. The relative intensity of the lobes can be easily controlled by simply changing the optical power of individual HoG modes. For a given experimental configuration, the central lobe size and range of the BB are inversely proportional to the base angle of the axicon. Therefore, it is not possible to increase the Bessel range while maintaining constant central lobe size. However, we can overcome this difficulty by pumping the axicon with two selective HoG beams.

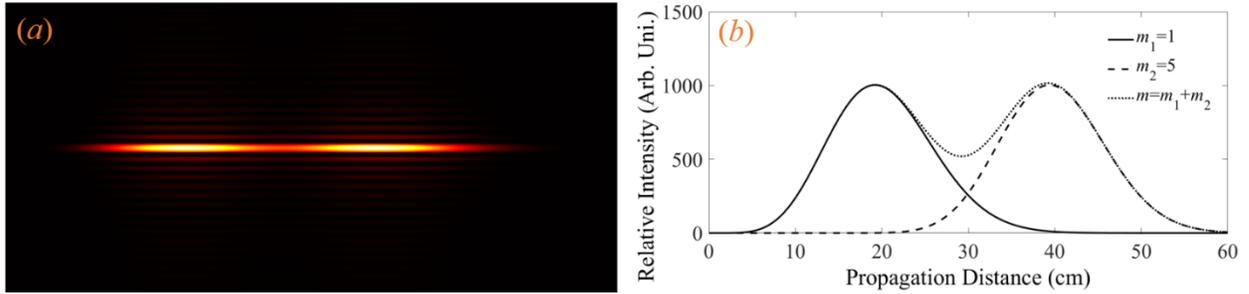

Fig. 4.4. Optical potential dipole created in the Bessel beam by pumping the axicon with superposed HoG modes of orders $m$ = 1 & 5. (a) The intensity distribution of the Bessel beam in the $xz/yz$ plane and its on-axis line profile along the propagation is given in (b).

As highlighted in this section and the previous section, the ability to tailor the properties of BB such as intensity, phase, OAM, range, and peak position either directly from the laser source or by using diffractive optical elements has paved the way for frontier technologies. The essence of these two sections will be appreciated and well understood when we discuss the applications of BB in section 9.

**5. Properties of Bessel beam**
The BBs have fascinating properties that we cannot perceive in most of the other generalized Gaussian beams. Non-diffraction and self-healing are the main properties of BB, and they make BB superior over other structured beams in some of the applications of light-matter interaction. This section provides a detailed analysis of the BBs' properties starting from their origin.

The origin and basic principle behind the BB can be understood as follows: Interference and diffraction are the two most fascinating features of light as a part of its wave nature. We encounter these two natural phenomena whenever we perceive or utilize light. In these two properties, while diffraction provides the divergence, interference can create intensity nulls (nodes) in the light. The phase at these intensity nodes is undefined. These nodes are either in points or lines depending upon the geometry followed by the individual light waves. The interference is used to encounter the diffraction to create a non-diffraction structure in light. Consequently, the light intensity formed in co-axial circular cylinders with a needle along their axis while the transverse profile of this structure follows the Bessel function.

### 5.1. Non-diffraction

As we know diffraction is a fundamental property of waves and optical beams formed by collection of electromagnetic waves. The diffraction of experimental optical beams in their free space propagation is understood with their Rayleigh length $z_R$, and is given by

$$z_R = \frac{\pi w_0^2}{\lambda}, \qquad (5.1)$$

here $\lambda$ is the wavelength and $w_0$ is the Gaussian spot size of the beam at its initial position where the beam has plane wave-front with zero diffraction. From the above Eq. 5.1, we can understand that the decrease in the Gaussian spot size makes its Rayleigh length shorter. Hence it is not possible to confine an optical beam within a small region in the order of its wavelength for certain long distances.

Even though, there are multiple theoretical attempts have been carried out on the creation of non-diffraction optical beams [189,190], the first optical beam theoretically demonstrated and experimentally realized with a non-diffraction nature is BB [191,192]. The non-diffraction nature of BB can be understood through the interference phenomenon as follows: One of the fundamental concepts of coherent light is that wherever you see light it is constructive interference and if not then it is destructive interference. This property of light allows us to generate various structured beams via phase engineering of coherent laser sources. As we discussed in the introduction, individual waves in the BB travel in a conical shape. Every wave interferes with other waves on the optical cone to produce constructive and distractive interference in the regular interval along the radial direction. This interference takes place along the optical axis of BB. As depicted in Fig. 5.1, the optical intensity in the constructive region is guided by the destructive region in the entire interference range (Bessel range). As a result, the BB acquires a non-diffraction nature without losing law of diffraction. Here, we must understand that the individual waves in the BB propagate with diffraction properties while their interference pattern is constant along the optical axis.

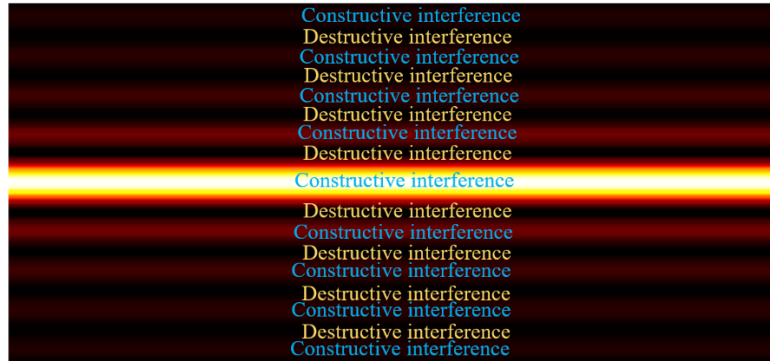

Fig. 5.1. The non-diffraction nature of the Bessel beam due to the interference phenomenon.

### 5.2. Self-healing

So far, we have discussed that the BB formation is due to a set of waves propagating on a cone and having a non-diffraction nature. It also possesses another interesting property, that of reconstruction or self-healing [193-196]. In the self-healing property, if we place an obstacle in the center of the beam, the waves that create the beam can move past the obstruction, casting a shadow into the beam, but ultimately reforming the intensity of the profile beyond the obstruction. The self-healing property of BB can be explained either by wave optics [197] or by ray optics [198,199]. However, here we demonstrate its self-healing property by ray optics owing to its simplicity. The ray diagram of BB in the presence of an obstacle is illustrated in Fig. 5.2. While the blocked rays due to an opaque object produce a shadow next to the obstacle, the unblocked rays reconstruct the BB. Based on geometric optics, the distance after which the BB can reconstruct is called the shadow length of obstacle $z_{sh}$ and is given by [200]

$$z_{sh} = \frac{a}{2\tan\theta} \approx \frac{ak}{2k_z}. \qquad (5.2)$$

Here, $a$ is the width of the obstacle. It is worth noticing that the self-healing effectiveness of BB is more prominent for circular obstacles owing to its cylindrical symmetry. As we discussed in our preceding sections, the Bessel range depends on the pump mode size at the diffractive optical element. So, the size of the obstacle must be smaller than the pump mode size. From Fig. 5.2, we can also understand that the self-healing of BB can be observed only when the optical shadow created by the obstacle is less than the Bessel range considered from the obstacle position. Furthermore, the 3D tunability of intensity distribution in BBs (discussed in the previous section) can be efficiently utilized in their self-healing property.

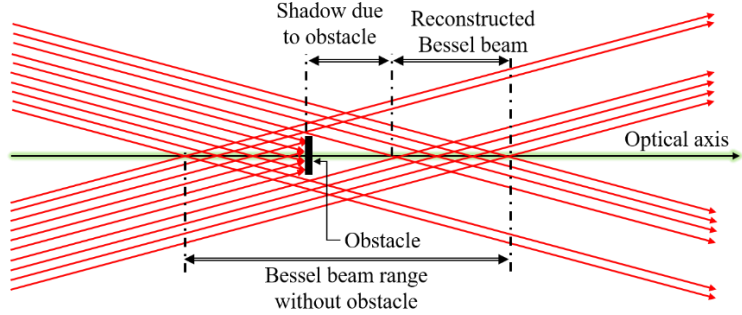

Fig. 5.2. The ray diagram of self-healing or reconstruction of the Bessel beam.

## 5.3. Gouy phase

The origin of the Gouy phase in the BB is due to the cross-propagation of its waves in a conical shape. In another way, we can also say that the origin of the Gouy phase in BB is because of the transverse component of the angular spectrum wave vectors which originates from its transverse confinement. The understanding of the Gouy phase in BBs is imperative for their utilization in the application of modern optics. The Gouy phase of BB can be analytically understood with reference to the plane wave through the difference of their propagation vectors along the propagation [201-203]. For a given monochromatic wavelength ($\lambda$), the reference plane wave and ideal BB have the propagation vectors of $k$ and $k_z$ along their propagation, respectively. Then the Gouy phase of BB is given by [204]

$$\phi_G = (k - k_z)z = k(1 - \cos\theta)z. \qquad (5.3)$$

The Gouy phase of BB primarily depends on the Bessel cone angle $\theta$. From Eq. 5.3, we can understand that the Gouy phase accumulated in BB is linearly increasing with its propagation distance $z$. It is worth noticing that the rate of change of the Gouy phase is independent of the order of the BB $l$. Further, the Gouy phase of BGB is different than the BB on account of the Gaussian envelop present in the BGB. The Gouy phase of BGB can be procured from Eq. 5.3 as [205]

$$\phi_G = \phi_{G1} + \phi_{G2}, \qquad (5.4)$$

where

$$\phi_{G1} = -\frac{k_r^2}{2k}z, \qquad (5.5a)$$

$$\phi_{G2} = -\frac{k_r^2 z^3}{2k(z^2 + z_R^2)}z - \arctan\left(\frac{z}{z_R}\right). \qquad (5.5b)$$

The total Gouy phase is the sum of the Gouy phase due to pure BB $\phi_{G1}$ and Gouy phase of Gaussian envelop $\phi_{G2}$. Also, the Gouy phase of the BGB in the presence of a focusing lens of focal length $f$ is given by

$$\phi_G = \frac{r_0^2 k}{2f^2}z\left(\frac{z^2}{(z^2 + z_R^2)} - 1\right) + z - \arctan\left(\frac{z}{z_R}\right). \qquad (5.6)$$

Here, $r_0$ is the ring radius of the annular beam used to generate the BGB through the lens which increases with the increasing topological charge of the annular beam. From Eq. 5.6, we can conclude that the rate of change of the Gouy phase of BGB under the focusing condition increases with its order and decreases with the focal length of the lens. The Gouy phase of any kind of BB in any circumstance can be understood by employing simple interferometric techniques. For instance, we can quantitatively measure and understand the Gouy phase of BB by interfering it with a collimated Gaussian beam in a free-space Mach-Zehnder Interferometer (MZI) [204].

## 5.4. Visibility

The quality of the BB primarily depends on its visibility. The ideal BB has constant visibility with $\eta=1$ everywhere in its propagation. However, this is not true in real BBs which are generated in an experimental laboratory. In laboratories, we generate BBs by using laser beams that have finite width with non-uniform intensity in beam cross-section. The BB formed as a result of self-interference of the incident beam through optical mode converters. The method of interference in BB generation is based on wave-front division. The non-uniform intensity over the wave-front produces spatial-dependent fringe visibility in interference. The non-uniform visibility in BB can be understood through ray optics analysis of BB formation in the presence of an axicon. In Fig. 5.3, we have illustrated the BB formation in the presence of a Gaussian beam and an annular beam pumping to an axicon. The color of the rays indicates the intensity along their path. The same color rays have the same intensity and the interference visibility $\eta = 1$ at their crossing position (white dot positions in Fig. 5.3) and different color rays produce interference with visibility $0 < \eta < 1$ at their crossing positions. The visibility at a given position depends on the relative amplitudes of interfering waves. In BBs, the fringe visibility is equal to one in the neighborhood of the optical axis and it decreases while increasing the radial position. Hence experimentally generated BBs can perfectly match with ideal BB close to their optical axis.

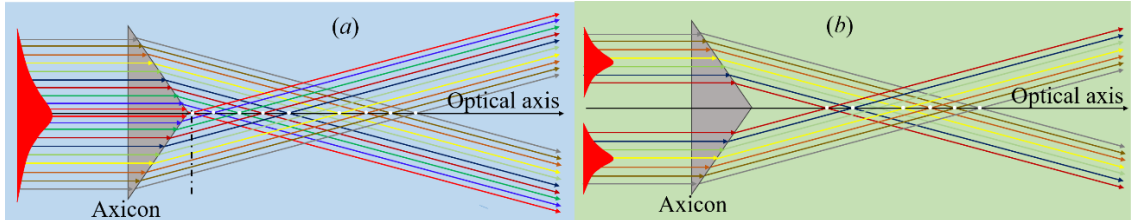

Fig. 5.3. The generation of Bessel beam with the aid of collimated (*a*) Gaussian beam pump and (*b*) annular beam pump to the axicon. Here, the color of each ray indicates its intensity. The white dots on the optical axis specify interference of waves with the same intensity to produce maximum visibility of $\eta=1$.

To further understanding, the intensity distribution of BGB produced from the paraxial Gaussian beam, in the *xz/yz* plane is depicted in Fig. 5.4. The parameters used in the simulation are collimated paraxial Gaussian spot size $w = 1$ mm, wavelength $\lambda=1064$ nm, axicon base angle $\alpha=1°$, and refractive index $n=1.5$. The BGB has dominant peak intensity at $z = 0$ and decreases with increasing its propagation [Fig. 5.4 (*a*)]. The fringe visibility in the longitudinal cross-section looks pretty good. However, this is not the exact dynamical properties of BGB. As shown in Fig. 5.4 (*b*), we can find the intensity distribution in the BGB cross-section as a function of propagation distance *z* in the *xz/yz* plane by normalizing its peak intensity along the propagation. As a function of propagation, BGB transformed into an annular beam. The fringe visibility in the longitudinal cross-section is position-dependent. The change in the fringe visibility of BGB is conspicuous in their line profiles provided in Fig. 5.4 (*c*). We can see that the visibility or the quality of the Bessel nature of BGB decreases with increasing its transverse and longitudinal position coordinates. The spatial dependency of fringe visibility as a function of *r* and *z* for any kind of real BB is unavoidable and it may not follow exactly the case of BGB.

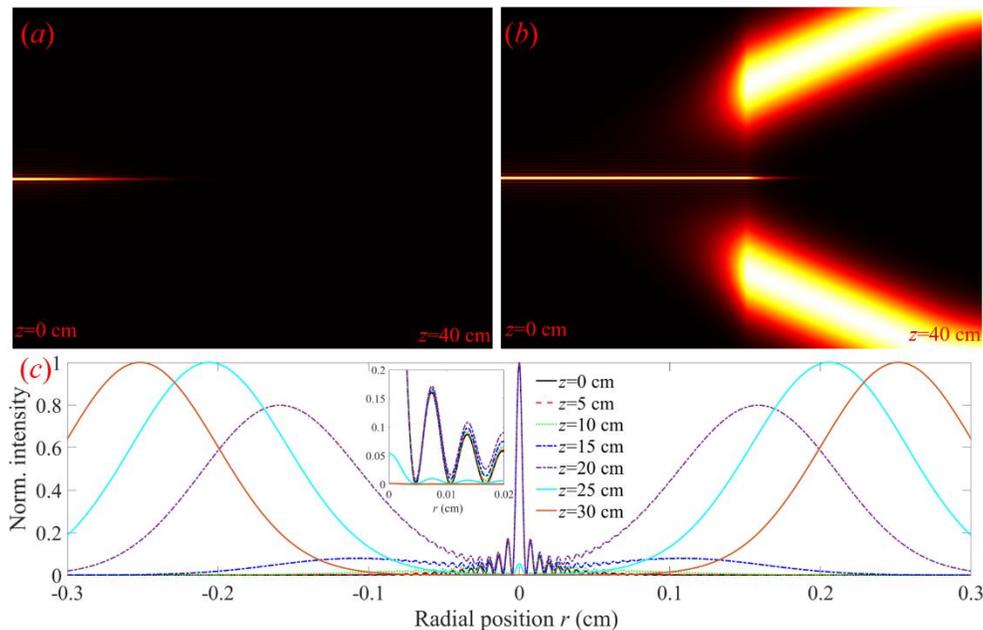

Fig. 5.4. The intensity distribution in the *xz/yz* plane of the Bessel Gauss beam (*a*) normalized with its peak intensity, and (*b*) normalized along the propagation. (*c*) The transverse axial intensity of the Bessel Gaussian beam normalized with their peak intensity at different propagation distances (the line profiles around the beam axis are shown in the inset for understanding the relative intensities).

*5.5. Polarization of Bessel beam*

Polarization is one of light's most salient features and plays a pivotal role in the applications. The state of polarization in the BBs is completely different with respect to other generalized Gaussian beams. This tendency arises due to the conical wave-front of BB and it creates 3D polarization distribution. The components $E_x$, $E_y$, and $E_z$ of the optical field $E$ are along the Cartesian coordinates $x$, $y$, and $z$ directions respectively. The explicit form of BB's amplitude in terms of Bessel cone half-angle $\theta$ and azimuthal angle $\phi$ is $E(x, y, z, \phi, \theta)$. For simple understanding, here, we can consider $r=(x^2+y^2)^{1/2}$, and the optical field $E$ is subtending an angle $\phi$ with the $x$-axis. For any state of linear polarization of BB, the Bessel cone angle $\theta$ splits its amplitude into $s$-polarized, $p$-polarized, transverse, and longitudinal components and all four components are function of azimuthal angle $\phi$. The $s$-polarized and $p$-polarized optical fields of BB are in the respective $E_s(r, \phi)$ and $E_p(r, z, \phi)$ forms. The transverse, and longitudinal components are given by $E_t(r, \phi, \theta)$ and $E_z(\phi, \theta, z)$ respectively. Moreover, the $E_s(r, \phi)$ is a purely transverse component of the BB, and it can be written as $E_s(r, \phi) = E_{st}(r, \phi)$. On the other side, the $E_p(r, \phi, \theta, z)$ has both transverse and longitudinal components. Hence, it can be further divided into transverse and longitudinal components as $E_p(r, z, \phi, \theta) = E_{pt}(r, \phi, \theta) + E_{pz}(\phi, z, \theta)$. From the above explicit forms, we can procure a relation between $E_s$, $E_p$, $E_t$, and $E_z$ as $E_t(r, \phi) = E_{st}(r, \phi) + E_{pt}(r, \phi, \theta)$ and $E_z(\phi, z, \theta) = E_{pz}(\phi, z, \theta)$. Further, the detailed theoretical analysis of the 3D polarization distribution of BB can be found in our recent work [206].

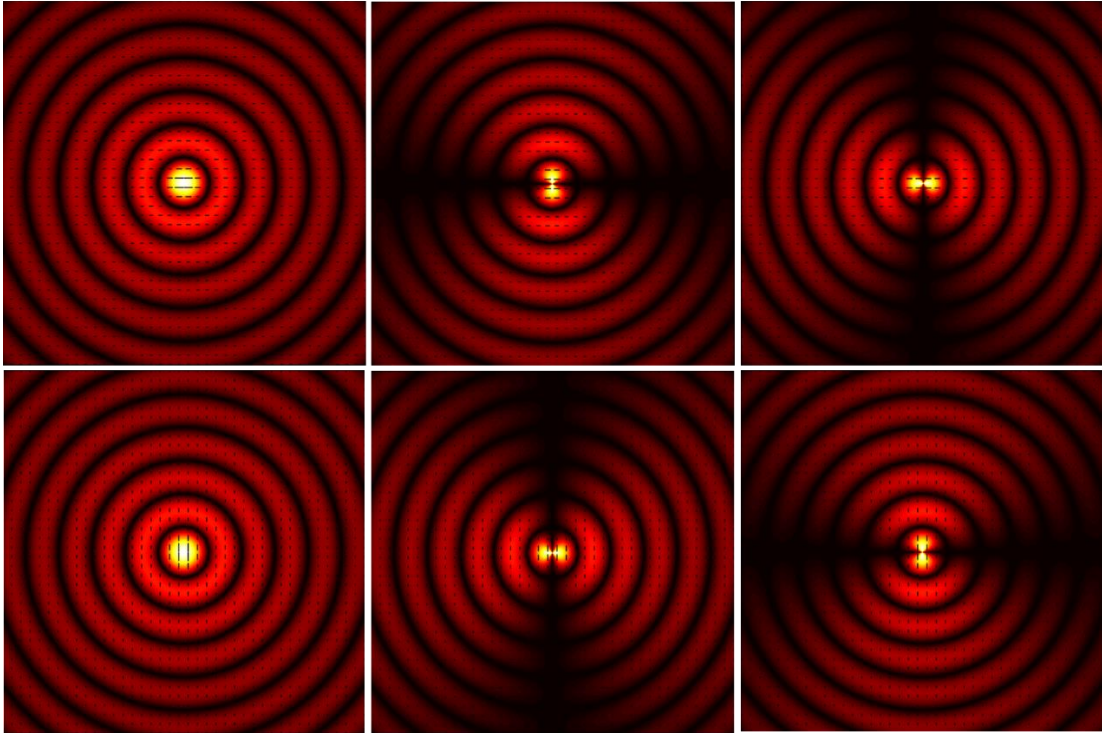

Fig. 5.5. The linearly polarized 0$^{th}$ order Bessel beam split into (second column) $s$-polarization and (third column) $p$-polarization components.

As shown in Figs. 5.5 and 5.6, for any arbitrary order of BB with uniform linear polarization, its optical energy is equally distributed between $s$-polarization and $p$-polarization. The petal structure created in $s$-polarization is orthogonal to the polarization direction while it is parallel to the polarization direction in $p$-polarization. A similar effect can also be seen in transverse, and longitudinal polarization components. The 3D polarization distribution created in BBs is prominent for large Bessel cone angles and it has multiple drawbacks in the applications. However, we can overcome this difficulty by using the superposition of two orthogonally linear polarized BBs [206] or by radially polarized BB [207].

Importantly, V-point polarization singular vector modes can bring into being in $s$-polarization and $p$-polarization by superposition of two identical BBs with the same order but orthogonally linear polarized [206]. Indeed, we can achieve all four types of vector modes that come under V-point polarization singular vector modes by providing the required phase delay between the two superposed Bessel modes. The vector modes formed in $s$-polarization and $p$-polarization can be understood as follows. The petal structures created in $s$-polarization and $p$-polarization, from the linearly polarized BB, are intimately related to first-order HG modes of $HG_{0,1}$ and $HG_{1,0}$. It is a well-established analysis that the superposition of orthogonally linear polarized $HG_{0,1}$ and $HG_{1,0}$ modes can produce vector LG mode. Similarly, we can generate vector BB in *s*-polarization and *p*-polarization by the superposition of two identical orthogonally linear polarized BBs. The schematic diagram of vector BB generation in *s*-polarization and *p*-polarization is depicted in Fig. 5.7 [206]. The vector modes created in *s*-polarization and *p*-polarization cannot be detected directly by using conventional methods [58]. However, one can perceive these vector modes by interacting them with polarization-sensitive optical materials.

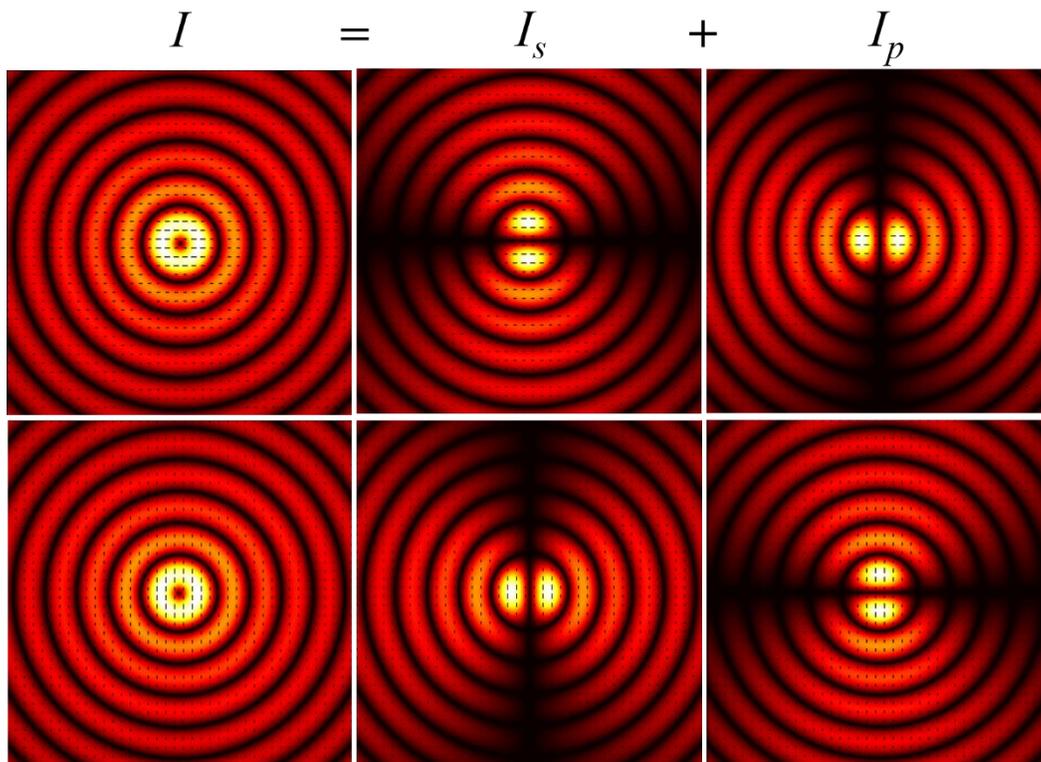

Fig. 5.6. The linearly polarized 1$^{st}$ order Bessel beam is written as the sum of the (second column) *s*-polarization and (third column) *p*-polarization components.

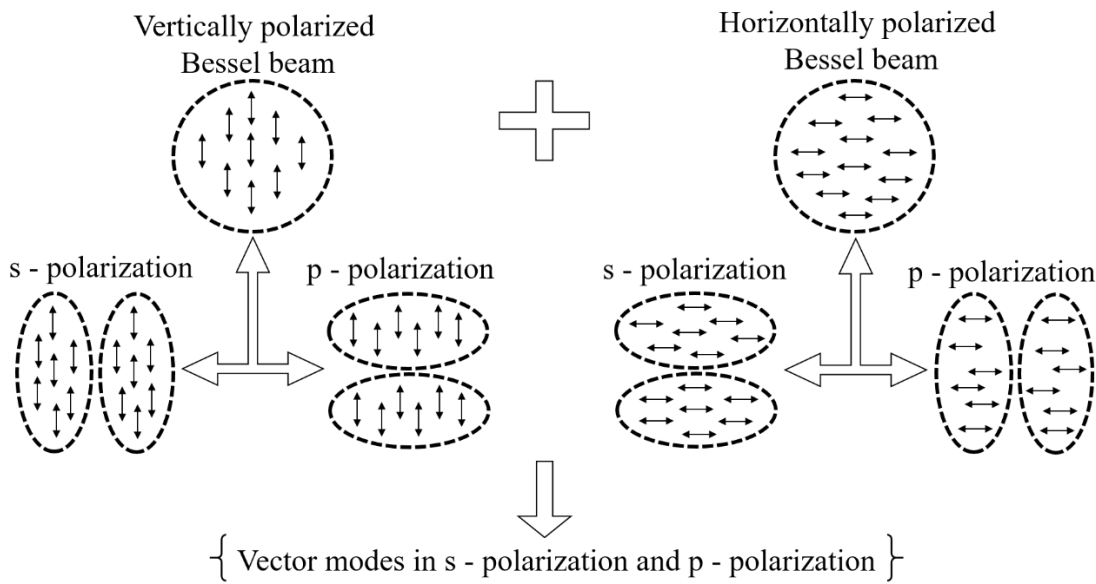

Fig. 5.7. Schematic flow-chat of vector modes generation in *s*-polarization and *p*-polarization by the superposition of two orthogonally linear polarized Bessel beams with the same order.

The 3D polarization distribution created in the BB plays a pivotal role in the light-matter interaction of polarization-sensitive materials and polarization-based data encryption. This 3D polarization distribution of BB predominates in a tight focusing configuration. It is also noted that despite enormous progress in the understanding of paraxial BBs' generation and characterization, the intrinsic properties of non-paraxial BBs remain somewhat elusive [207].

*5.6. Reflectance at interface*

As we discussed in earlier sections, the central lobe size ($r_0$) and range ($z_{range}$) of BB depends on the Bessel cone angle ($\theta$) as $r_0=2.405/(k\sin\theta)$ and $z_{range}=w_0/\sin\theta$. The Bessel cone angle changes while it is propagating in multiple optical mediums due to the refraction occurring at the interface of two mediums. For instance, as shown in Fig. 5.8, when an optically denser medium is inserted normally into the BB propagation direction in air, the cone angle of BB changes from $\theta_a$ to $\theta_m$ with $\theta_m<\theta_a$. As a consequence, the central lobe size and range of BB decreases while it is propagating in the denser medium. Even though the BB has a normal incidence at the interference, the individual waves in BB always propagate in a conical shape and make an angle with the interface. Thus, the propagation properties of BB change whenever it crosses the interface of two mediums with different refractive indices. The change in the angle of the Bessel cone can be estimated by Snell's law, where $\theta_a$ and $\theta_m$ are related by $n_a\sin\theta_a = n_m\sin\theta_m$. Here, $n_a$ and $n_m$ are refractive indices of air and denser medium respectively. It is also noted that the oblique incidence of BB on any medium results in an asymmetry in its shape because the incidence angle at the interface changes in the azimuthal direction.

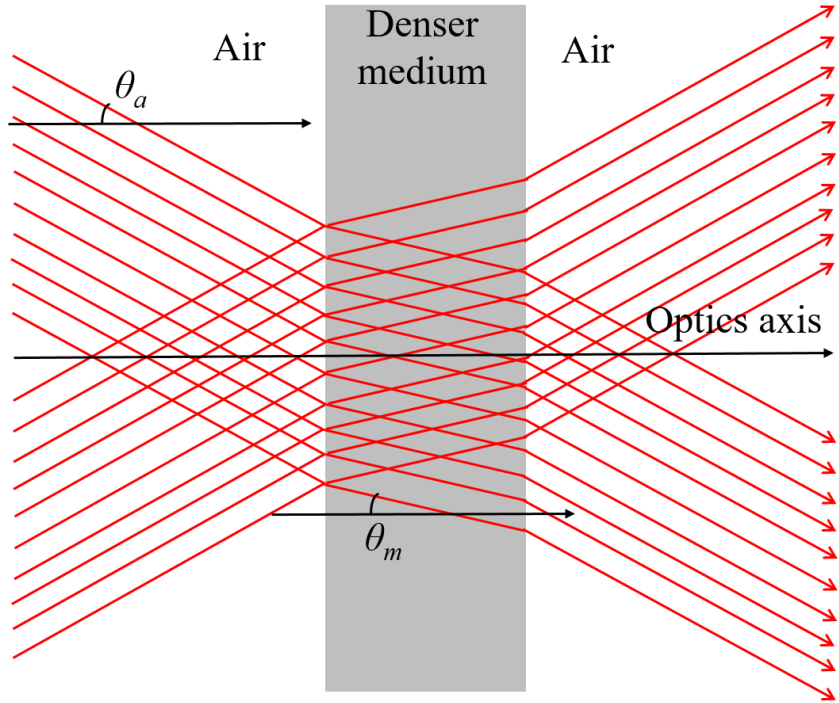

Fig. 5.8. Ray diagram of Bessel beam while it is propagating from air to denser and denser to air mediums. The Bessel cone angle in air is $\theta_a$ and in denser medium it is $\theta_m$.

*5.7. $M^2$ value*

As like other structured beams, the quality of experimentally generated BBs can be measured with propagation/quality factor $M^2$. The quality factor of BGB of order $l$ can be expressed in an Eq. form as [208]

$$M_l^2(\xi) = \sqrt{\left\{\left[1+n+\xi\frac{I_{n+1}(\xi)}{I_n(\xi)}\right]-\xi^2\right\}}. \qquad (5.7)$$

Here, $\xi= k_r^2\,w_0^2/4$ and $I_n(.)$ being the $n^{th}$ order modified Bessel function of the first kind [209]. The quality factor $M_l^2$ tends to the value $l+1$ when $k_r w_0$ approaches zero and the BGB transforms into a LG beam with zero radial index (LG beam with zero radial index and $l$ azimuthal index have the quality factor $M_l^2 = l+1$). This is due to the fact that $1/k_r$ is much larger than Gaussian spot size $w_0$ (i.e., the Gaussian profile is much narrower than the central lobe of the Bessel function). On the other side, with increasing the $w_0$ the quality factor $M_l^2$ increases and the dependency on the order $l$ decreases. Therefore, the BGB will be transformed to a BB for large values of beam waist $w_0$ and $M_l^2 = k_r w_0/2$.

## 6. Other structured beams related to Bessel beam

In the preceding sections, we provided a detailed analysis on the theories and experiments carried out for the realization of BBs and their characterization. Furthermore, we can generate diverse structured beams whose transverse profile follows the Bessel function by incorporating several amendments in BB's generation techniques. The structured modes generated similar to BB or from the BB have unique properties with unconventional propagation dynamics with reference to other structured modes. Hence, they have several advantages in light-science applications.

### *6.1. Bessel-like beams*

As mentioned earlier before, the quasi-BBs are experimentally realized from the mod conversion of the collimated laser beam. The *k*-vectors of individual waves along the *z*-direction ($k = k_z$) transformed into a conical shape with $k = k_z + k_r$. While the BB direction is provided by $k_z$, the radial propagation *k*-vector $k_r$ produces standing interference with intensity distribution in the form of the Bessel function which is perpendicular to the direction of BB. The properties of BB can be quantitatively understood uniquely by $k_r$ with limited propagation. To overcome this, consider a finite width of $k_r$ ($\Delta k_r$) instead of single $k_r$, thereby allowing for long-range propagation with shape-invariant. The new kind of beam produced owing to the spread in the $k_r$ will not have diffraction-free propagation. As such, they are often referred to as Bessel-like beams. The range of a Bessel-like beam can be larger or smaller than the range of the corresponding BB. The properties of a Bessel-like beam predominantly depend on $\Delta k_r$. In recent times, the Bessel-like beam propagation through diffractive optical elements and free space has been well investigated via ABCD matrices with numerical calculations and these calculations can be easily applied in any application of Bessel-like beams [210,211].

The Bessel-like beams in the wide range of electromagnetic spectrum can be produced by different techniques. Here, we have given some of the techniques used to generate Bessel-like beams. As shown in Fig. 6.1, we can generate high peak power Bessel-like beams with maximum mode conversion efficiency by pumping an axicon with a diverging laser beam instead of a collimated one [212, 213]. As discussed in section 3, the diverging laser beam has a finite width in the $k_r$ and the $k_r$ radially increases by following the condition $k_r(r = 0) = 0$. The finite spread in the $k_r$ ($\Delta k_r$) of the pump beam creates propagation-dependent Bessel cone angle. As a result, the Bessel cone angle decreases with increasing the propagation and it leads to the divergence in the Bessel-like beam. Therefore, the central lobe size of BB increases with increasing the propagation distance. Similarly, we can also produce a Bessel-like beam by pumping the axicon with a converging beam [214,215]. However, in this case the central lobe size of the beam decreases while we are scanning the beam away from the axicon. For given specifications of the axicon and pump beam, the range of the Bessel-like beam produced by the converging pump beam is smaller than that of the range of conventional BB while the range of the Bessel-like beam in the presence of the diverging pump beam is larger than the range of conventional BB. The properties of a Bessel-like beam (range, cone angle, and size) can be widely tuned in a controlled manner by varying the diverging angle of the incident pump beam. The propagation-dependent changes in the transverse phase profile of Bessel-like beams are similar to the transverse phase profile of paraxial Gaussian vortex beams.

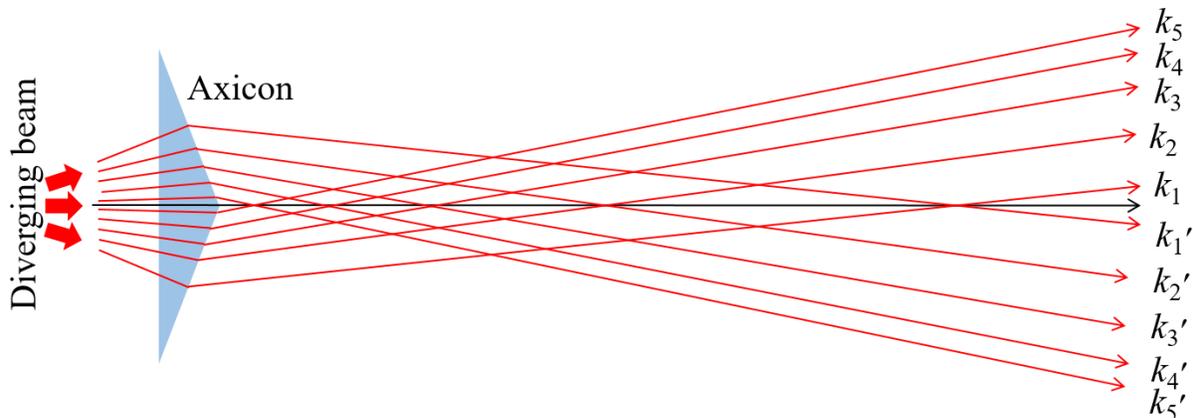

Fig. 6.1. The ray diagram of Bessel-like beam generation by axicon in the presence of diverging laser beam pump. For distinguishing the direction of each k-vector, we named each with a different name but all having same magnitude.

The Bessel-like beams can also be produced by other techniques like a superposition of multiple Airy beams [216], a combination of positive and negative axicons with convex lens [217], cross-phase modulation [218], flat acoustic lens [219], and SLM [220]. In the above-mentioned methods, SLM-based method is easy to implement, requires no special alignment of multiple optical elements, and easily tunable the characteristics of the Bessel-like beam by changing input parameters assigned to the hologram. The transmission function $T(r)$ assigned to the optical phase element projected on the SLM screen for Bessel-like beam generation is given by

$$T(r) = \exp[ik(ar^n + br^m)], \quad (6.1)$$

and

$$b = -a\frac{n}{m}(r_I)^{n-m}. \quad (6.2)$$

Here $r_I$ may be interpreted as the clear aperture of the phase-only optic. By providing selective values for parameters $a$, $n$, and $m$, we can easily generate suitable Bessel-like beams for any kind of application. For example, Fig. 6.2 shows the intensity distribution in the first order Bessel-like beam in its longitudinal beam cross-section for $a = 0.1$ $n = 2$, $m = 3$, and $r_I = 6$ μm.

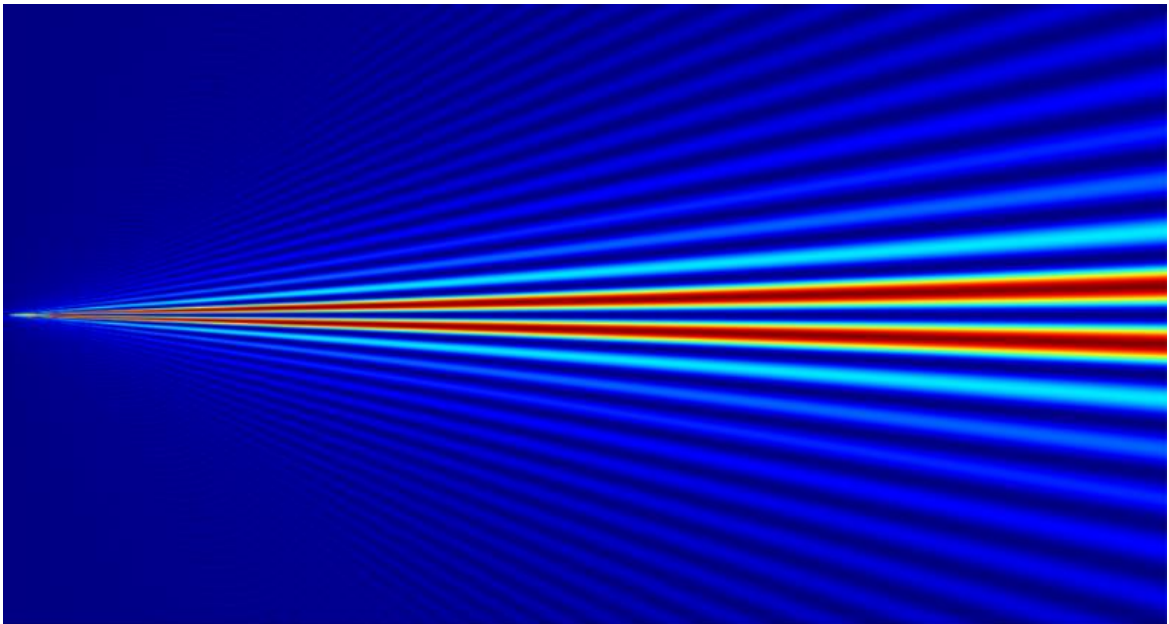

Fig. 6.2. Longitudinal cross-section of Bessel-like beam in *xz/yz*-plane.

On top of the above-mentioned techniques, we can also generate these modes directly from the laser cavity. For example, by inserting an axicon within the laser cavity [221] or by intra-cavity thermal effects [222], we can generate Bessel-like beams with high-mode quality. The strong-intra-cavity spherical aberration can be used to generate a tunable Bessel-like beam [223]. Fibers generally have high mode purity and can also be used to generate Bessel-like beams by fabricating micro-axicons on fiber tip [224,225] or by multimode interference within the fiber [226]. Further, a Bessel-like beam array can be generated from the periodical arrangement of the polymeric round-tip micro-structures [227]. Even though Bessel-like beams are diffraction beams, still they have self-healing like BB and it can be detected in a similar way we did for BB.

*6.2. Bessel bottle beams*
Laser beams with central zero or low intensity surrounded by high-intensity regions are called optical bottle/bubble beams [228]. The optical bottle beams can be experimentally synthesized easily in laboratories by superpositioning two structured laser modes. Generally, the optical bottles created in laser beams result from the Gouy phase difference created between the two superposed beams. The bottle created within the laser beams has a 3D structure as shown in Fig. 6.3. The intensity distribution in the transverse beam cross-section at the bottle center and neck/cap are there in respective doughnut shape and Gaussian-like shape.

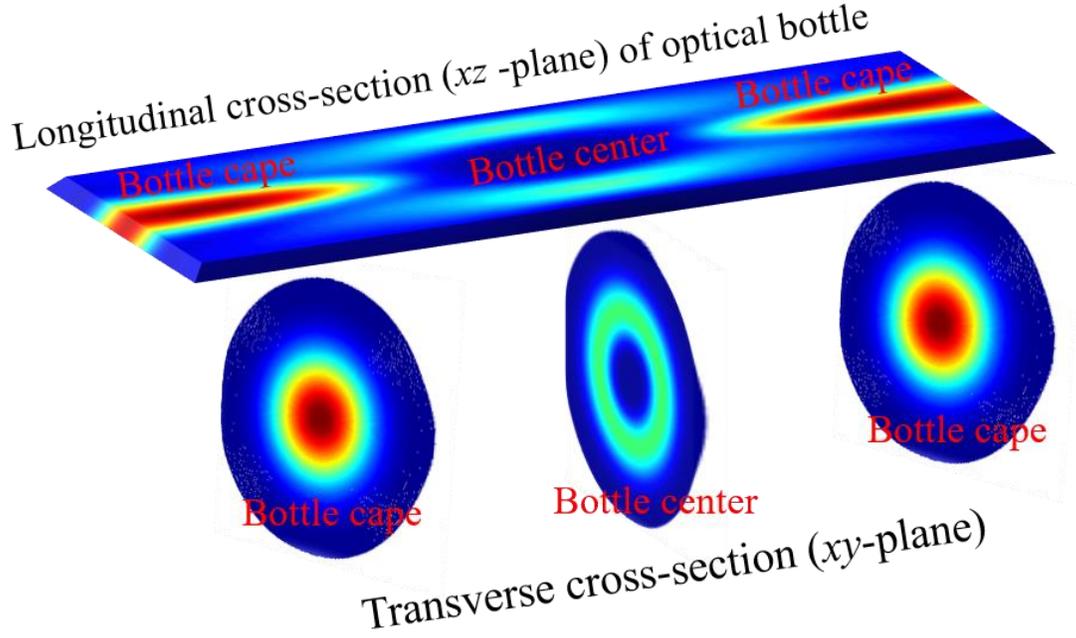

Fig. 6.3. Three-dimensional structural properties of optical bottle beam in its longitudinal and lateral cross-sections.

We can generate micro-size optical bottles with excellent quality in BBs. The bottle structures created in BBs are called Bessel bottle beams (BBBs). These beams can be created by the superposition of two zeroth order BBs with different $k_z$ but the same $k$ [229]. Here, the Gouy phase difference occurs between the two superposed BBs as an outcome of $k_{z1}- k_{z2}$. Here, $k_{z1} = (k^2- k_{r1}^2)^{½}$ and $k_{z2} = (k^2- k_{r2}^2)^{½}$ are longitudinal propagation vectors for respective radial vectors $k_{r1}$ and $k_{r2}$. The two ideal BBs with these parameters can be written in simple forms as

$$E_1(r,z) = \exp[i(k_{z1}z + \varphi_1)] J_0(k_{r1}r), \qquad (6.3a)$$

$$E_2(r,z) = \exp[i(k_{z2}z + \varphi_2)] J_0(k_{r2}r). \qquad (6.3b)$$

Then the resultant intensity distribution in the interference is given by

$$I = J_0^2(k_{r1}r) + J_0^2(k_{r2}r) + 2J_0(k_{r1}r)J_0(k_{r2}r)\cos[(k_{z1} - k_{z2})z + \Delta\varphi]. \qquad (6.4)$$

As depicted in Fig. 6.4, we can generate a chain of optical bottles along the beam axis in the direction of propagation. In BBBs, the optical bottle structure is repeated at a regular interval along the propagation, and we often call this kind of phenomenon as Talbot effect. The position of the bottles along the propagation can be easily controlled by employing relative phase difference between the two interfering beams $\Delta\varphi=\varphi_1-\varphi_2$. The oscillation period or the Talbot length created in the BBB is nothing but the length of the optical bottle and it is given by $\Lambda=2\pi/(k_{z1}- k_{z2})$. Further, the bottle width is governed by $k_{r1}- k_{r2}$. The two images shown in Fig. 6.4 are the same size, however, we can see the change in the dimensions of the optical bottle chain created in BBB on account of the change in base angles of axions used to generate the interfering BBs. These optical bottle beams are completely different from other bottle beams owing to their non-diffraction and self-healing properties inherently acquired from parent BBs.

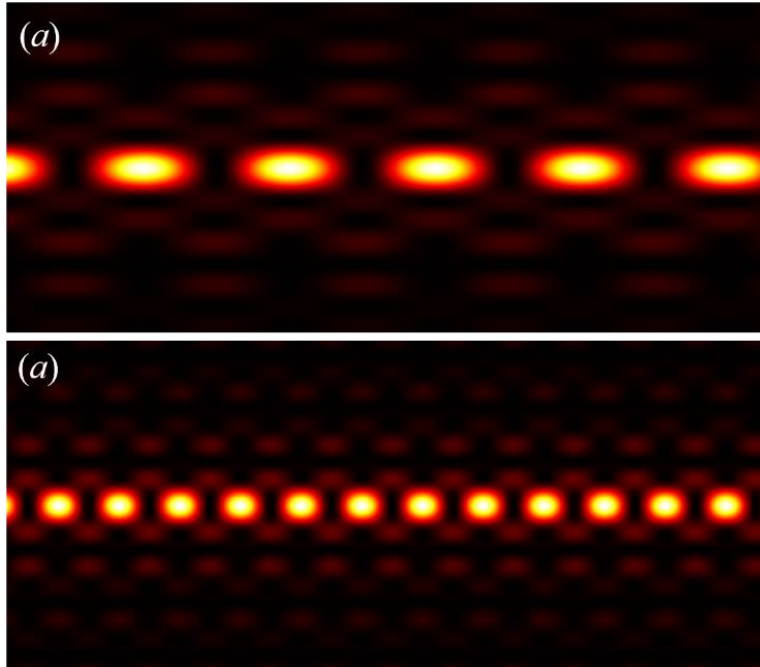

Fig. 6.4. The Bessel bottle beams are created by the superposition of two Bessel beams. The axion base angles used here are (*a*) $\alpha_1 = 2°$ $\alpha_2 = 3°$ and (*a*) $\alpha_1 = 2°$ $\alpha_2 = 4°$.

The BBBs can be experimentally generated in different ways by using a different kind of diffractive optical elements, *viz.*, MZI with two different angle of axicons [230], combination of the axisymmetric acoustic-optic device and the spatial filtering enabled by a mask or a digital micro-mirror device [231], two concentric annular apertures with different radii with a simple convex lens in Durnin technique [232], computer generated double concentric annular slits with different radii on SLM [233], passing a LG beam, with a radial mode index of $p > 1$, through an axicon [234], precise amplitude and phase manipulation of BGBs by means of DMD [235], single beam passing via two co-axially aligned axicons [236], optical feedback with a single axicon [237], focusing a π-phase shifted multi-ring HoG beam using a lens with spherical aberration [238]. However, here, we discuss the experimental generation of BBBs by MZI with two different angles of axicons [Fig. 6.5 (*a*)] and a single beam passing through two co-axially aligned axicons [Fig. 6.5 (*b*)]. By inserting each of the two axicons in each arm of the MZI, we can generate two BBs independently and these two BBs interfere after passing via the second beam splitter of MZI. The delay line in one of the arms of MZI is required to perfectly overlap the laser pulses of pulsed laser or overlapping of two laser beams within the coherence length. In addition, the delay line can control the relative phase between the two BBs $\Delta\varphi$.

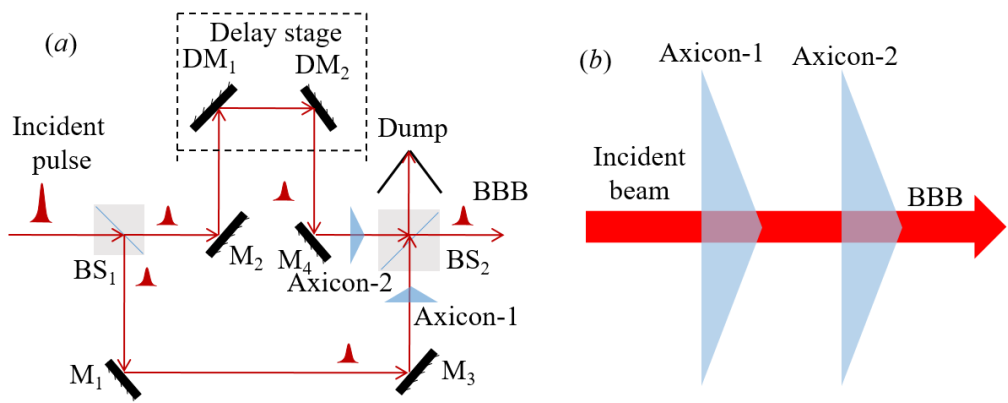

Fig. 6.5. Bessel bottle beams generation by using two axicons with different base angles in (*a*) Mach-Zander interferometer and (*b*) single beam configurations. $M_i$ is mirror, $DM_i$ is delay mirror, $BS_i$ is beam splitter, and dump is power dumper.

The generation of BBB by passing a single laser beam through two co-axially aligned axicons is a little tricky to understand as compared with MZI-based BBB generation but it is a robust technique. The BBB generation, by passing a single optical beam through two axicons, can be understood by employing its geometrical ray diagram provided in Fig. 6.6. As shown in Fig. 6.6(*a*), collimated laser beam passing through the first axicon with base angle, $\alpha_1$ produces BB with single radial *k*-vector. The Bessel cone created by the first axicon splits into two Bessel cones after it passes via the second axicon with a base angle of $\alpha_2$. The radial *k*-vectors of two Bessel cones are given by $k_{r1}= k \sin[(n-1)(\alpha_2+\alpha_1)]$ and $k_{r2} = k \sin[(n-1)(\alpha_2-\alpha_1)]$. The individual waves passing through the same side of two axicons with reference to the beam axis have the radial *k*-vectors $k_{r1}$ (full lines in Fig. 6.6) and waves propagating through different sides of the axicons have second kind of radial *k*-vectors $k_{r2}$ (dotted lines in Fig. 6.6). In this case, the bottle dimensions in BBB is constant along the propagation. Further, we can generate variable size optical bottle chains along the propagation by creating optical bottles in a Bessel-like beam. As depicted in Fig. 6.6 (*b*), we can produce optical bottles in a Bessel-like beam by simply replacing the collimated beam with a diverging beam as a pump source [237]. The cone angle in Bessel-like beams varies with propagation distance, *z*. Therefore, the dimensions of optical bottles created in a Bessel-like beam change with position. Further, the intensity modulation depth and intensity distribution in BBB can be tuned by carefully modulating the transverse intensity distribution of pump modes to the respective axicons. Moreover, the position and size of the optical bottle created within the BB can be tunable in a controlled manner for the comfortability of applications by changing the divergence of pump beams.

To further understand how the characteristics of BBB depend on axion, we made a numerical calculation on the dimensions of optical bottles created in the Bessel profile. The parameters used in the calculations are the refractive index of the axicon is 1.449, the wavelength of the laser source is 1064 nm, and the collimated Gaussian spot size at the first axicon is 1.5 mm. The achieved numerical results are presented in Fig. 6.7 and 6.8. As shown in Fig. 6.7, the properties of BBBs can be changed in a controlled way in experimental configurations provided in Fig. 6.5 by changing the base angle of one of the axicons while the other one is fixed. The dimensions of the optical bottle, the number of bottles, and the range of BBB decrease with increasing the difference in the base angles of two axicons. Similar way we can also change the characteristics of BBB in optical feedback with a single axicon technique [237]. For instance, we have shown in Fig. 6.8 how we can alter the properties of optical bottles created in the Bessel profile by changing the angle of a single axicon.

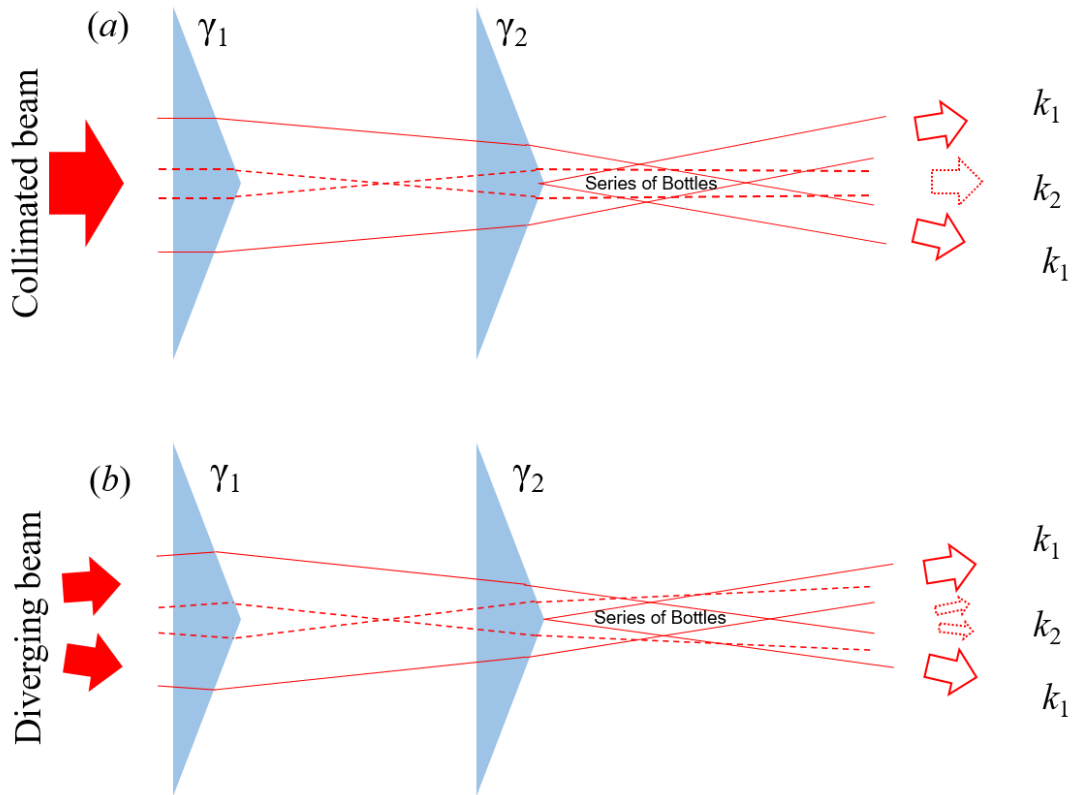

Fig. 6.6. The ray diagram of Bessel bottle beams generation by employing two axicons with apex angles $\gamma_1$ and $\gamma_2$ in the presence of (*a*) collimated laser beam and (*b*) diverging laser beam.

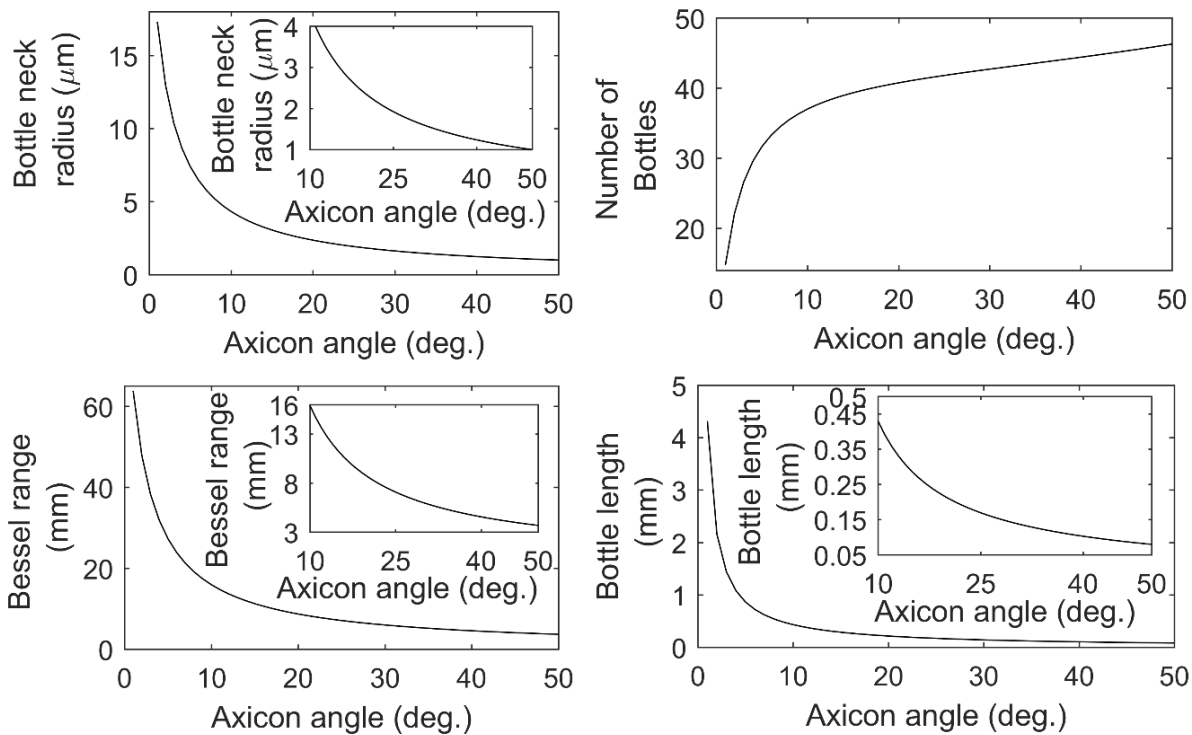

Fig. 6.7. Characteristics of Bessel bottle beam generated from two identical axicons with different base angles pumped with 1.5 mm collimated Gaussian spot size. Here, the first axicon's base angle was fixed to 2° and the base angle of the second axicon was variable.

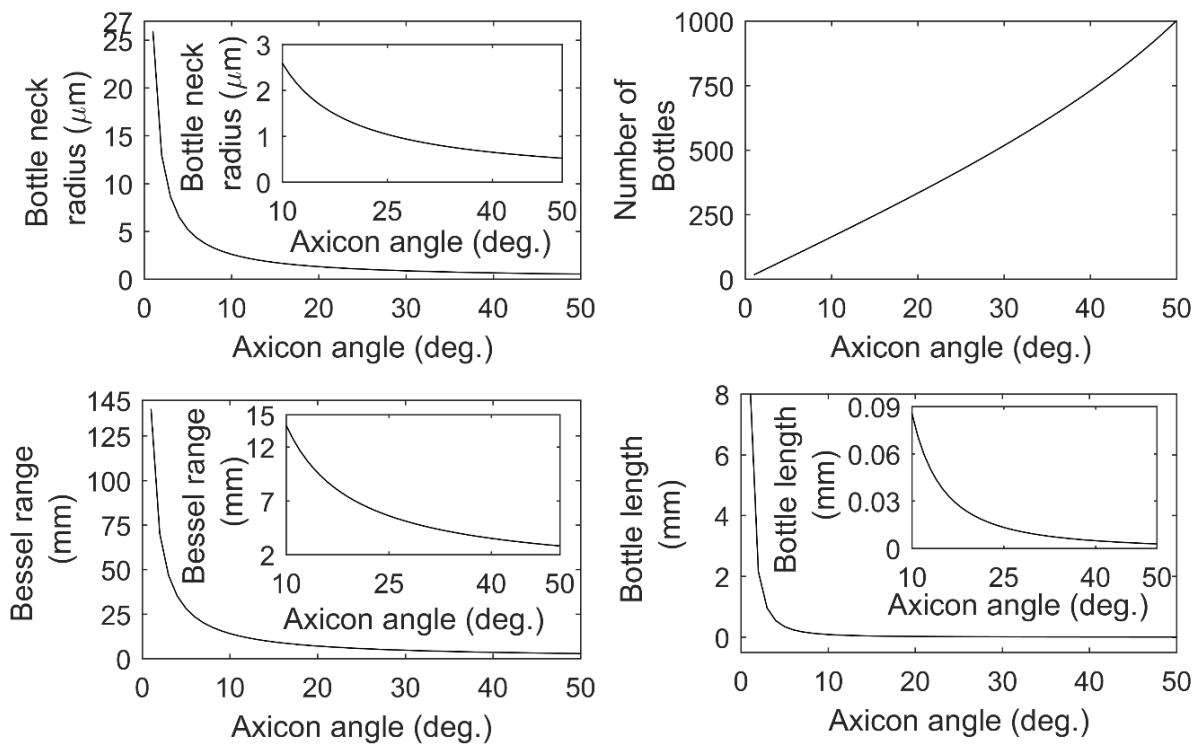

Fig. 6.8. Characteristics of Bessel bottle beam generated from two identical axicons with the same base angle pumped with 1.5 mm collimated Gaussian spot size.

## 6.3. Perfect vortex

The conventional vortex beams created in any generalized Gaussian beam profile have topological charge-dependent size. The annular size of the vortex increases with its topological charge. To overcome this limitation in laser beams, Ostrovsky and co-workers have introduced a concept of a perfect vortex beam whose size is independent of topological charge [239,240]. The Perfect vortex is a Fourier transformation of BB. This Fourier transformation can be carried out experimentally by a simple convex lens of focal length, $f$. The mathematical expression for Fourier transformation is given by [241]

$$E(r,\phi) = \frac{k}{i2\pi f}\int_0^\infty \int_0^{2\pi} E(r',\phi')\exp\left(\frac{k}{if}r'r\cos(\phi-\phi')\right)r'dr'd\phi'. \quad (6.5)$$

Fourier transformation of ideal BB [Eq.2.6] can be acquired by substituting it into Eq. 6.5 as [242,243]

$$E(r,\phi) = \frac{i^{l-1}}{k_r}\delta(r-r_r)\exp(il\phi). \quad (6.6)$$

Here $\delta(r)$ is the Dirac delta function and the perfect vortex ring radius is $r_r = k_r f/k$. Ideal BB transformed into perfect vortex with an infinitely small ring width given by the Dirac delta function. In similar fashion, we can generate a finite thickness perfect vortex experimentally by the Fourier transformation of real BB. For example, the Fourier transformation of BGB is a perfect vortex of the form as [242]

$$E(r,\phi) = B_0 \frac{w_0 i^{l-1}}{W_0}\exp(il\phi)\exp\left(-\frac{r^2+r_r^2}{W_0^2}\right)I_l\left(\frac{2r_r r}{W_0^2}\right). \quad (6.7)$$

Here, the ring thickness is given by $W_0 = 2f/kw_0$. $w_0$ is Gaussian spot size and $B_0$ is constant. $I_l(\cdot)$ is the modified Bessel function. For reference, the perfect vortex with different topological charges is plotted in Fig. 6.9. The perfect vortex forms in the Fourier plane of the lens. The size of the perfect vortex we can control with the axicon parameter $a = k(n-1)\alpha$. The ring radius of the perfect vortex decreases with decreasing $a$ irrespective of its topological charge. It is also noted that the slope of $I_l(\cdot)$ decreases with $l$, which shifts the radius where the exponentially decreasing Gaussian term and the exponentially increasing Bessel term intersect. As a result, the radius of the perfect vortex slightly changes with topological charge, $l$. In recent times, J. Pinnell et. al., have successfully removed this effect by individually adjusting $r_r$ [244]. By this technique, they could experimentally produce a constant ring size perfect vortex of order up to 50. The perfect vortex radius can also be controlled by adjusting the separation between the lens and axicon phase element [245]. The recent developments in perfect vortex allows to generate concentric [246] and array perfect vortices [247,248].

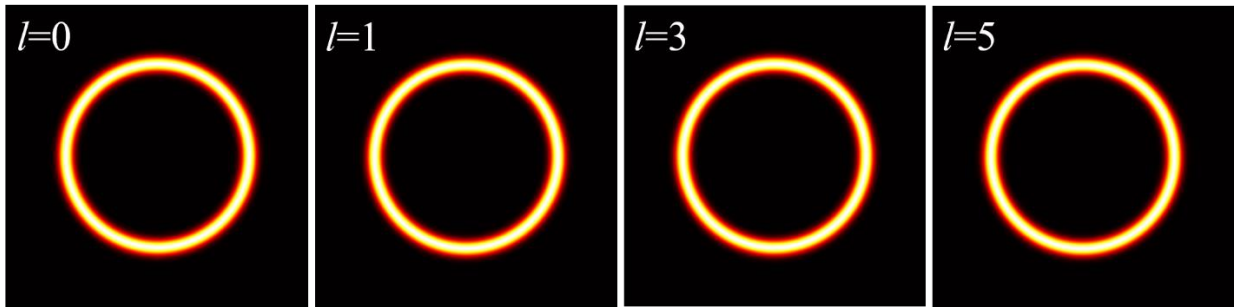

Fig. 6.9. Perfect vortices with different topological charges.

Along with the above three types of structured modes originating from BB, there are other structured modes generated by implementing further modifications in the generation processes for enhancing the applications of non-diffraction modes. For instance, Array of BBs [249], evanescent BBs [250-252], partially coherent BBs [253,254], vector space-fractional BBs [255], spiraling BBs [256], curved BBs [257], and multi-singularities BBs [258].

## 7. Vector Bessel beams

The vector BBs are the solution to the vector Helmholtz wave equation and have the non-diffraction and self-healing natures [259-264]. These beams have a non-uniform polarization distribution in their beam cross-section. Similar to vector LG modes, we can quantitatively understand the vector BBs by classifying them into L-lines, C-points, and V-points [265,266]. Here, the L-lines are lines of linear state of polarization at which the handedness of the polarization ellipse is undefined and by which the right and left ellipses are usually separated [265]. The C-points are points of a circular state of polarization in which the orientation of the major axis of the polarization ellipse is undefined [267]. These modes are also called Poincare modes (PMs). The C-point singularities of PMs can have maximum or zero intensity and hence these modes can further be classified into bright Poincare modes (BPMs) and dark Poincare modes (DPMs). The PMs can have polarization structures such as lemon, monstar, and star. In recent times, PMs have also been referred to as hybrid-order Poincare sphere beams. Finally, V-points occur in linear polarized light fields at which the orientation of the electric vector becomes undefined and light intensity is zero [268]. The V-point vector modes are further divided into cylindrical vector modes (CVMs), and π vector modes (πVMs). The CVMs and πVMs have polarization distribution in radial, azimuthal, spiral, flower, and spider web shapes. Furthermore, V-point polarization singularities are referred to as higher order Poincare sphere beams.

A monochromatic vector Bessel mode with in-homogeneously polarization can be written as superposition of two orthogonally circular polarized BBs as

$$\psi(\theta,\varphi) = E_{l_1}\cos(\theta/2)e^{-i\varphi/2}|R\rangle + E_{l_2}\sin(\theta/2)e^{i\varphi/2}|L\rangle, \quad (7.1)$$

where $\theta \in [0, \pi]$ and $\varphi \in [0, 2\pi]$ are the polar and azimuthal angles in the spherical coordinates of the Poincare sphere. $|R\rangle$ and $|L\rangle$ are right-circular (P) and left-circular (Q) polarizations. Depending on the azimuthal index sign and amplitude, the vector mode provided by Eq.7.1 produces different kinds of vector modes like cylindrical vector modes (CVMs), and π vector modes (πVMs) with V-point singularity, and Poincare modes (PMs) with C-point singularity. The V-point and C-point singularities are quantitatively understood with the help of their respective characteristic parameters of the Poincare-Hopf index ($\eta$) and the singularity index ($I_C$). As shown in Eq. 7.2, these polarization indices can be quantitatively estimated in a similar fashion of the phase singularity of scalar modes.

$$\eta = \frac{1}{2\pi}\oint \nabla\gamma \cdot dl, \quad (7.2a)$$

$$I_C = \frac{1}{2\pi}\oint \nabla\gamma \cdot dl. \quad (7.2b)$$

Here, $\gamma$ is the azimuth of the polarization ellipse. In the superposition technique, $\eta$ and $I_C$ are given by $(l_2 - l_1)/2$ [8]. Here, $l_1$ and $l_2$ are OAM of LG modes involved in the superposition. The Eq. 7.1 produces V-point vector modes if $l_1 = -l_2$ and C-point vector modes whenever it satisfies the condition of $l_1 \neq -l_2$.

The PMs with different singularity index ($I_C$) generated in ideal BBs by Eq. 7.1 are presented in Fig. 7.1. Here, the red and blue polarization ellipses indicate the right and left circular polarizations. The BPMs given in the first row are obtained by the superposition of $0^{th}$ order BB with an arbitrary order BB ($l \neq 0$). As shown in the second row, the DPMs can be achieved by utilizing the superposition of two non-zero order BBs. It is conversant in BBs that the width and radius of the Bessel rings depend on their order and the PMs generated by the superposition of BBs with different orders. Hence, the PMs generated in BB have a completely different intensity distribution than the scalar BBs and they may not have local intensity nulls/nodes. The intensity mismatch between the superposed modes taken place in the radial direction creates periodic polarization change between right circular and left circular. The helical phase created from the phase difference between the two superposed modes produces rotation of polarization in azimuthal direction.

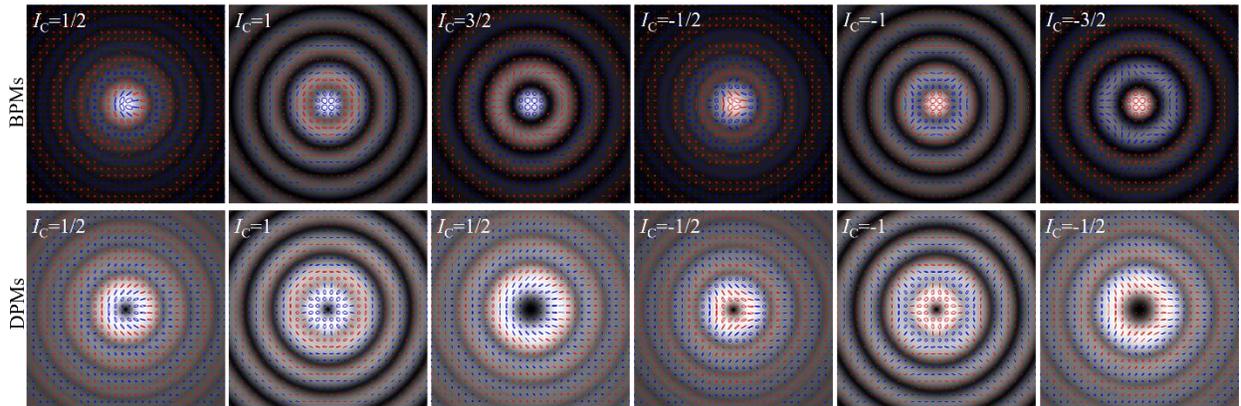

Fig. 7.1. Poincare vector modes generated in Bessel beams by the superposition of orthogonally circular polarized Bessel beams. Bright and dark polarization singularity Poincare modes are given in the respective first and second rows.

Moreover, experimentally realization of invariant PMs in the Bessel profile is very difficult due to the superposition of different order BBs [269,270]. As we discussed in sections 2, 3, and 4, the properties of BB like range, onset position, and offset position depend on its order. It could lead to PMs with propagation-dependent polarization distribution. However, we can overcome this snag in PMs by perfect longitudinal mode matching in the presence of SLM and axicon. For example, we can generate perfect BPMs with axicon in the presence of HoG and Gaussian vortex beams as input modes. However, this technique cannot be utilized for the generation of propagation invariant DPMs.

The V-point vector modes, generated with Eq. 7.1 by considering the superposition of two orthogonally circular polarized BBs with same order but opposite sign, are depicted in Fig. 7.2. The CVMs and πVMs are given in respective first and second rows of Fig. 7.2. While CVMs have the positive Poincare-Hopf index ($\eta$), πVMs have negative values. The perfect transverse mode matching in the superposition provides the well-defined circular intensity nulls/nodes. In V-point vector modes, the polarization distribution in the transverse plane is independent of longitudinal position owing to perfect longitudinal mode matching and we could see propagation-independent polarization distribution. Hence, these modes are called propagation-invariant polarization modes. Multiple techniques were efficiently generated these vector modes, *viz*: Pancharatnam–Berry phase optical elements [271], axicon tipped fibers [272,273], metasurfaces [274,275], and SLM [276,277].

From the above discussion we can understand that BBs have a 3D non-uniform polarization distribution i.e., polarization in the transverse and longitudinal cross-sections depends on the intensity distribution of superposed BBs in their transverse and longitudinal cross-sections. Hence, in the generation of VBBs, the 3D intensity distribution of superposed BBs is a vital parameter. Moreover, we can control the 3D polarization distribution independently at any point in the 3D structure of vector BB by simply controlling the intensity distribution of individual BBs of superposition. The required intensity modulations in BBs can be easily synthesized by multiple techniques which we discussed in sections 3 and 4. In recent years, multiple research works focused on the generation of propagation variant vector textures in VBBs by utilizing various techniques. Generation of a Bessel-type vector beam with a spatial polarization, oscillating along the optical axis when propagating in free space by two SLM-based holographic axicons [278-281]. Further, we can modulate the polarization of BPMs, DPMs, CVMs, and πVMs presented in Fig. 7.1 and 7.2 in a controlled manner by polar and azimuthal angles of Eq. 7.1.

Even though we can generate all vector modes in LG beam and BB, the vector modes in BB have multiple advantages over the former one like non-diffraction, self-healing, and long depth of focus with sub-wavelength mode size. Therefore, in some of the applications, vector BBs are more suitable than the vector LG modes.

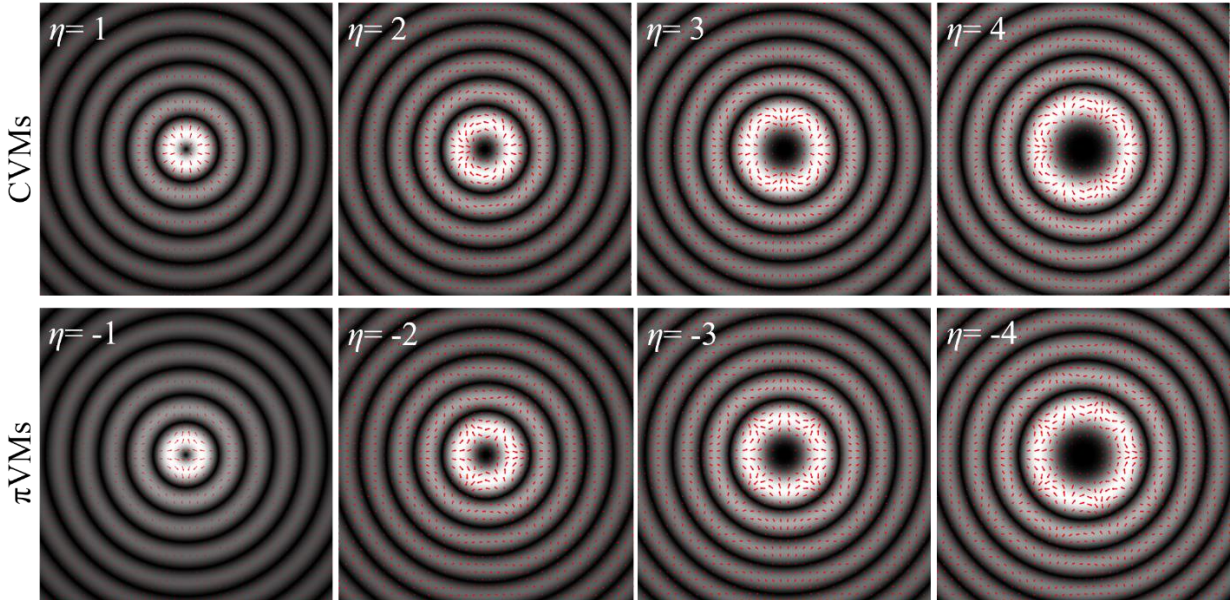

Fig. 7.2. Vector modes with different order are created in Bessel profiles by the superposition of orthogonally circular polarized Bessel beams. Cylindrical vector modes and π vector modes are provided in the respective first and second rows.

## 8. Nonlinear optics of Bessel beam

As we discussed in the second section, BBs can be viewed as a superposition of an infinite number of equally weighted plane waves whose wave vector lies on a conical surface centered on the propagation axis. Hence, the nonlinear optical effects promoted by BBs are manifested with significant differences as compared with other Generalized Gaussian beams [282-285]. The BBs generated either directly from the laser cavity or by diffractive optical elements have wavelength tuning limitations. Despite that, we can tune their wavelength through nonlinear wave-mixing for their selective wavelength-dependent applications. Therefore, the wavelength tuning of BB via nonlinear wave-mixing is its fundamental aspect. The BB can provide a self-organized phase matching for a wide range of refractive index variations, which may automatically compensate for the spatially non-uniform phase mismatch created by the optical Kerr effect [286,287]. This Self-phase matching is expected to be very efficient for frequency tripling of broadband and/or ultrashort pulses, as well as for high-order harmonic generation [288-290]. A truncated Bessel intensity profile can create low-divergence, high-brightness harmonic emission [291]. The use of pump beams with a Bessel intensity profile allows significant improvements in the spatial quality of the harmonic beam. These higher harmonics are very useful for the applications of radiation, where extreme temporal resolution and high brightness are desired.

In nonlinear wave-mixing, a single Gaussian beam follows only collinear phase matching whereas a single BB can have collinear as well as non-collinear phase matching. The phase matching involved in the nonlinear wave-mixing of BB can be understood very well in Second Harmonic Generation (SHG) [292]. The phase matching conditions involved in SHG of BB can be visualized with the $k$-vector diagram of BB. For example, the collinear and non-collinear phase matching of BB in SHG in terms of $k$-vectors are shown in Fig. 8.1. In collinear phase matching of BB, the wave vectors with the same radial direction participate in SHG and produce Bessel cone at new wavelength. Therefore, fundamental BB transformed into SHG BB. Here, the phase mismatch between two fundamental waves is $\Delta k_\omega \approx 0$ $m^{-1}$. Whereas in non-collinear phase matching of BB, wave vectors with the opposite radial components on the fundamental Bessel cone generate second-harmonic output along the beam axis, thus the output mode becomes Gaussian-like mode. The phase mismatch in the nonlinear SHG is given by $\Delta k_\omega \approx k_r^2/k$. The phase matching condition for both cases can be achieved for a single BB in a single nonlinear crystal either by critical or noncritical phase matching. The window between these two phase matching conditions is $k_r^2/k$ and it decreases with decreasing the Bessel cone angle, $\theta$.

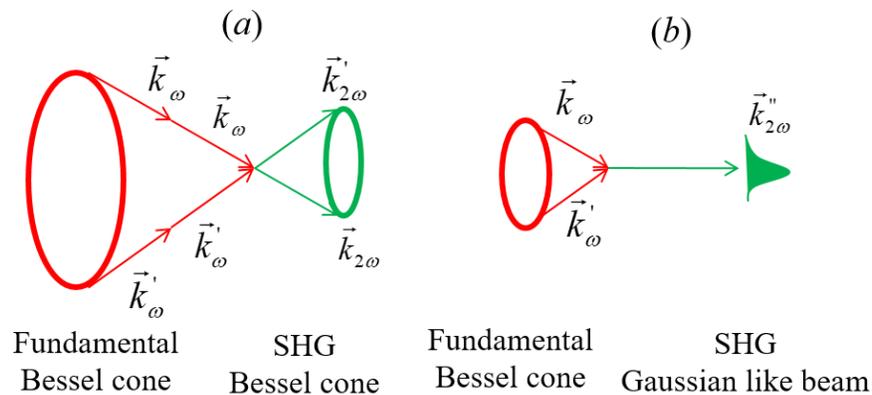

Fig. 8.1. (*a*) Collinear phase matching and (*b*) non-collinear phase matching of Bessel beam in second harmonic generation.

The nonlinear wave-mixing of BBs was successfully investigated in gas and solid mediums. Multiple numbers of lower order harmonics have been generated from BBs by the nonlinear wave-mixing in the gas medium [293,294]. The phase matching of the nonlinear wave-mixing in gases is achieved by the mixture of atomic vapor with the positively dispersive buffer gas [295]. The efficiency of the harmonic generation depends on the interactive length of the gas medium and the range of the BB. By matching the gas medium length to the range of the BBs, conversion efficiencies considerably higher than those can be obtained from Gaussian beams [296,297]. Also, the variation of the harmonic conversion efficiency with the conical half-angle of BB depends on the properties of the medium only through the macroscopic dispersion [298]. In the case of solids, the phase mismatching between fundamental and newly generated nonlinear signal beams is compensated on account of the birefringence of nonlinear crystal [299]. The nonlinear frequency conversion of BB can have increased by using longer-length nonlinear crystal owing to its non-diffractive nature [300,301]. However, this nonlinear conversion efficiency in crystals is always less than for a Boyd-Kleinman-focused Gaussian beam of the same power [292]. The reason behind this effect can be understood as follows. In BBs, total optical power distributed among all Bessel rings with nearly equal and the peak intensity decreases with increasing the ring size. As a result, the intensity-dependent nonlinear frequency conversion drastically decreases while we are moving radially outward from the Bessel center. It is worth noticing that BGB has a larger nonlinear frequency conversion efficiency than the ideal BB because optical power distributed in rings of BGB decreases with increasing ring size. In nonlinear frequency conversion of BBs in solids, with due consideration of their characteristics like their smaller size, longer non-diffracting propagation range, and very less intensity distribution in outer rings, periodically poled nonlinear crystals are the best choice to use for the nonlinear frequency conversion. The schematic diagram for SHG of BB through periodically poled nonlinear crystal is shown in Fig. 8.2. Periodically poled crystals can have adequate length to cover the Bessel range without phase mismatching. By properly controlling the peak position, $z_{peak}$ and range, $z_{range}$ of BB with reference to the nonlinear crystal length, we can enhance the nonlinear conversion efficiency.

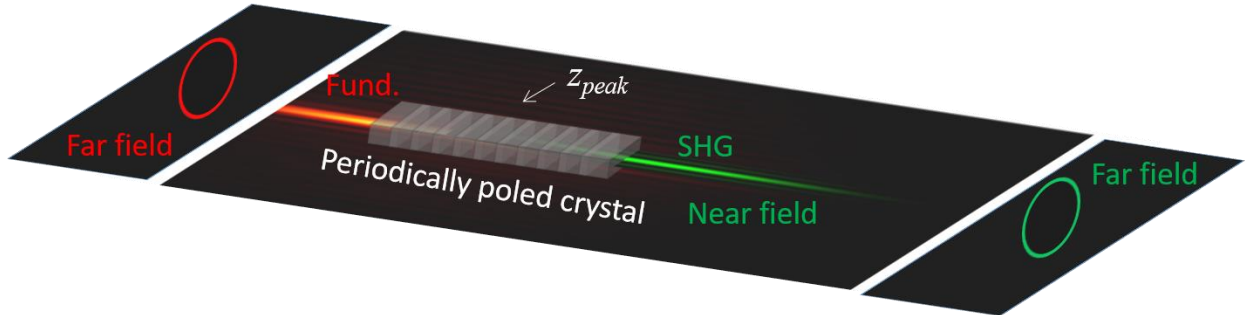

Fig. 8.2. Schematic diagram of second harmonic generation of Bessel beam through periodic poled crystals.

The wavelength of Bessel-like beams can also be tuned by employing nonlinear wave-mixing with low conversion efficiency in comparison to BB due to its radial propagation vector change with the propagation distance. The conversion efficiency decreases with increasing the divergence of BB. The change in the radial $k$-vector $\Delta k_r$ must be within the acceptance bandwidth of nonlinear crystal to frequency conversion of full beam. Depending upon the divergence of the Bessel-like beam and the acceptance bandwidth of the nonlinear crystal, we have to choose a suitable length of the nonlinear crystal.

In the case of SHG of BBB, we have to independently frequency double two pump $k$-vectors: $k_{P1}$ and $k_{P2}$ present in the fundamental BBB (Fig.8.3). The dimensions of optical bottles formed in the BBB are constrained by $k_{P1}$ and $k_{P2}$. In the SHG process, the nonlinear conversion efficiency must be the same for both the pump $k$-vectors to maintain the quality of frequency doubled BB. This can be achieved on account of temperature or angular tuning of phase matching. The dimensions of optical bottles in the frequency doubling change according to the new wavelength. The acceptance bandwidth of nonlinear crystal must be greater than the $\Delta k_P = k_{P2} - k_{P1}$. Also, the phase matching condition must be collinear.

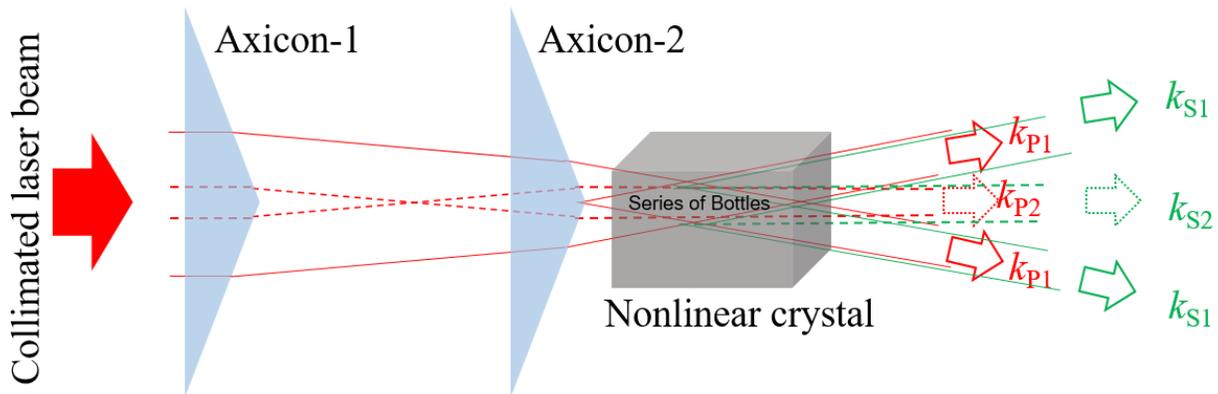

Fig. 8.3. Frequency doubling of Bessel bottle beams via second harmonic generation.

## 9. Applications of Bessel beam

The unique features of BB have gotten much attention in the optical community and indeed many applications in multiple research fields. The credentials of BB with reference to Gaussian, LG, and HG beams are the smallest spot size with the longest focal depth and self-healing. The BBs carry helical wave-front quantified with OAM with non-diffraction and self-healing. These advantages of BB are useful for light applications in modern science technology [302]. As discussed in Section 3, even though we have several ways to generate BB, each one has its advantages and disadvantages in its functioning. Hence, from an application point of view, we have to choose a proper method of generation of BB based on demand. Further, we can use the tuning of the intensity distribution in BB (discussed in section 4) to make it more user-friendly. Out of all light properties, wavelength plays a major role in light-matter interaction. The wavelength of BB is restricted by virtue of discrete emission lines of lasers in the limited region of the electromagnetic spectrum due to the characteristics of available laser gain mediums. Also, the method of generation and diffractive optical elements used for the generation of BB are wavelength-dependent. To overcome this difficulty in the application of BB, we can use nonlinear wave-mixing to tune the wavelength of BB (section 8). The versatility in the generation and tuning of BB's properties has opened a wide range of applications in multiple scientific fields. Hoverer, the majority of applications of BBs belong to light-matter interaction. Here, we discussed some of the major applications of BB in recent years and the details are given below.

*9.1. Material processing*
One of the key trends in laser material processing is the structured light-matter interaction. In that LG beams are the most popularly known tools in structured light-based material processing. However, more exotic beams with unique properties like BBs can be beneficial too. Also, the influence of focusing aberrations due to convex lenses in BBs is much smaller than that of the LG beams. The energy to the central lobe of BB is provided by virtue of its side rings along the propagation and this property plays a pivotal role in its material applications.

High-power ultra-fast BBs are generally accessible through axions and have seen tremendous advantages in material processing. In materials, they can create micro and nanoscale complex structures with utmost precision and reduce collateral damage in 2D and 3D [303-306]. The properties of these structures can be produced in the required direction by means of proper selection of diffractive optical elements and characteristics of the ultra-fast laser beam. The ultrashort time envelope with micro-size needle structure of BB via single and multi-photon absorption processes can fabricate various micro and nanostructures in the materials [307,308]. The structures formed in the materials are considerably smaller than the beam waist, enabling structuring below the diffraction limit and making accessible the nanoscale on account of cavitation and material rupture due to stress gradients [309-311]. Also, micron-sized and sub-micron-sized structures can be patterned on material surfaces, with the depth adjustable by the beam localization with respect to the surface with the aid of BB's range and position tuning [312]. The helical wave-front of higher order BBs can imprint chiral structures in the interacting material and this chirality can be controlled in a required amount by the selection of BB's order. The circular intensity distributions with central dark core in higher order BBs agglomerate the material in the dark core region to create micro-sized needle structures.

The high-power BBs have been used for laser cleaving and cutting [313,314]. The long depth of focus of BB provides sharp cutting of thick materials in a required shape with high quality. The micro-size central lobe of BBs can be efficiently used in material joining and welding on microscales, where energy must be precisely transported at the interface [315,316]. The non-diffracting and self-healing petal structure, formed by the superposition of two BBs of a different order, creates the rotation in its petal structure due to the Gouy phase difference. The petal structure can further rotate in a controlled manner either by a time-dependent hologram of SLM or by providing variable phase delay between the interfering BBs. The rotating petal beam resembles a spinning mechanical drill and is called an optical drill beam. These optical drill beams can be used for micro-size drills within an optically active medium. The characteristics of the optical drills like length, radius, and winding period can be easily controlled by means of input parameters of the petal beam. In recent years, fabricated hybrid structures with high quality for 3D photonics by BBs [317-319]. The BBs with femtosecond pulsing are used for high aspect ratio taper-free microchannel fabrication [320-322].

High-quality microfabrication based on polymerization is carried out in photosensitive materials with single-photon and two-photon absorption in the presence of BB [323-325]. Photosensitive materials have high photo-responsive quality. These materials exhibit a chemical/physical reaction while they are illuminated with a light beam. Therefore, we can easily fabricate the required structure within them with low-power BBs. The non-diffracting nature of BB forms photo-polymerized fibers, hollow fibers, or rotationally symmetric structures with micron-range diameter. Self-trapping and self-focusing of the propagation-invariant nature of higher order BBs in a photopolymer provides the fabrication of extended helical micro-fibers with a length scale of a centimeter [323]. This length is more than an order of magnitude larger than the range of the BB. These micro-fibers rotate at a rate proportional to the incident optical power, but the periodicity of the helical structures remains constant. In the presence of two-photon polymerization, flexible fabrication of flowerlike micro-structures and micro-tubes with slits are created by illuminating the material with superposed higher order BBs [326]. Further, multiple photo-polymerization structures can be created in a single shot by using a BB array [327].

*9.2. Trapping*
Optical trapping is a process of light-matter interaction to manipulate micro-objects by virtue of momentum transfer from the light to particles [328,329]. The optical trapping or tweezing is established by forces caused by reflection, refraction, and absorption of the laser beam at the particle. A high gradient of the optical power density of laser beams at the focus produces adequate gradient force to trap the particles. In the trapping technique, we can hold and move microscopic particles in a controlled manner by modulating the interacting light shape. For example, the structural modulations produced in BBs (discussed in sections 3 and 6) offer several types of particle trapping and manipulation.

In trapping and manipulating particles, the focused Bessel trap offers highly superior capabilities over conventional Gaussian beams for manipulating individual glass beads in 3D [330,331]. The line of focus formed by $0^{th}$ order BB can rotationally align rod-like particles along the beam axis to form a stack of particles within its non-diffraction region [332]. Spatially periodic optical potentials created by Bessel rings can be used to sort dielectric micro-particles with reference to their size, shape, or refractive index [333]. Frozen waves, created by the superposition of multiple co-propagating BBs with the same frequency and order, were used to build stable optical trapping with greater stability [334]. Frozen waves have been used for trapping and guiding micro-particles in different transverse planes along the beam propagation. While $0^{th}$ order BB provides stable trapping of particles, higher order BBs produce particle trapping and rotation owing to their helical wave-front [335,336]. In the case of higher order vector BBs, the total angular momentum comes from the combination of Spin Angular Momentum (SAM) and OAM, transferred to the dielectric particles [337,338]. The rotation period can be controlled with the optical power of BB. The on-axis intensity modulations created by various techniques, discussed in section 3, in the required manner, can be potentially used in the controlling periodicity of rotation.

Optical trapping of micro-size particles in multiple optical planes along the beam axis has been carried out using BBs owing to their non-diffraction and self-healing properties [339]. The schematic diagram of particle trapping in two parallel planes which are perpendicular to the beam propagation is shown in Fig. 9.1. The distortion in BB due to the particles in the first plane (chamber-I) is self-healed and acquires its original intensity profile after a certain propagation distance, which coincides with the separation distance of the cells. Again, the BB can be utilized to trap the micro-particles in the next plane (chamber-II). The separation between the two planes can be procured by measuring the self-healing distance of BB for a given particle size. The chamber separation $d$ must be greater than the self-healing length $z_{sh}$ of BB. Similarly, we can use BBs in other applications where the BB interaction and self-healing lengths are smaller than the Bessel range.

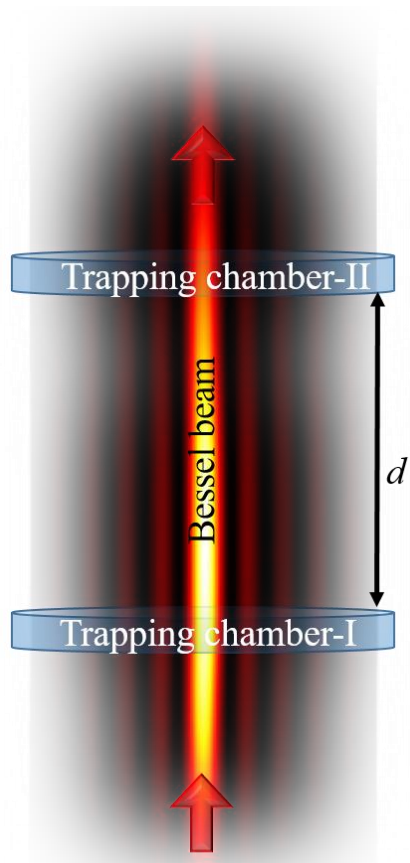

Fig. 9.1. Schematic diagram of trapping of particles in multiple parallel planes along the propagation of Bessel beam. The separation between the chambers is $d$ which must be greater than the self-healing length of the Bessel beam $z_{sh}$.

An optical conveyor belt has been developed by a standing wave created from two counter-propagating non-diffracting beams where the phase of one of the beams can be changed. The conveyor belt provides trapping and subsequent precise delivery of several submicron particles over a distance of hundreds of micrometers [340].

Using a chain of optical bottles with self-healing and non-diffraction created in the Bessel profile can efficiently 3D trap dielectric spheres with a refractive index smaller than that of their surrounding medium in the multiple planes along the beam propagation [341-343]. The BBBs have self-healing and non-diffraction properties inherited from superposed BBs (full details can be found in section 6). Therefore, the series of 3D intensity nulls and bright spots along the propagation of BBB can create 3D trapping potentials for low and high refractive index particles in multiple planes [344].

As we discussed in section 6, perfect vortex modes generated by the Fourier transformation of BBs possess annular intensity profiles independent of topological charges. In optical manipulation with perfect vortex modes, the particles continuously move along the annulus due to the scattering force from the inclined wave-front and it is completely new from other trapping beams [345,346]. For instance, for a fixed order of perfect vortex, a single trapped particle motion shows the same local angular velocity independent of its azimuthal position.

### 9.3. Guiding

The long-range micro-size optical potentials created by BBs and Bessel-like beams can used to trap and guide micro-particles and atomic gas along the beam axis [347-351]. The experimental configuration shown in Fig. 9.2, was suggested by J. Arlt et al. [352] and can be used for particle guiding. As depicted in Fig. 9.2(*a*), bright optical potential can be created by pumping the axicon with a hole by a HoG beam. Depending on the size of the hole present in the axicon, we have to control the HoG mode size by simply changing its order to reduce the diffraction effects due to the edges of the hole. The atomic gas or micro-particles can be injected into the optical potential using the holes created on the laser beam guiding mirror and axicon. The position and range of optical potential created by the HoG beam can be controlled by its order (section 6). In a similar procedure, we can create dark optical potentials for atomic guiding by replacing the HoG beam with a Gaussian vortex beam [Fig. 9.2(*b*)]. In recent times it has been demonstrated that the transport of a microscopic dielectric particle distance in excess of 2 mm by an all-fiber Bessel-like beam generator [353].

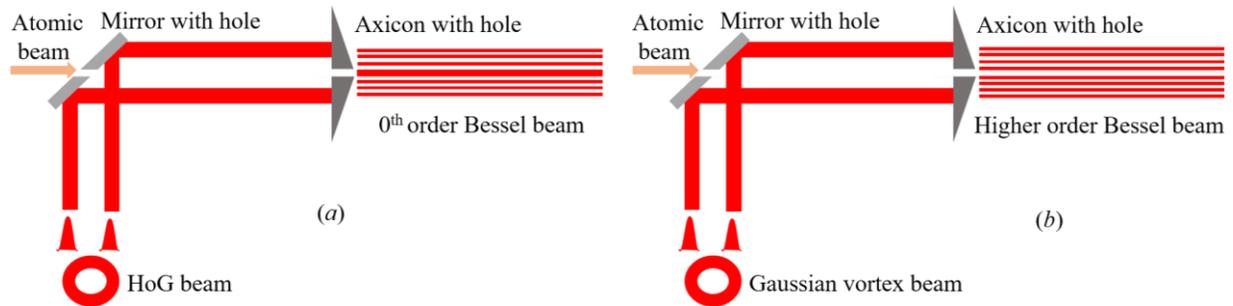

Fig. 9.2. Atomic guiding through (*a*) bright optical potential and (*b*) dark optical potential generated by respective Hollow Gaussian beam and Gaussian vortex beam pumping to the axicon with a hole.

### 9.4. Bio-imaging

Imaging technology based on laser beams in biology has seen tremendous applications. Various types of structured beams have been used in bio-imaging. However, BBs show excellent benefits over other structured modes in bio-imaging [354-356]. For instance, the BBs can prevent the effect of spherical aberration in imaging which generally occurs in a focused Gaussian beam in the presence of a convex lens. The position of BB formation can be easily controlled within the biological tissue. The signal to background noise created by the side lobes of BB can be reduced by engineering the BBs' generation technique. The BBs with high quality and structure tunability are widely available in the therapeutic window of bio-imaging. The distortion in BB due to scattering occurring in the imaging tissue can be healed through its reconstruction property. The non-diffraction of BB has an advantage in the deep imaging of biological tissues.

The model experimental configuration of microscopy generally used in imaging of biological tissues in the presence of BB is shown in Fig. 9.3. First, the collimated laser beam from the laser source will be divided into two parts: one is for Gaussian source and other is for Bessel source. The laser beam corresponds to the Bessel source will pass through the diffractive optical element (ex: axicon, SLM) which is designed to produce the required BB. Based on the requirement we can use one of the techniques used to generate BB which are discussed in sections. 3 and 4. The flip mirror FM present next to the axicon, can switch the probing beam between the Gaussian beam directed from two mirrors and BB. Further, multiple plano-convex lenses will be used for the Fourier transformation of BB and to project it with the required dimension on the biological sample. A single galvanometric mirror (GM) will be used for sample scanning in a required region. Before and after the illumination of the sample, two micro-objective lenses were used. The first one is to illuminate BB on the sample and the second one will collect the spontaneously emitted fluorescence signal from the sample. A couple of high-pass and low-pass filters can separate the fluorescence signal from the residual pump and detect it by a photomultiplier tube.

The long depth of field and self-healing property of the BB enable an increased penetration depth of the focused beam in tissues compared to a conventional Gaussian beam. These properties of BB improved the quality of deep-tissue microscopy in highly scattering and heterogeneous media. It is well demonstrated in fluorescence microscopy that low numerical aperture BBs exhibit reduced beam-steering artifacts and distortions compared to Gaussian beams and are therefore potentially useful for microscopy applications in which pointing accuracy and beam quality are critical, such as dual-axis confocal microscopy [357]. The BBs have been used to demonstrate high-resolution phase measurement of jewel scarabs, clarifying the light shaping in the cuticle [358]. In Single Photon Microscopy (SPM) with BB illumination, we encounter a significant signal-to-background ratio due to Bessel rings. However, we can suppress these side rings by engineering the Bessel hologram or by superposition of two suitable BBs [359].

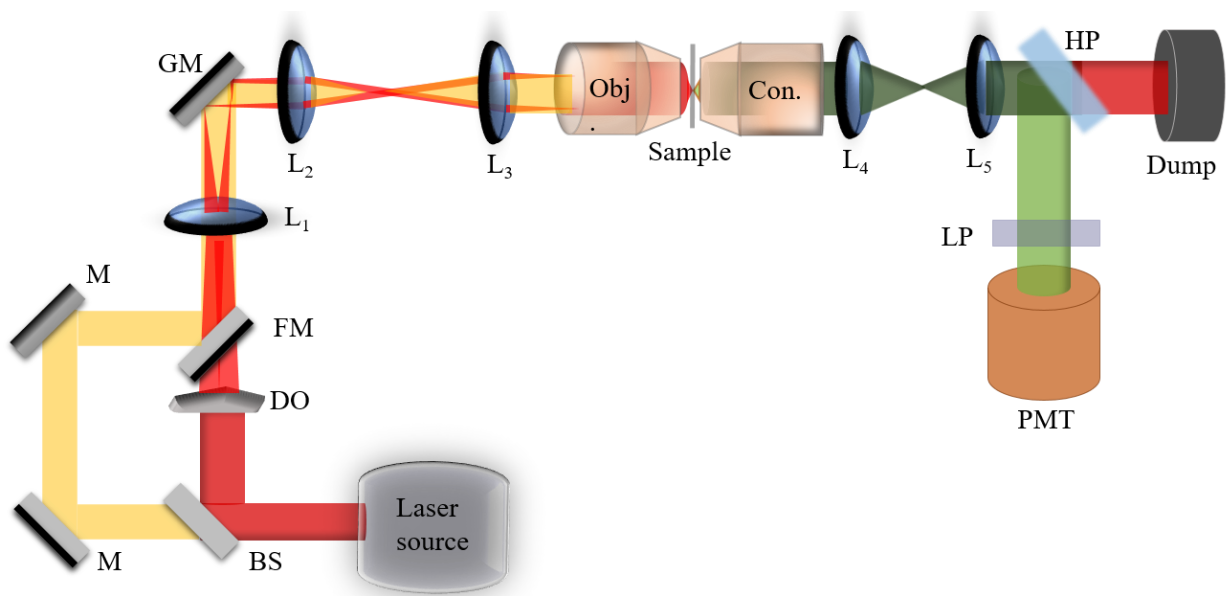

Fig.9.3. Schematic diagram of bio-imaging using Bessel beam as probing laser source in fluoresce microscopy. In the diagram, BS is the beam splitter, M is the mirror, DO is the diffractive optical element, FM is a flip mirror, $L_i$ is the plano-convex lens, GM is a galvanometric mirror, Obj. is the objective lens, Con. is a condenser, LP is a low-pass filter, HP is a high-pass filter, Dump is the power dumper, and PMT is a photomultiplier tube. Here red and yellow color laser beams represent Bessel and Gaussian beams.

The imaging quality of biological tissues in fluorescence microscopy can be further enhanced through nonlinear interaction. In nonlinear optical microscopy, image quality can be improved by simultaneous absorption of more than one photon through virtual energy levels and it is called Multi-Photon Microscopy (MPM). The MPM has unique advantages over SPM in bio-imaging applications as it can provide high-resolution images at depth with minimal background noise, and long penetration depth [360-362]. However, the image signal strength decreases with increasing its order, and to overcome this problem we have to pump with high optical power which can produce severe damage to the sample. Considering this experimental artifact, Two-Photon Microscopy (TPM) [363,364] is suitable for most biological tissues and has gained much attention in recent years. High axial resolution becomes a significant merit of the TPM owing to its highly localized fluorescent emission. Further, the drawbacks of MPM can be overcome by utilizing BB as a laser source owing to its reconstructing and non-diffraction propagation. As shown in Fig. 9.4(*a*), the BB has a longer needle-shaped intensity distribution with low-intensity side lobes. In MPM, the long depth of focusing with low-intensity side rings of BB has improved the image quality in its way owing to the low absorption of side lobes. The two-photon fluorescence contribution from the side lobes of BB called signal to background noise is very negligible [365] which we can see in Fig. 9.4(*b*). Further, we can suppress the signal to background noise due to Bessel rings via Three-Photon Microscopy (ThPM) [Fig. 9.4(*c*)]. The absorption cross-section of MPA decreases with increasing the order of the nonlinear absorption and consequently, corresponding fluorescence emission decreases. From this, we can infer that the fluorescence due to low-intensity side rings of BB is negligible and we can successfully suppress it completely by increasing the order of MPA. For instance, the three-photon fluorescence in the 1$^{st}$ ring is approximately 5.3% that in the central lobe, while the two-photon fluorescence in the 1$^{st}$, 2$^{nd}$, and 3$^{rd}$ rings is approximately 27.2%, 15.2%, and 10.6% that in the central lobe, respectively [365]. It is also noted that the low absorption of Bessel rings leads to the enhancement in the depth of volumetric imaging. Therefore, the central needle structure of BB in the MPM produces a 3D volumetric image of biological tissues without any multiple scanning along the sample thickness [366,367]. Hence, MPM with BB as a laser source can provide high-quality 3D images of biological samples in a short time of scanning [368-370]. By controlling the position and range of BB via intensity tuning, we can procure 3D images within the sample in its selective region. Moreover, the size of the volumetric image can be increased by increasing the range of BB through the superposition of two BBs shown in Fig. 4.4. Depending upon the sample nonlinear absorption cross-section, thermal conductivity, emission cross-section, we must selectively choose the order of the MPA and find a suitable technique of BB generation.

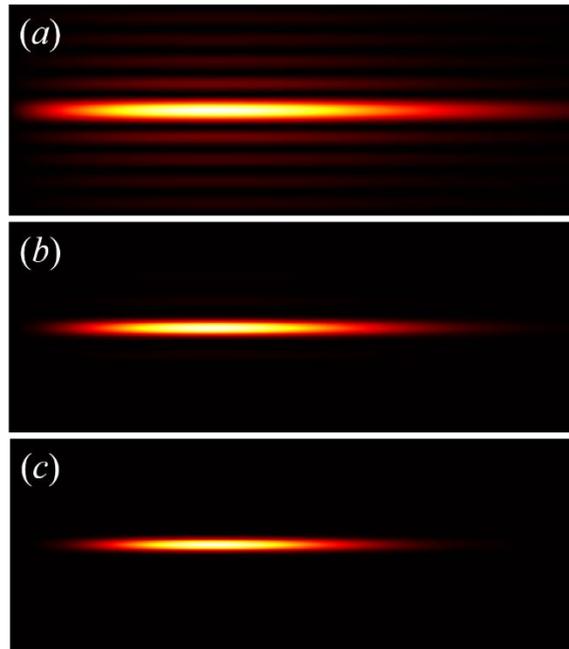

Fig. 9.4. (*a*) Incident intensity profile of Bessel beam in *xz*-plane. Point spread function of Bessel beam in (*b*) two-photon microscopy and (*c*) three-photon microscopy in *xz*-plane.

Another nonlinear microscopy is Second Harmonic Microscopy (SHM) which works based on SHG signal [371,372]. The SHM can be used as a 3D-resolved laser scanning technique by a tightly focused Gaussian beam scanned across the sample while SHG signals are recorded. By replacing the Gaussian beam with BB, we can achieve a 3D image without multiple number of scannings. In SHM, we can successfully prevent the signal-to-background ratio due to Bessel rings and it works in a similar way to MPM. This technique can be used only for the materials that are actively participating in SHG. N. Vuillemin et al. demonstrated that BB excitation combined with spatial filtering of the harmonic light in wave vector space can be used to probe collagen accumulation more efficiently than the usual Gaussian excitation scheme [373].

The BBs have also been utilized in photoacoustic microscopy for imaging [374]. The short focal depth of a conventional Gaussian beam limits the volumetric imaging speed of photoacoustic microscopy due to its short focal depth. This problem can be easily overcome by utilizing the long depth of focus of BBs. In optical resolution photoacoustic microscopy, the Bessel rings deteriorate image quality when the BB is directly employed to excite photoacoustic signals. However, a nonlinear approach based on the Grueneisen relaxation effect can be used to suppress the Bessel rings' artifacts in photoacoustic imaging [375]. Y. Zhou et al. have developed a hardware-software combined approach by integrating Bessel-beam excitation and conditional generative adversarial network-based deep learning. Side-by-side comparison of their approach against the conventional Gaussian-beam multi-parametric photoacoustic microscopy shows that the new system enables high-resolution, quantitative imaging of $C_{Hb}$, sO$_2$, and cerebral blood flow over a depth range of ~600 $\mu$m in the live mouse brain, with errors 13–58 times lower than those of the conventional system [376].

Next, Light-Sheet Microscopy (LSM) with BB as a probing laser source can also provide high-quality images in linear and nonlinear regimes [377-379]. The LSM has been widely used in high-speed fluorescence imaging with low phototoxicity. The BB-based LSM greatly enhances the light-sheet length with the self-reconstruction ability. The advantage of BBs is that they can self-reconstruct their initial beam profile even in the presence of massive phase perturbations and can propagate deeper into inhomogeneous media. This ability has crucial advantages for light sheet-based microscopy in thick media. Similar to other BB-based imaging techniques, the side rings of BB will create a significant out-of-focus background in LSM, and it will lead to limiting the axial resolution of the imaging system. However, H. Jia et al. overcome this issue by scanning the sample twice with zeroth-order BB and another type of propagation-invariant beam, complementary to the zeroth-order BB, which greatly reduces the out-of-focus background created in the first scan [380]. Also, B. Xiong et al. have introduced a photobleaching imprinting technique in BB-based LSM. By extracting the non-linear photobleaching-induced fluorescence decay, they could get rid of the large concentric side lobe structures of the BB to achieve uniform isotropic resolution across a large field of view for large-scale fluorescence imaging [381]. F. O. Fahrbach et al. proposed a detection method that minimizes the Bessel ring's negative effect on the image quality. They block the photons emitted from the BB's ring system in the fluorescence detection process. To remove background light from the BB's rings, they record the image line-wise so that the final image contains only that part of the object illuminated by the BB's bright central lobe. By this method, they could improve the image quality by nearly 100% relative to the standard light-sheet techniques using scanned Gaussian beams [382]. Two-photon fluorescence excitation can be used in BB-based LSM to reduce the effect of side lobes on image quality and can lower the phototoxicity with improved penetration depth [383] but its narrow excitation range restricts its applications. However, this difficulty was successfully overcome by proposing simple illumination optics, a lens-axicon triplet composed of an axicon and two convex lenses, to generate longer extent BBs [384].

Besides, another interesting imaging technique that is based on the coherence of illuminating light is Optical coherence tomography (OCT). The OCT can produce 2D and 3D images within the optical scattering media in the presence of a low-coherence light beam and it can have micrometer resolution. It offers a non-invasive, micrometer solution in bio-imaging. The BB has multiple benefits over the Gaussian beam in OCT due to its long depth of focus with diffraction-limited spot size. Multiple works have been carried out to enhance CT imaging with BB [385-390].

The final one is Stimulated emission depletion (STED) optical microscopy, formed by simultaneous probing of the Gaussian beam and doughnut shape beam, which breaks the barrier of the Abbe diffraction limit, enabling the visualization of subcellular structure [391]. The STED optical microscopy can suppress the spontaneous emission from the outer region of the Gaussian beam due to the excited state absorption created by the doughnut beam. Deep imaging STED microscopy can be created using a Gaussian beam for excitation and a higher order BB for depletion. W. Yu et al. have used this scheme and showed an improved imaging depth of up to about 155 $\mu$m in a solid agarose sample, 115 $\mu$m in polydimethylsiloxane, and 100 $\mu$m in a phantom of gray matter in brain tissue with consistent super-resolution, while standard STED microscopy shows a significantly reduced lateral resolution at the same imaging depth [392]. L. Gao et al. have demonstrated that rapid 3D dynamics can be studied beyond the diffraction limit in thick or densely fluorescent living specimens over many time points by combining ultrathin planar illumination produced by scanned BBs with super-resolution structured illumination microscopy [393].

*9.5. Optical communication*

Optical communication systems have been a popular tool in modern human life for transmitting information due to their high reliability and high usability. Information transfer through optical communication is limited by the wavelength bandwidth of the light. This can be overcome by OAM in the optical beams. The orthogonality nature of the OAM can extend the capacity of the information transfer and the capacity of communication can be increased by increasing the number of OAM states in the optical beams. The LG beam and BB have quantized OAM and are well-suitable for optical communication with an extended capacity of data encryption. The data capacity can be further enhanced by the combination of OAM and SAM. The self-healing and long-distance propagation without any diffraction of BB have excellent advantages in optical communication. The non-diffraction nature of BB is suitable for information encryption in optical communication because it does not rely on the distribution of the amplitude, phase, or even polarization of the spatial frequency range [394,395]. Also, BBs are shown to have several benefits over other structured modes when they propagate through atmospheric turbulence [396-399]. Optical beam wander is one of the most important issues for free-space optical communication. In BB-based optical communication, it has been shown that higher order BBs have the advantage of mitigating the beam wander in OAM multiplexing free-space communication [400]. For optical communication, we must create long-range propagation BBs. In recent years, attempts have been made to create long-range BBs. For example, long-range BBs with a phase-only element encoded on a spatial light modulator and imbued them with OAM generated and demonstrated their ability in optical communication in a real turbulent atmosphere [401]. A ring-shaped annular lens and a spherical lens in 4*f*-configuration were used and successfully produced long-range BBs [402].

## 10. Conclusion

This review article is written on the ideas and implementations used in BBs understanding by various researchers from the past 36 years. This tutorial review covers the development history from fundamental theories to tunable methods of generations used for BBs and then to widespread scientific and industrial applications. We have given a detailed conceptual analysis of the origin of BBs in experimental realization and theoretical concepts. Practically all the investigations and analyses utilized here to understand the structural properties of BB are almost exclusively based on ray optics and wave optics. The focused optical waves by means of the axicon phase are considered to be a set of secondary plane waves rather than the secondary spherical waves which are generally used to understand the focused waves from conventional lens systems.

In this review article, we first reviewed the basic concept, theory, and experiments of BB and emphasized the unique properties related to non-diffraction, self-healing, phase singularities, and polarization singularities. Then, we reviewed the recent advances in tunable BBs, where the tunability includes not only wavelength tunability but also 3D spatial structure tunability. We have bestowed detailed information on how the versatility in the experimental generation of BB has gained an advantage in its applications in multiple scientific and industrial fields. We discussed focused applications of BB. Finally, the conclusions that are drawn from our tutorial review are illustrated below.

The basic principle behind the experimental realization of BB is encountering the diffraction effects with interference to produce a propagation-independent transverse beam profile. Indeed, inevitable consequences of the diffraction effect in the propagation of light have been successfully suppressed by self-interference to achieve a finite range of self-similarity in its structure. Overall in BB generation techniques, axicon-based BB generation is most suitable for several applications owing to its simplicity, robustness, high-power handling, and wavelength versatility. The beam shaping with conventional axicons generally encounters difficulty, which is the processing of the tip, which cannot be infinitely sharp. The axicon defect at its apex position generates deleterious oscillations of the on-axis intensity. However, we can overcome this defect by the Gaussian beam transforming into a HoG beam before the axicon. The SLM is the suitable diffractive optical element for the generation of arbitrary order BB containing low optical power. The most popularly used techniques to generate BBs are based on either axicon with SPP or SLM.

When it comes to the properties, BBs appear to be immune to diffraction over finite propagation distances due to constructive and distractive interference of individual optical waves in the '×' shape/conical shape propagation. Non-diffraction and self-healing properties of BBs have made them strong in several applications over other structured modes. 1D Cosine beams are equivalent to BBs in one-dimensional non-diffraction. Overall structured modes, LG beams, and BBs have helical wave-front with well-defined OAM. Moreover, the non-diffraction nature of helical wave-front in BB has shown a unique advantage in scientific and industrial applications which we cannot fulfill with LG beams. The well-developed diffractive optical elements can be successfully used to modulate the properties of BB to make it a suitable optical probe for investigation of material properties.

In addition to conventional BBs, we can generate various kinds of Bessel-like beams, by various techniques, which can have some of the properties of BB. The Bessel-like beams will have convergence if the radial propagation vector $k_r$ increases with propagation distance, thus, the Bessel spot size decreases with the propagation. In the case of a divergence Bessel-like beam, the $k_r$ decreases with propagation distance and it leads to an increase in Bessel spot size as we move farther from its origin. Generally, for given fixed experimental parameters, the converging Bessel-like beams have a smaller range of propagation as compared with conventional BBs, and on the other side, the diverging Bessel-like beam has a larger propagation range with reference to the conventional BBs. Superposition of two zeroth order BBs can be used to generate a series of non-diffracting and self-healing optical bottles along the propagation. The dimensions of these bottles can be controlled by changing the properties of interfering BBs. Order-independent size vortex modes called perfect vortex modes can be achieved by the Fourier transformation of BB.

The vector modes generated in Bessel profiles will have non-diffraction and self-healing properties. The CVMs and πVMs come under V-point polarization singularities have the gradient polarization change in their cross-section and have a stable polarization structure in their entire propagation due to the superposition of BBs with the same mode size. The C-point polarization singularity modes also called PMs formed in BBs have the propagation-dependent polarization distribution in the conventional superposition technique.

The wavelength of BBs can be tuned to any particular wavelength for demanded application through nonlinear wave-mixing. The phase matchings involved in the nonlinear wave-mixing of BBs have a completely different nature from their involvement in other structured modes. The way the phase matching is involved in the nonlinear wave-mixing of BBs is completely different from the other structured beams.

The non-diffraction, self-healing, and long depth of focus with sub-wavelength central lobe of BBs have unique applications in Material and bio-science applications which cannot be fulfilled with other structured optical beams.

It is always difficult for new researchers to remember the basic concepts and formulae in any research field. Multiple theories and experiments, different kinds of notations, styles of writing, and analyses used in well-developed research fields create ambiguity for the new researchers in their understanding. To overcome these circumstances in the understanding of BB and preventing ambivalence, below we have given detailed information on the BBs in key points from our review.

1. BBs are self-interfered modes.
2. BBs have non-diffraction and self-healing.
3. BB keeps both the transverse field amplitude and the phase distribution almost constant while it is propagating.
4. The wavelength of a BB, measured along the optical axis, differs from that of a usual collimated diffracting beam of the same optical frequency.
5. BB carries an equal amount of energy in all the lobes.
6. To experimentally generate the BB, the necessary condition is to pump the corresponding diffractive optical element with a plan wave-front laser beam which is generally obtained in the experiment with collimation in the limit of paraxial approximation.
7. The on-axis intensity distribution of BB along its propagation depends on the radial intensity distribution of the pump mode used in its generation.
8. Ideal BBs have infinite energy with an infinite range of spatial extension and non-diffraction propagation. However, experimentally we can achieve quasi-BB which has these properties in a finite range.
9. The optical phenomenon behind all non-diffracting beams is interference along the beam propagation axis.
10. The experimental realization of BB is based on wave-front division interference.
11. In the BB, individual optical waves propagate in the shape of "X" and the crossing area is BB.
12. 1D non-diffracting beams formed by Fresnel biprism are cosine beams and 2D non-diffracting beams produced by axicon are BBs.
13. Ideal BB intensity is constant throughout its propagation [$I_l(r, z_1) = I_l(r, z_2)$]. Experimentally realized BB's intensity along their propagation is attributed to the radial intensity of the optical beam used to generate it [$I_l(r, z_1) \neq I_l(r, z_2)$].
14. In the $0^{th}$ order BB, the peak intensity ratio of the central lobe and the first ring is 6.
15. In $0^{th}$ order BB, half width of the central peak is approximately $k_r^{-1}$, and the transverse skirt of distribution decays as $r^{-1}$ (intensity profile).
16. The intensity distribution in the beams' cross-section for BB depends on the propagation vector ($k_r$), and for BGB it depends on $k_r$ and Gaussian beam waist ($w_0$).
17. The first experimental demonstration of BB is carried out by Durnin with an annular aperture technique.
18. Annular aperture technique cannot be used to generate higher order BBs.
19. The radial propagation vector, $k_r = k\sin\theta$ and longitudinal propagation vector, $k_z = k\cos\theta$ (tan $\theta = k_r/k_z$). Here, propagation vector $k = 2\pi/\lambda$, and the half-angle of Bessel cone $\theta = d/(2f)$, $d$ is the annular slit mean diameter, $f$ is the focal length of the Fourier lens and $\lambda$ is the wavelength of light.
20. Bessel range formed in annular aperture method is limited to $z_{range} = \lambda f^2/\Delta r_{12}$. Here, $\Delta$ and $r_{12}$ are the width and mean radius of the annular aperture.
21. In annular aperture-based BB generation, the on-axis intensity modulation is negligible within a finite output aperture $R$ (lens radius), provided that width of the annular slit $\Delta d \ll \lambda f/R$.
22. According to Durnin's geometrical optics prediction, the Bessel range, $z_{max}=R/\tan\theta$; $R$ is the smaller of either the radius of the lens or the effective radius of the diffraction pattern that is cost onto the lens.
23. The radius of the central lobe (minimum intensity reference i.e., zeros of BB) of $0^{th}$ order BB is $r_0 = 2.405/k_r$.
24. Radial phase retardation used in SLM to generate BB is exp($-i2\pi r'/r_0$) and $r_0$ is constant.
25. The Bessel offset position, $z_{max}$ for SLM-based hologram is $z_{max} = r_0 R/\lambda$ and $R$ is the radius of the Bessel hologram.
26. SLM can produce any arbitrary order BB with low optical power.
27. Axicons are of two types: 1) refractive axicon and 2) reflective axicon. The reflective axicon is most commonly known as axicon due to so many advantages over reflective axicon.
28. Axicons alone can generate high-power $0^{th}$ order BBs with conversion efficiency but not higher order.

29. In case of Gaussian mode pumped axicon based $0^{th}$ order BB generation, the range of $0^{th}$ order BB is given by $z_{max} = w_0 (k/k_r) \approx w_0/\theta$ from the axicon tip. Here, $\theta = (n-1)\alpha$, $n$, and $\alpha$ are the refractive index and opening angle (base angle) of the axicon and $w_0$ is the Gaussian spot size of the laser beam at the axicon. The axicon base angle $\alpha$ and apex angle, $\gamma$ are related by $2\alpha + \gamma = 180$.
30. The phase retardation due to the axicon is $\exp(-ik_r r)$
31. The peak position of the BB along the beam axis is given by $z_{peak} = (2l+1)^{½} z_{max}/2$. Here, $l$ is the order of the BB.
32. The focal length of axicon changes with the order of BB and it is given by $f_a = z^G_{max}(2l+1)^{1/2}/2$.
33. HoG beam pumped axicon is the best way to generate the high-power $0^{th}$ order BBs with tunable range with no on-axis intensity modulations.
34. The peak position of the $0^{th}$ order BB in the presence of the HoG beam pumping is $z_{peak} = (4m+1)^{½} z_{max}/2$. Here, $m$ is the order of the HoG beam.
35. The onset position ($z^m_{min}$) and offset position ($z^m_{max}$) of BB are given by $z^m_{min} = r_1/(n-1)\alpha$ and $z^m_{max} = r_2/(n-1)\alpha$ respectively
36. The minimum self-healing distance after the obstacle beam reconstruction is $z_a = ak/2k_z$, here $a$ is the width of the obstruction from the beam axis.
37. The fringe visibility in the real BB is not a constant and it is a function of transverse and longitudinal positions.
38. BB has the 3D polarization distribution even in the case of the pump source is linearly polarized.
39. BB polarization distribution, spot size, range, and shape can change while it is propagating through different refractive index mediums.
40. Bessel-like beams can have a propagation range which either larger or smaller than the conventional BBs for a given experimental configuration.
41. Bessel-like beams are diffraction beams.
42. Optical bottles created in Bessel profiles are self-healing and non-diffracting.
43. Fourier transformation of BB is a perfect vortex.
44. The significant properties of perfect vortex depend on the transverse intensity distribution of BB.
45. The polarization texture created in VBBs has self-healing and non-diffraction.
46. BBs can be used for long-depth of micro-size material processing.
47. BBs can be used for trapping particles in multiple parallel planes along the propagation.
48. By using BB, we can achieve a 3D volumetric image of biological tissues without any multiple scanning.